%% file: main.tex
\documentclass[
11pt, 
english, 
singlespacing, 
headsepline, 
]{MastersDoctoralThesis} 

\usepackage[utf8]{inputenc} 
\usepackage[T1]{fontenc} 
\usepackage{pdfpages}
\usepackage{mathpazo} 

\usepackage[style=numeric, natbib=true, sorting=none]{biblatex} 

\usepackage[autostyle=true]{csquotes} 

\usepackage{graphicx}

\usepackage{caption}
\usepackage{xcolor,colortbl}

\usepackage{amsmath}

\usepackage{soul}

\DeclareMathOperator*{\argmax}{argmax}

\thesistitle{Studies of Critical Phenomena in Causal Dynamical Triangulations on a Torus} 
\supervisor{Prof. dr hab.  Jerzy \textsc{Jurkiewicz} \\ Dr. hab.  Jakub \textsc{Gizbert-Studnicki}} 
\examiner{} 
\degree{Doctor of Philosophy} 
\author{Dániel \textsc{Németh}} 
\addresses{} 

\university{\href{https://en.uj.edu.pl/}{Jagiellonian University}} 
\department{\href{https://fais.uj.edu.pl/en_GB/the-institute-of-physics}{Jagiellonian University}} 
\group{\href{.}{Faculty of Physics, Astronomy and Applied Computer Science}} 

\faculty{\href{https://fais.uj.edu.pl/}{Faculty of Phyiscs, Astronomy and Applied Computer Science}} 

\AtBeginDocument{
\hypersetup{pdftitle=\ttitle} 
\hypersetup{pdfauthor=\authorname} 
\hypersetup{pdfkeywords=\keywordnames} 
}

\usepackage{xpatch} 
\makeatletter
\xpatchcmd{\@bibitem}
  {\item}
  {\item[\@biblabel{\changekey{#1}}]}
  {}{}
\xpatchcmd{\@bibitem}
  {\the\value{\@listctr}}
  {\changekey{#1}}
  {}{}
\makeatother

\ExplSyntaxOn
\cs_new:Npn \changekey #1
 {
  \str_case:nVF {#1} \g_changekey_list_tl { ?? }
 }
\cs_new_protected:Npn \setchangekey #1 #2
 {
  \tl_gput_right:Nn \g_changekey_list_tl { {#1}{#2} }
 }
\tl_new:N \g_changekey_list_tl
\cs_generate_variant:Nn \str_case:nnF { nV }
\ExplSyntaxOff

\setchangekey{pub1}{1}
\setchangekey{pub2}{2}
\setchangekey{pub3}{3}
\setchangekey{pub4}{4}
\setchangekey{pub5}{5}
\setchangekey{pub6}{6}

\begin{document}

\frontmatter 

\pagestyle{plain} 


\begin{titlepage}
\begin{center}

\vspace*{.06\textheight}
{\scshape\LARGE \univname\par}\vspace{1.5cm} 
\textsc{\Large Doctoral Thesis}\\[0.5cm] 

\HRule \\[0.4cm] 
{\huge \bfseries \ttitle\par}\vspace{0.4cm} 
\HRule \\[1.5cm] 
 
\begin{minipage}[t]{0.4\textwidth}
\begin{flushleft} \large
\emph{Author:}\\
\href{http://www.johnsmith.com}{\authorname} 
\end{flushleft}
\end{minipage}
\begin{minipage}[t]{0.4\textwidth}
\begin{flushright} \large
\emph{Supervisors:} \\
\href{http://www.jamessmith.com}{\supname} 
\end{flushright}
\end{minipage}\\[3cm]
 
\vfill

\large \textit{A thesis submitted in fulfillment of the requirements\\ for the degree of \degreename}\\[0.3cm] 
\textit{in the}\\[0.4cm]
\groupname\\\deptname\\[2cm] 
 
\vfill

 
\vfill
\end{center}
\end{titlepage}


\begin{declaration}
\addchaptertocentry{\authorshipname} 
\vspace{1cm}
Ja, niżej podpisany, \authorname, (nr indeksu: 1159847), doktorant Wydziału Fizyki, Astronomii i Informatyki Stosowanej Uniwersytetu Jagiellońskiego oświadczam, że przedłożona przeze mnie rozprawa doktorska pt. "Studies of Critical Phenomena in Causal
Dynamical Triangulations on a Torus" jest oryginalna i przedstawia wyniki badań wykonanych przeze mnie osobiście, pod kierunkiem prof. dr hab. Jerzego Jurkiewicza i dr hab. Jakuba Gizbert-Studnickiego. Pracę napisałem samodzielnie.\vspace{0.35cm} \\

Oświadczam, że moja rozprawa doktorska została opracowana zgodnie z Ustawą o prawie autorskim i prawach pokrewnych z dnia 4 lutego 1994 r. (Dziennik Ustaw 1994 nr 24 poz. 83 wraz z późniejszymi zmianami). \vspace{0.35cm}\\

Jestem świadomy, że niezgodność niniejszego oświadczenia z prawdą ujawniona w dowolnym czasie, niezależnie od skutków prawnych wynikających z ww. ustawy, może spowodować unieważnienie stopnia nabytego na podstawie tej rozprawy.
\vspace{1cm}\\
 
\noindent Kraków, ............................... \hfill .......................................................... \\
\hspace{1.7cm}(data)\hfill (podpis doktoranta)
 
\end{declaration}

\cleardoublepage






\begin{abstract}
\addchaptertocentry{\abstractname} 
This document contains my publications and results based on  research done as a member of the Causal  Dynamical Triangulations (CDT) group at the Jagiellonian University during my PhD studies. The field of my research was the four-dimensional CDT, which is a lattice regularization of the theory of quantum gravity, based on the formalism of Regge Calculus and Feynman path integrals. Due to  mathematical complexity, analytical solutions to the model exists only in two dimensions. The four-dimensional theory is analyzed by numerical simulations. Earlier discoveries include  dynamically emergent quantum de Sitter universes with emergent four-dimensional properties, scale-dependent spectral dimensions and a complex phase structure in which first- and higher-order phase transitions were shown to exist. The document describes the nature of previously not yet analyzed phase transitions, new ways to analyze triangulations and the impact of classical and dynamical (quantum) scalar fields in four-dimensional CDT with toroidal spatial topology. The main results of the dissertation are the six publications attached to the last chapter. This document is intended as an introduction to CDT and serves as a guide to the papers comprising the doctoral thesis. 
\nocite{pub1,pub2,pub3,pub4,pub5,pub6}
\end{abstract}

\begin{abstract}
\addchaptertocentry{\abstractname} 
Niniejszy dokument zawiera moje publikacje i wyniki oparte na badaniach prowadzonych jako członek grupy Causal Dynamical Triangulations (CDT) na Uniwersytecie Jagiellońskim podczas studiów doktoranckich. Obszarem moich badań był czte- rowymiarowy model CDT, który stanowi sieciową regularyzację teorii kwantowej grawitacji, opartą na formalizmach rachunku Regge i całek po trajektoriach Feynmana. Ze względu na złożoność matematyczną, rozwiązania analityczne tego modelu istnieją tylko w dwóch wymiarach. Czterowymiarowa teoria jest analizowana przez symulacje numeryczne. Wcześniejsze odkrycia obejmują dynamicznie pojawiające się kwantowe wszechświaty de Sittera z emergentnymi właściwościami czterowymiarowymi, zależne od skali wymiary spektralne oraz skomplikowaną strukturę fazową, w której istnieją przejścia fazowe pierwszego i wyższego rzędu. Dokument zawiera opis natury nie analizowanych dotychczas przejść fazowych, nowych sposobów analizy triangulacji oraz wpływu klasycznych i dynamicznych (kwantowych) pól skalarnych w czterowymiarowym CDT o toroidalnej topologii przestrzennej. Głównymi wynikami rozprawy jest sześć publikacji załączonych w ostatnim rozdziale. Dokument ten ma na celu wprowadzenie do CDT i stanowi przewodnik po artykułach składających się na rozprawę doktorską.
\end{abstract}


\tableofcontents 




\begin{abbreviations}{ll} 

\textbf{ADM} & \textbf{A}rnowitt-\textbf{D}eser-\textbf{M}isner\\
\textbf{AS} & \textbf{A}symptotic \textbf{S}afety\\
\textbf{CDT} & \textbf{C}ausal \textbf{D}ynamical \textbf{T}riangulations\\
\textbf{CFT} & \textbf{C}onformal \textbf{F}ield \textbf{T}heory\\
\textbf{EDT} & \textbf{E}uclidean \textbf{D}ynamical \textbf{T}riangulations\\
\textbf{GFT} & \textbf{G}roup \textbf{F}ield \textbf{T}heory\\
\textbf{HLG} & \textbf{H}ořava–\textbf{L}ifshitz \textbf{G}ravity\\
\textbf{IR} & \textbf{I}nfra \textbf{R}ed \\
\textbf{LCDT} & \textbf{L}ocally \textbf{C}ausal \textbf{D}ynamical \textbf{T}riangulations \\
\textbf{LQCD} & \textbf{L}attice \textbf{Q}uantum \textbf{C}hrono \textbf{D}ynamics \\
\textbf{LQG} & \textbf{L}oop \textbf{Q}uantum \textbf{G}ravity\\
\textbf{MC} & \textbf{M}onte \textbf{C}arlo \\
\textbf{NCG} & \textbf{N}on-\textbf{C}ommutative \textbf{G}eometry \\
\textbf{OP} & \textbf{O}rder \textbf{P}arameter \\
\textbf{QCD} & \textbf{Q}uantum \textbf{C}hrono \textbf{D}ynamics \\
\textbf{QFT} & \textbf{Q}uantum \textbf{F}ield \textbf{T}heory \\
\textbf{QM} & \textbf{Q}uantum \textbf{M}echanics \\
\textbf{RG} & \textbf{R}enormalization \textbf{G}roup \\
\textbf{ST} & \textbf{S}tring \textbf{T}heory \\
\textbf{UV} & \textbf{U}ltra \textbf{V}iolet \\
\textbf{UVFP} & \textbf{U}ltra \textbf{V}iolet \textbf{F}ixed \textbf{P}oint \\

\end{abbreviations}













\dedicatory{To my daughter who cannot read yet\ldots} 


\mainmatter 

\pagestyle{thesis} 


\include{Chapters/Chapter0}
\include{Chapters/Chapter1}
\include{Chapters/Chapter2} 
\include{Chapters/Chapter3}

\include{Chapters/Chapter4} 
\include{Chapters/Chapter5}


\begin{acknowledgements}
\addchaptertocentry{\acknowledgementname} 
First of all I would like to thank for all the support to prof. Jerzy Jurkiewicz, who gave me the possibility to join to the research group and supervised me throughout the years, shared with me many of his thoughts and ideas which definitely influenced the way I see my field of research. Also to dr. Jakub Gizbert-Studnicki for all the discussions during my PhD. and for his patience and the significant amount of time he spent with advising my thesis. Special thanks to dr. Andrzej Görlich also for the discussions and constant technical help with the computers, codes and simulations. I would like to thank also to prof. Jan Ambjorn for all the fruitful discussions throughout our regular CDT meetings, as he shared his deep knowledge with us, my knowledge and understanding also improved. Additionally, I would like to express my gratitude to my wife, dr. Anna Francuz for her presence in my life, and for her support during my studies and her patience and understanding in those times, when I work in late hours instead of being there. Last but not least, I would like to thank to my daughter Eszter Eleonóra, for letting me sleep, sometimes....
\end{acknowledgements}


\appendix 

\include{Appendices/AppendixB}

\include{Appendices/AppendixA}



\printbibliography[heading=bibintoc]


\end{document}

%% file: Chapters/Chapter0.tex

\chapter{Motivation to the study of parallels, random geometry and quantum gravity} 

\label{Chapter0} 


\newcommand{\keyword}[1]{\textbf{#1}}
\newcommand{\tabhead}[1]{\textbf{#1}}
\newcommand{\code}[1]{\texttt{#1}}
\newcommand{\file}[1]{\texttt{\bfseries#1}}
\newcommand{\option}[1]{\texttt{\itshape#1}}

\definecolor{Gray}{gray}{0.85}
\definecolor{LightCyan}{rgb}{0.88,1,1}

\newcolumntype{a}{>{\columncolor{Gray}}c}
\newcolumntype{b}{>{\columncolor{white}}c}

\newcommand{\mc}[2]{\multicolumn{#1}{c}{#2}}
\textit{"You must not attempt this approach to the parallels. I know this way to the very end. I have traversed this bottomless night, which extinguished all light and joy of my life. I entreat you, to leave the science of parallels alone. For God’s sake, please give it up. Fear it no less than the sensual passion, because it, too, may take up all your time and deprive you of your health, peace of mind, and happiness in life. I thought I would sacrifice myself for the sake of truth. I was ready to become a martyr who would remove the flaw from geometry and return it purified to mankind. I accomplished monstrous, enormous labors: my creations are far better than those of others and yet I have not achieved complete satisfaction. I turned back when I saw no man can reach the bottom of this night. I turned back unconsolidated, pitying myself and all mankind. Learn from my example: I wanted to know about parallels. I remain ignorant, this has taken all the flowers of my life and all my time from me...."} - \textbf{A letter of Bolyai Farkas to his son Bolyai János}\\\\

Geometry, the mathematical study of shapes, always interested humans. From the ancient Greeks till today's science various topics related to geometry are key aspects to mathematics and natural sciences. Euclid laid down five axioms, which became the foundations of mathematics. At that time, mathematics was postulated in terms of words and rarely graphics, but not equations. The postulates of Euclid~\cite{postulates}, based on his axioms, defined geometry until the $19^{th}$ century. His postulates were:

\begin{itemize}
    \item  A straight line segment may be drawn from any given point to any other.
    \item A straight line may be extended to any finite length.
    \item A circle may be described with any given point as its center and any distance as its radius.
    \item All right angles are congruent.
    \item If a straight line intersects two other straight lines, and so makes the two interior angles on one side of it together less than two right angles, then the other straight lines will meet at a point if extended far enough on the side on which the angles are less than two right angles.
\end{itemize}

None dared to question the truth of these postulates as any sane person could check their truths by drawing those lines and not finding any which doesn't fit. This was true until some questioned whether it is possible to draw triangles on various non-flat shapes such that the sum of their inner angles is different than that of $\frac{\pi}{2}$. This is exactly what Bolyai Farkas is talking about in his letter to his son. He discovered that parallels can meet sometimes, but did not manage to describe the phenomena in its entirety even though he worked in that field for his whole life. Thus he warned his son not to pursue geometry and its parallels. But his son, János had his own ideas, and years later he constructed the basics of non-Euclidean geometry. Had he listened to his father, the topic of my doctoral thesis would be probably significantly different. Bolyai pursued a non-mainstream topic of mathematics and reached success with it. Later Riemann based his work on the work of Bolyai (and others), which was then used by Einstein when he worked out the general theory of relativity. The science of fundamental physics brings forth our knowledge of nature, if we wouldn't walk off-road in the theory space, but only follow the mainstreams, we wouldn't be able to solve the hardest problems of science.\\

At the beginning of the twentieth century, the appearance of two theories gave us an enormous leap toward understanding nature. The paradigm shift which is related on one hand to the curving relativistic four-dimensional spacetime described by general relativity (\textbf{GR}) and on the other hand to the discreteness of nature as it is seen by quantum mechanics turned science into science-fiction in the eye of the scientifically not educated people. Math and physics, needed in order to understand it, started to be so complex and demanding,  that scientific results became non-trivial. Many physical theories are validated or falsified via mathematical derivations and many cannot be accessed because of their mathematical complexity. The work presented in this thesis belongs to a similar off-road field, which is strongly related to parallels and geometry. Quantum gravity is the field where quantum mechanics and general relativity meets. Quantum mechanics is the theory that describes the smallest scales, the tiny fluctuations of matter, and the rules of nature that escape everyday experience, and gravity is the theory that describes the physics of the largest scales, the orbiting of planets, and even the earliest history of the Universe. Their intersection should be the theory of quantum gravity, the theory which describes how the attraction between bodies behaves on the smallest scales, on the scales where other forces of nature dominate and bodies fall apart to their components. As we advanced in our understanding of the world and the Universe it turned out that quantum gravity could potentially explain also the largest scales and the earliest moments of history. It could tell us why we have such a large-scale structure of galaxies that we see, could explain why visible matter constitutes only four percent of everything, could hint at whether we live in a closed or an open Universe, and foreshadow a potential cold death at the end of times. The quantum theory of gravity has the potential to explain the nature and the structure of space-time, to resolve singularities of GR, and furthermore to explain or disprove the theories regarding dark matter and dark energy.\\

After Einstein introduced GR many scientists tried to find the theory of quantum gravity without success. The first attempt to describe quantum gravity was a naive application of perturbative methods of QFT to GR, but it failed. The treatment of infinities by perturbative renormalization techniques, which can be used in the case of the standard model physics, cannot be applied to gravity, which turned out perturbatively non-renormalizable \cite{non_renorm_g}. However, S. Weinberg conjectured that gravity may adhere to an Asymptotic Safety (\textbf{AS}) scenario \cite{Weinberg, asym, asym_critiq}, where using Renormalization Group Flow (\textbf{RG}) techniques one may find a \textit{fixed point}, where there exist only a finite number of coupling constants needed to describe the full quantum  theory in a non-perturbative way. In a lattice formulation of a quantum theory, fixed points are typically connected to phase transitions, and the hypothesis is that there is at least one non-trivial fixed point for gravity related to the ultraviolet (\textbf{UV}) regime, which necessarily requires the existence of a higher order phase transition. Such a phase transition can be typically recognized from the diverging correlation length  and related scaling exponents.\\

By the end of the $20^{th}$ century, with the increasing available computational power, numerical algorithms became widely used. One of the most notable computer-based techniques in physics is related to Lattice Quantum Chromodynamics (LQCD) \cite{lqcd1,lqcd2}, which was developed in parallel with the physical experiments. The basic idea is to discretize the continuum theory such that the field variables are located at the vertices of a regular $D$-dimensional lattice ($D$ depends on the dimensionality of the discussed model). The lattice spacing $a$, which is the length between two adjacent vertices of the lattice, should be sent to zero while keeping the relevant physical observables constant, in order to reach the continuum limit within the numerical simulations. Since the beginning of the development of lattice theories, many physically relevant observations were derived from numerical simulations, e.g., related to phase transitions \cite{lqcd_chiral}, physical masses of particles \cite{lqcd_masses}, and many other phenomena. In contrast to  the LQCD, lattice quantum gravity is special in the sense that the lattice connectivity itself encodes the geometric degrees of freedom and therefore provides information about the distinct features of gravitational physics on the quantum level. In order to create a physically relevant model of  lattice quantum gravity one also has to be able to include matter fields, e.g., scalar fields or gauge fields \cite{2d_cdt_m,gauge_2d_cdt}.\\

This document is a guide to a collection of articles published in the past years and constituting my doctoral thesis. All of the publications were published in peer-reviewed journals. \\ 

The structure of this document is as follows: The introduction to Causal Dynamical Triangulations (CDT), which is a non-perturbative approach in the quest of quantizing gravity, is the topic of chapter two. In chapter three, I discuss some details of numerical implementation and Monte Carlo simulation methods used to study CDT. The fourth and fifth chapters discuss respectively the results of my studies obtained for empty Universes (pure gravity) and Universes with matter content (gravity coupled to scalar fields). Afterward, all publications which constitute my thesis are briefly discussed in chapter six, together with information about my contribution to them. The published papers are attached at the very end of chapter seven in the following order:

\begin{enumerate}
    \item[\cite{pub1}]  J. Ambjorn G. Czelusta et al. “The higher-order phase transition in toroidal CDT”. In: J. of High Energ. Phys. 2020 (5), p. 30.\\DOI: 10.1007/JHEP05(2020)030
    \item[\cite{pub2}] J. Ambjorn et al. “Towards an UV fixed point in CDT gravity”. In: Journal of High Energy Physics 2019 (7), p. 166.\\ DOI: 10.1007/JHEP07(2019)166
    \item[\cite{pub3}]  J. Ambjorn et al. “Topology induced first-order phase transitions in lattice quantum gravity”. In: Journal of High Energy Physics 2022 (4), p. 103.\\ DOI: 10.1007/JHEP04(2022)103.
    \item[\cite{pub4}] J.Ambjorn et al. “Cosmic voids and filaments from quantum gravity”. In: The European Physical Journal C 81 (8 2021), p. 708.\\ DOI: 10.1140/epjc/s10052-021-09468-z
    \item[\cite{pub5}] J. Ambjorn et al. “Matter-Driven Change of Spacetime Topology”. In: Phys. Rev. Lett. 127 (16 Oct. 2021), p. 161301.\\ DOI: 10.1103/PhysRevLett.127161301
    \item[\cite{pub6}]  J. Ambjorn et al. “Scalar fields in causal dynamical triangulations”. In: Classical and Quantum Gravity 38 (19 Sept. 2021), p. 195030.\\ DOI: 10.1088/1361-6382/ac2135
\end{enumerate}

%% file: Chapters/Chapter1.tex

\chapter{Causal Dynamical Triangulations} 

\label{Chapter1} 


\section{Introduction to Quantum Gravity}

\textit{"The beauty and clearness of the dynamical theory, which asserts heat and light to be modes of motion, is at present obscured by two clouds..."} - \textbf{Lord Kelvin}\\\\

Lord Kelvin wrongly predicted the end of physics in the late nineteenth century. The two clouds mentioned were the problem of heat and radiation, more precisely the theorized material that fills everything called "ether" and the black body radiation. When we mention modern physics, we refer to the time when the solutions to these two "clouds" were presented in the form of special relativity and quantum mechanics. The start of the twentieth century brought us an explosion of physical theories, as special relativity led to general relativity, which is extensively studied today in relation to astrophysical and cosmological models or technologies, such as GPS tracking devices. At the same time, quantum mechanics evolved into quantum field theory, and later our technological advancements led to the ability to measure the properties of particles. The standard model of particle physics is one of the greatest achievements in physics, as it gives an explanation of the fundamental nature of matter. The biggest problem of modern physics is that the theory of matter and the theory of gravity cannot be matched into a unified framework together. Many physicists tried in the past hundred years to describe the theory of quantum gravity, which led to many different research projects, such as Loop Quantum Gravity (LQG), String Theory (ST), Causal Sets (CS), Group Field Theory (GFT), Non-Commutative Geometry (NCG), Canonical Quantum Gravity (CQG), Hořava–Lifshitz Gravity (HLG), Asymptotic Safety (AS), Euclidean Dynamical Triangulations (EDT), Causal Dynamical Triangulations (CDT) and many other approaches. \\

\subsection{(Non-)renormalizability of quantum gravity and the need for non-perturbative approaches}

Merging the quantum theory with gravity is not a trivial task. Quantum field theory (QFT) predicts fluctuations of  fields, and according to GR and Einstein's field equations, where there is energy density, there is curvature. These fluctuations can at very high energies produce such a large energy density in a small volume that the naive application of Einstein's equations would predict the appearance of black holes\cite{scale_grav,scale_grav2}. The problems with UV-completion of quantum gravity become apparent in the perturbative expansion of a QFT based on GR. Such a formulation is perturbatively non-renormalizable \cite{non_renorm}, which means, that the naive application of the perturbation theory would result in infinitely many parameters and coupling constants appearing in the theory, that cannot be eliminated via renormalization thus yielding the theory to be un-predictive. \\

It is well known, that the couplings appearing in QFTs are scale-dependent, this scale dependence is referred to as "running of the couplings". In the case of the full theory, where one integrates from zero to infinite momenta (or alternatively zero distances) many models exhibit infinite divergences, the solution to which is provided by some cutoff $\Lambda$ introduced to the high energy regime. Up to this cutoff, the theory is predictive, and the aim is to remove the cutoff and avoid the appearance of infinities. The UV completeness of a QFT is provided by the existence of fixed points of the renormalization group flow in the coupling constant space: as the energy scale changes, the running coupling constants approach some fixed point value. The microscopic theory is defined in such fixed points, thus finding them is a crucial part of any theory based on QFT language. Let  $g$ be a coupling constant of a given theory, then the so-called "beta function" $\beta(g)$ will define the scale dependence, or running of the coupling. The fixed points are defined by  zeros of $\beta(g)$, which can result in a free or interactive theory. The free theory is achieved when the zero of the beta function corresponds to zero values of the couplings, which is called "asymptotic freedom" and such a fixed point is called trivial or Gaussian. If instead zeros of the beta function are achieved for a finite number of non-zero couplings, it is called "asymptotic safety", where one has non-trivial fixed points corresponding to an interactive theory \cite{nontriv_fp}. A fixed point corresponding to high energy, or short scale, is called the "ultraviolet" (UV) fixed point, while the "infrared " (IR) fixed point will correspond to the low energy, or large scale theory.\\

A QFT description of GR means, that one treats the metric tensor $g_{\mu\nu}$ as the field of gravitation and defines an action in terms of geometric invariants obtained from the metric tensor, such as e.g., $R, R^2, R_{\mu\nu}R^{\mu\nu}$, etc. The most important couplings in the case of gravity are Newton's coupling $G$, and the cosmological constant $\Lambda$. The theory of gravity is perturbatively non-renormalizable, as applying perturbation theory in every order one has to introduce  infinitely many counter-terms and the corresponding new couplings, which renders the theory to be non-predictive. Nevertheless, according to the "asymptotic safety" conjecture, formulated by Steven Weinberg \cite{Weinberg},  most of the (potentially infinitely many) couplings appearing in such a theory become irrelevant at the non-trivial UV fixed point, and there will be only a finite number of relevant couplings rendering the theory non-perturbatively renormalizable, i.e., UV-complete and predictive to arbitrarily large energy scale. Therefore a non-perturbative description of quantum gravity is needed which can be done with the help of numerical simulations. The non-perturbative approach discussed in this thesis  is called Causal Dynamical Triangulations (CDT) and it is based on Regge calculus and Feynman path integral formulation.

\subsection{Regge calculus}

Before jumping into the description of CDT, it is necessary to discuss the mathematical formulation that led to it. This formulation was introduced by Regge and is called Regge calculus \cite{Regge}. The aim of Regge was to approximate space-times, which are solutions to the Einstein field equations, via piecewise-flat manifolds.\footnote{Often the name {\it piecewise-linear manifold} instead of {\it piecewise-flat manifold}  is used.} The approximation is done with the help of internally flat triangular building blocks (simplices) glued together in a non-trivial way, hence the name "triangulation". The simplices in a 2-dimensional triangulation are triangles, in 3-dimensions are tetrahedra, and in 4-dimensions are pentachora. All simplices in a triangulation are glued to each other via their $(d-1)$ dimensional faces (links for $d = 2$, triangles for $d = 3$, and tetrahedra for $d = 4$). These $(d-1)$ dimensional sub-simplices are also connected via "hinges", also called  "bones", which are $(d-2)$ dimensional objects. The hinges play a crucial role, as  curvature can be defined there locally. The curvature is related to the angular difference (deficit angle) at a given bone. Let's imagine a triangulation consisting of $n$ equilateral triangles glued together along edges (links) around a single point (vertex). If $n = 6$ then one can place it on a flat 2-dimensional surface. If $n = 5$ then one can place it only if it is cut along one edge, and it will be visible that a triangle is "missing". The angle associated with the missing (or for $n > 6$ extra) triangles is the deficit angle.\\

Let us consider the simplest (nontrivial) case of a three-dimensional Riemannian manifold which is well approximated by a fine triangulation. Following the approach of Regge \cite{intro_regge_calc}, the discretized curvature is obtained by considering the parallel transport of a vector around a bone. Many simplices (in this case tetrahedra) touch each other at the bone forming a bundle $p$. One can associate the number of simplices in the bundle with bone density $\rho$ at $p$, which is equal to the number of simplices divided by a unit area. The deficit angle ($\epsilon_p$) associated with the bone is a measure of a dihedral angle:

\begin{equation}
    \epsilon_p = 2\pi - \sum_n \theta_n,
\end{equation}
$\theta_n$ being the dihedral angle of the $n$-th simplex at the bone. One can alternatively define $\epsilon = \frac{1}{N}\epsilon_p$, which is the deficit angle of the bone smeared on its simplices. Now, let's take a loop $a$ with area $\Sigma$ around the bundle and parallel transport a vector $\Vec{A}$ around the loop. If $n_\Sigma$ is a unit vector orthogonal to $\Sigma$, then one can define:

\begin{equation}
    \Vec{\Sigma} = \Sigma n_\Sigma,
\end{equation}
which is an area vector associated with the loop. Parallel transporting a vector around the bundle will rotate $\Vec{A}$ by an angle $\sigma$ due to the process of the parallel transport. One can associate a vector of length $\sigma$ to the rotation, and let this vector  be parallel to the bone, so it  will be defined by: 

\begin{equation}
    \Vec{\sigma} = \sigma n,
\end{equation}
where $n$ is the unit vector parallel to the bone. Rotating $\Vec{A}$ by an angle $\sigma$  will produce the vector $\Vec{A'}$ $=\Vec{A}+\delta \Vec{A}$. The infinitesimal change $ \delta \Vec{A}$ will be equal to the product $ \delta \Vec{A} = \Vec{\sigma} \times \Vec{A}$. The rotation angle $\sigma$ is proportional to the number of simplices ($N$) visited by the loop $a$ circumventing the bone $p$, thus :

\begin{equation}
     \sigma = N \epsilon, 
\end{equation}
where $N$ can be expressed in terms of the
bone density $\rho$, the oriented area vector $\Sigma$ and the unit vector parallel to the bone $n$:

\begin{equation}
    N = \rho \, n \cdot \Vec{\Sigma}.
\end{equation}
Putting all the expressions together the infinitesimal change $\delta \vec{A}$ is given by:

\begin{equation}
    \delta \Vec{A} = \rho \epsilon (n \Vec{\Sigma}) \cdot (n \times \Vec{A}).
\end{equation}
Using coordinate (vector component) notation:

\begin{equation}
\delta A_\mu = \rho \epsilon (n^\nu \Sigma_\nu) ( \varepsilon_{\mu \alpha\beta} n^{\alpha}  A^\beta), 
\end{equation}
where:  $\varepsilon_{\mu \alpha\beta}$ is the Levi-Civita symbol. Now, one can express the $n$ and $\Vec{\Sigma}$ vectors in the dual space, i.e., the space of two-forms:

\begin{equation}
    n_\nu = \frac{1}{2}\varepsilon_{\nu\rho\sigma}n^{\rho\sigma}, 
\end{equation}
and 
\begin{equation}
    \Sigma_\nu = \frac{1}{2}\varepsilon_{\nu\alpha\beta}\Sigma^{\alpha\beta}.
\end{equation}
Using the fact that $n^{\nu\lambda} = -n^{\lambda\nu}$, the infinitesimal change $\delta \vec A$ can be now written as:

\begin{equation}
    \delta A_\mu = \frac{1}{4}\rho \epsilon  (\varepsilon_{\nu \rho\sigma} n^{\rho \sigma} \frac{1}{2}\varepsilon^{\nu\alpha\beta} \Sigma_{\alpha \beta}) (2n_{\gamma\mu}) A^\gamma = \frac{1}{2}(\rho \epsilon n_{\alpha \beta} n_{\gamma \mu})\Sigma^{\alpha\beta}A^\gamma.
\end{equation}
Using the continuous counterpart of the same equation with the help of the Riemann tensor one can write:

\begin{equation}
    \delta A_\mu = \frac{1}{2}{R^\gamma}_{\mu\alpha \beta} \Sigma ^{\alpha \beta} A_\gamma.
\end{equation}
Comparing the two equations one can recognize the discretized Riemann curvature tensor. The Ricci tensor can be then defined by index contraction:

\begin{equation}
    {{R^\alpha}_{\mu\alpha\nu}} = R_{\mu\nu} = \rho \epsilon (\delta_{\mu\nu} - n_{\mu} n_{\nu}),
\end{equation}
where we switched back to the unit vector $n$. And with further index contraction, one can get the Ricci scalar:

\begin{equation}
    R = {R^{\alpha}}_{\alpha} = \rho \epsilon ({\delta^\alpha}_\alpha - n^\alpha n_\alpha) = 2 \rho \epsilon,
\end{equation}
which gives a direct connection between the curvature of continuous Riemannian manifolds and their discretized approximations. The above formula can be generalized to more  dimensions as well as to pseudo-Riemannian manifolds. \\

Using Regge calculus, the  Regge action $S_{R}$, i.e., the gravitational action for a piecewise-flat triangulation, can be formulated. The starting point of this is the Einstein-Hilbert action:

\begin{equation}
    \frac{1}{16\pi G}\int d^dx \sqrt{-g} (R - 2\Lambda), 
\end{equation}
where $G$ is the Newton's constant, $R$ is the scalar curvature and $\Lambda$ is the cosmological constant. Writing the curvature in terms of Regge calculus one gets the form:

\begin{equation}
\frac{1}{16\pi G}\int d^dx\sqrt{-g}R = \frac{1}{8\pi G}\int d^dx\sqrt{-g} \rho \epsilon = \kappa \sum_{n_{(d-2)}} k_n\epsilon_n,   
\label{eq:curvature}
\end{equation}
where $\kappa=(8 \pi G)^{-1}$ is the  (inverse) bare  gravitational constant, $k_n$ denotes the volume of the $(d-2)$-dimensional hinge, $\epsilon_n$ is the deficit angle associated with  the hinge and the summation is over $(d-2)$-dimensional simplices, denoted by $n_{(d-2)}$. The term including cosmological constant reads:

\begin{equation}
\frac{1}{16\pi G}\int d^dx\sqrt{-g} (-2 \Lambda) = \frac{-2\Lambda}{16\pi G}\int d^dx\sqrt{-g} = \lambda \sum_{n_d} V_{n_d}, 
\end{equation}
where $\lambda=-\Lambda \kappa$ is the bare cosmological constant, $V_{n_d}$ is the volume of the $d$-dimensional simplices building up the triangulation and the summation is over $d$-dimensional simplices. This leads to the full Regge action:

\begin{equation}
    S_R = \kappa \sum_{n_{(d-2)}} k_n\epsilon_n + \lambda \sum_{n_d} V_n,
    \label{eq:curvature2}
\end{equation}
which holds in any dimension.  
One should note that the Regge form of the gravitational action (\ref{eq:curvature2}) is not expressed in terms of the metric tensor, but in terms of numbers of simplices and sub-simplices. Expressing the Regge action for a particular triangulation can lead to a complicated form, however applying certain constraints can simplify the expressions. 

\section{Causal Dynamical Triangulations}

\textit{"The more success the quantum theory has, the sillier it looks. How nonphysicists would scoff if they were able to follow the odd course of developments!"} - \textbf{Albert Einstein}\\\\

Following the ideas of Weinberg and assuming the existence of a UV fixed point for gravity the properties of quantum gravity can be analyzed using non-perturbative methods. As fixed points were found in other QFT-based theories, such as Quantum Chronodynamics (QCD) \cite{as_freed}, theorists turned towards lattice formulations (e.g. Lattice Quantum Chronodynamics (LQCD)). The simplest lattice theory of GR is called Dynamical Triangulations (DT). In DT, one can use the Regge action straight away. The spacetime is constructed by gluing $d$-dimensional simplicial building blocks: triangles, tetrahedra, and pentachora. The triangulation does not play a role in the physics of the model, as it serves the purpose of regularization, providing a UV  cutoff related to lattice spacing $a$, which should be removed from the continuum limit if it exists. A huge difference between the DT approach from other techniques based on the Regge calculus, such as Quantum Regge Calculus \cite{quantumregge} or some versions of LQG \cite{lqg_regge}, is that the edge length ($a$) of all the simplices is kept fixed and thus piecewise-flat manifolds are constructed from identical equilateral simplices. Transforming the metric signature with the Wick rotation one gets a Euclidean description which allows studying the (regularized) path integral of quantum gravity using statistical methods. In the  DT there is no difference between space and time, however, CDT twists the picture via the introduction of a foliation and thus the notion of time is restored as the causal evolution of the leaves of the foliation. The decomposition of the four-dimensional space-time into space and time is similar to that of the Arnowitt-Deser-Misner (ADM) formalism \cite{adm_f}. Thus, the 4-dimensional space-time is assumed to be globally hyperbolic and each ($d-1$)-dimensional hypersurface ("leaf" of the foliation) has the same fixed topology.  The word "causal" in the name of CDT refers to the time-slicing of the triangulation, as opposed to usual DT, and "dynamical" points at the difference between CDT and traditional lattice approaches, as in CDT the lattice connectivity is not fixed and it encodes  the gravitational degrees of freedom. For example, in LQCD there is a fixed and regular lattice, on which the theory is defined, but in CDT the different lattice configurations correspond to the different trajectories (histories) in the gravitational path integral. Therefore a single configuration (single trajectory) is non-physical, and one has to compute a suitable average over an ensemble of such configurations. \\

In a $d$-dimensional CDT triangulation, by construction, every (sub-)simplex lies in a $d$-dimensional {\it slab} (part of the triangulation) between lattice (discrete) time $t$ and $t+1$. Different types of simplicial building blocks (simplices) $s_{\alpha \beta}$ can be defined by indicating the number $\alpha$ of their vertices in $t$ and the number $\beta$ of  vertices in $t+1$. In 2 dimensions there are two types of building blocks, i.e., triangles: $s_{21}$ and $s_{12}$. In 3 dimensions there are three different types of building blocks, i.e., tetrahedra: $s_{22}$, $s_{31}$, and the mirror reflection $s_{13}$. Finally, in 4 dimensions there are 4 types of such simplices: $s_{41}$ with its mirror-reflection $s_{14}$ and $s_{32}$ with its mirror-reflection $s_{23}$. Due to this construction and the symmetry of the action, as we will see, CDT exhibits a time reflection symmetry as well. Thanks to a small number of different categories of simplices appearing in the four-dimensional CDT and due to  topological constraints of the triangulated manifolds, see  Appendix \ref{AppendixA}, the Regge action (\ref{eq:curvature2}), which governs the dynamics of the model, can be expressed in terms of these $4$-dimensional simplices and  vertices in a triangulation $\cal T$ \cite{nonperturb}:

\begin{equation}
S_{R} = - (\kappa_0 + 6\Delta) N_0 + \kappa_4 (N_{41} + N_{32}) + \Delta N_{41},
\label{eq:ation_kappa}
\end{equation}
where $N_0= \sum s_{10}$ is the total  number of vertices, while $N_{41}= \sum (s_{41} + s_{14})$ and $N_{32} = \sum (s_{32} + s_{23})$ are the total numbers of the various types of simplices in the triangulation $\cal T$. The three bare coupling constants are $\kappa_0$, the bare inverse  Newton constant, $\kappa_4$, the bare cosmological constant, and $\Delta$, related to the asymmetry between  lengths of space-like and {time-like links in the lattice}. From now on we will refer to $N_0$, $N_{41}$ and $N_{32}$ as \textit{global numbers}.\\

The path integral of quantum gravity is formally defined as:

\begin{equation}
    \mathcal{Z}_{QG} = \int D[g_{\mu\nu}]e^{iS_{EH}[g_{\mu\nu}]} \to^{reg}\to \sum_\mathcal{T} \frac{1}{C_\mathcal{T}}e^{iS_R[\mathcal{T}]} = \mathcal{Z}_a,
\end{equation}
where $D$ is the measure term, which enables one to integrate over geometries, i.e., diffeomorphism invariant equivalence classes of smooth metrics $g_{\mu \nu}$, and $S_{EH}$ is the Einstein - Hilbert action. After the lattice regularization ($\to^{reg}$) the path integral is replaced by a sum over all possible triangulations with a measure $1/C_\mathcal{T}$, the size of the automorphism group of $\mathcal{T}$. The index $a$ in $\mathcal{Z}_a$ refers to the lattice regulator, which is the edge length of the simplices and $S_R$ is the Regge action (\ref{eq:ation_kappa}), which is the lattice-regularized version of the Einstein-Hilbert action. The distinction of space and time introduced by the foliation is also present in the edge lengths, as the time-like edge lengths $a_t$ and the space-like edge lengths $a_s$ are not necessarily the same, which gives rise to a degree of freedom, called the asymmetry parameter $\alpha$, where $-\alpha a_t^2 = a^2_s$ in the Lorentzian setting. The aim of CDT is to define the gravitational path integral, or at least approximate it as close as it gets.  All  possible triangulations  $\mathcal{T}$ include only such triangulations which respect the foliation structure and some additional topological constraints. To be able to treat the model with methods of statistical physics a Wick rotation has to be applied to the partition function to change the metrics from Lorentzian to Euclidean signature. Due to the imposed global foliation, the  "Euclideanization" of the path integral via the Wick rotation is well defined and is related to the analytic continuation of the Regge action to negative values of $\alpha$  in the lower half of the complex $\alpha$ plane. Performing it one turns the path integral into the partition function: 

\begin{equation}
\mathcal{Z}_{R} = \sum_{\mathcal{T}} \frac{1}{C_\mathcal{T}}e^{-S_{R}[\mathcal{T}]} ,
\label{eq:partfun}
\end{equation}
where, for a simpler notation, we kept the same symbol $S_R$ for the (now) Euclidean Regge action. The Wick rotation allows for the application of statistical physics methods on the model, for example, one can compute the expectation values of observables as:

\begin{equation}
\langle \mathcal{O}\rangle = \frac{1}{\mathcal{Z}} \sum_\mathcal{T} \frac{1}{\mathcal{C}_{\mathcal{T}}}\mathcal{O} e^{-S_{R}[\mathcal{T}]}    .
\end{equation}

One of the benefits of the Wick rotation is that the model became suitable for numerical Monte Carlo (MC) simulations, where the partition function can be approximated by an ensemble of configurations generated in such simulations. The past twenty years of  numerical studies of the 4-dimensional CDT model led to many interesting and important results.

\subsection{Most important previous results of CDT}

\begin{figure}[ht!]
\centering
\frame{\includegraphics[width=0.23\textwidth]{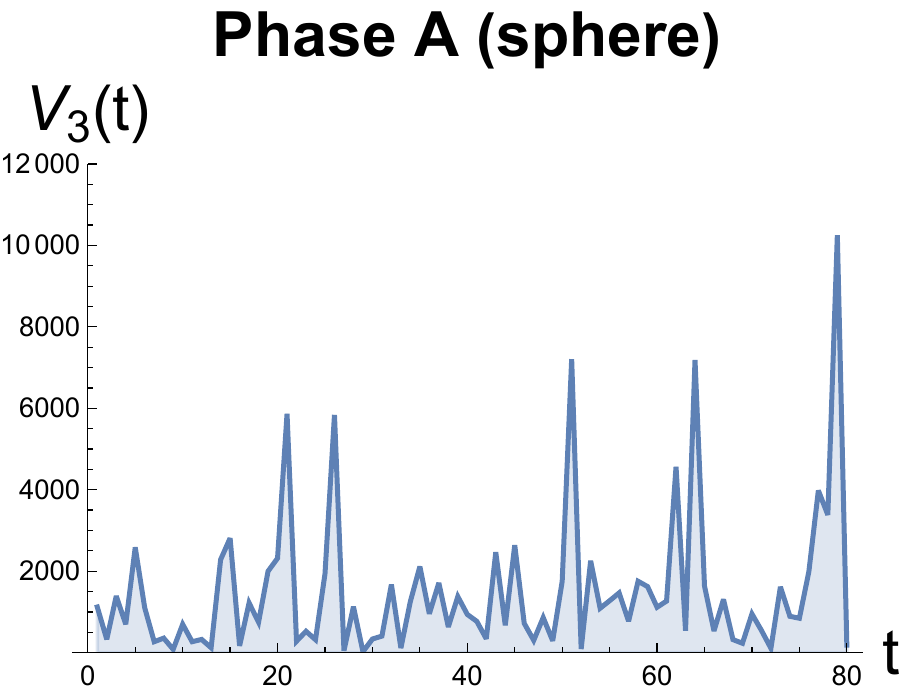}}
\frame{\includegraphics[width=0.23\textwidth]{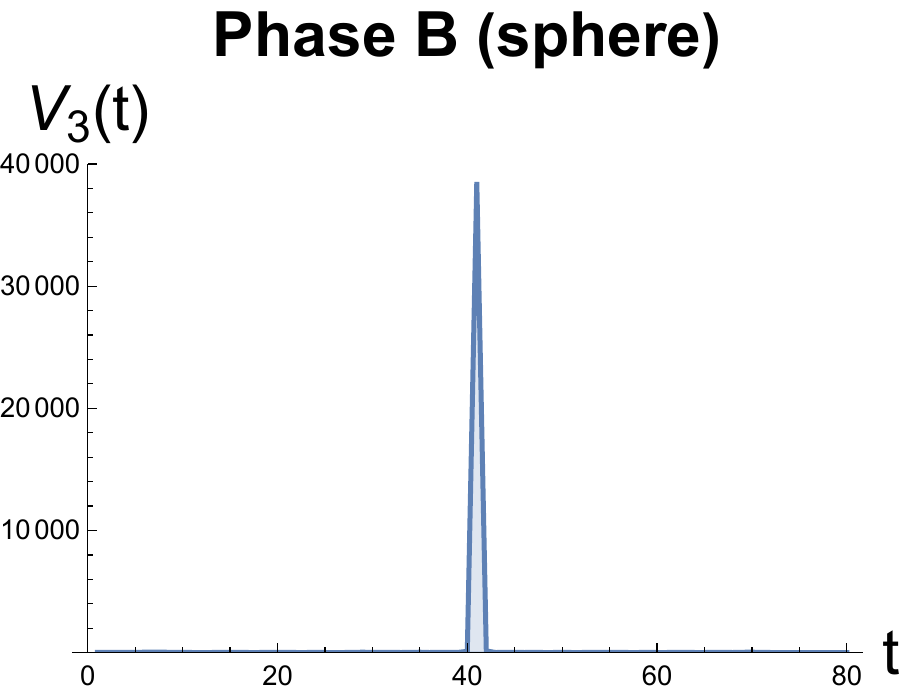}}
\frame{\includegraphics[width=0.23\textwidth]{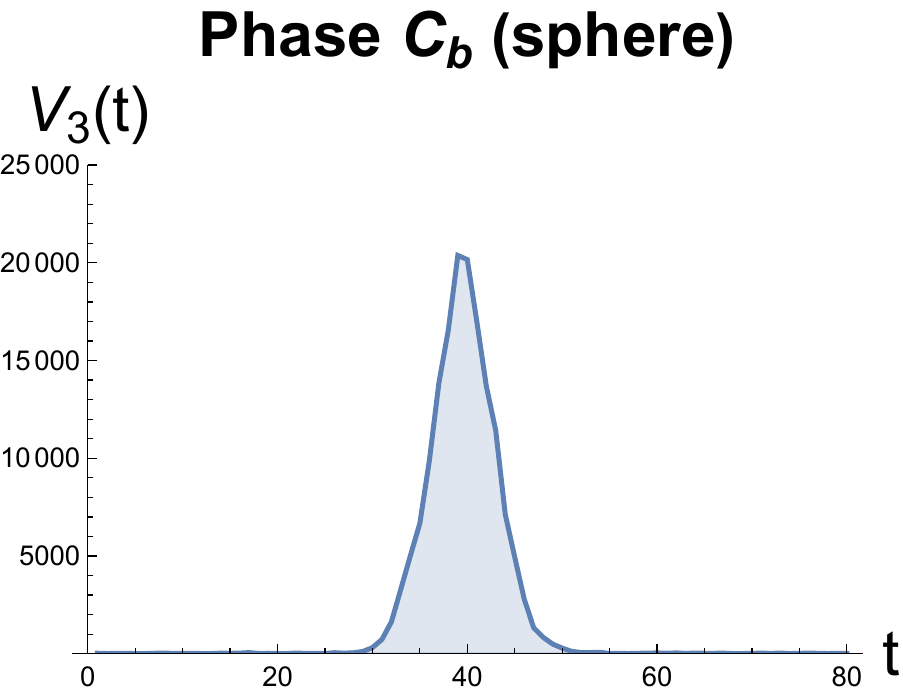}}
\frame{\includegraphics[width=0.23\textwidth]{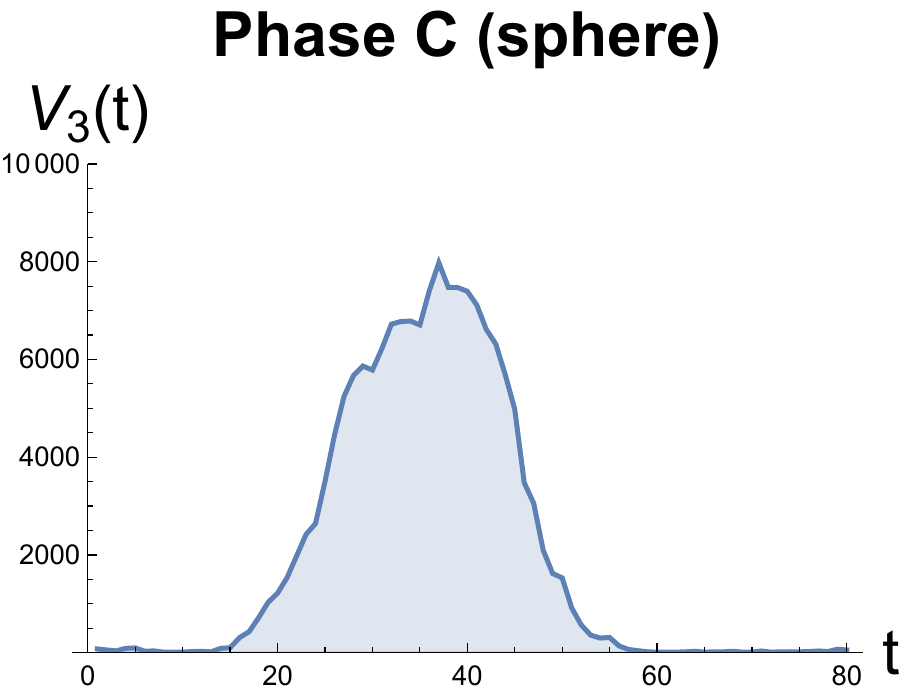}}\\
\frame{\includegraphics[width=0.23\textwidth]{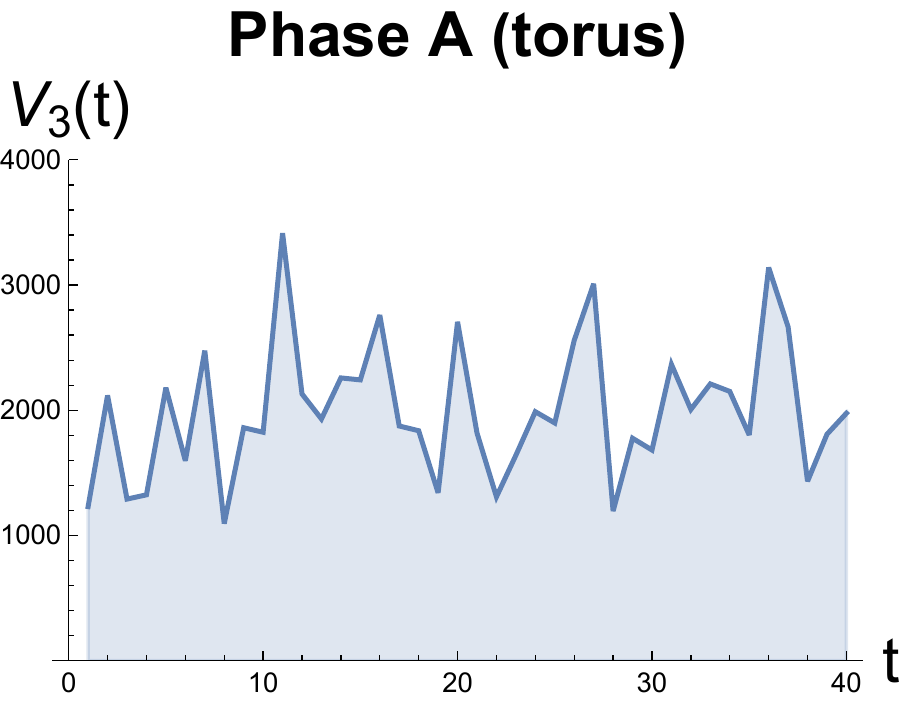}}
\frame{\includegraphics[width=0.23\textwidth]{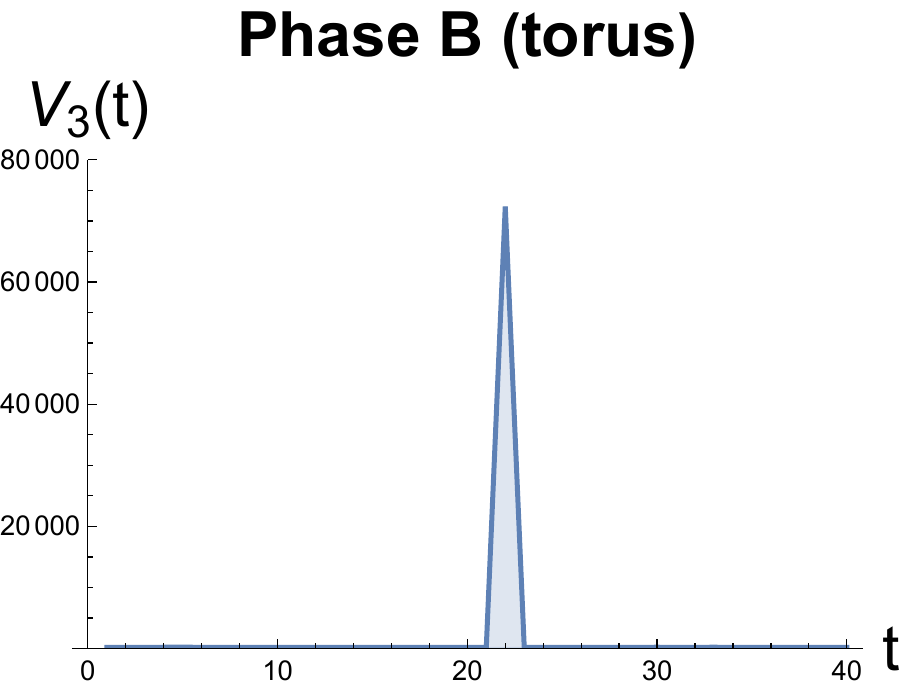}}
\frame{\includegraphics[width=0.23\textwidth]{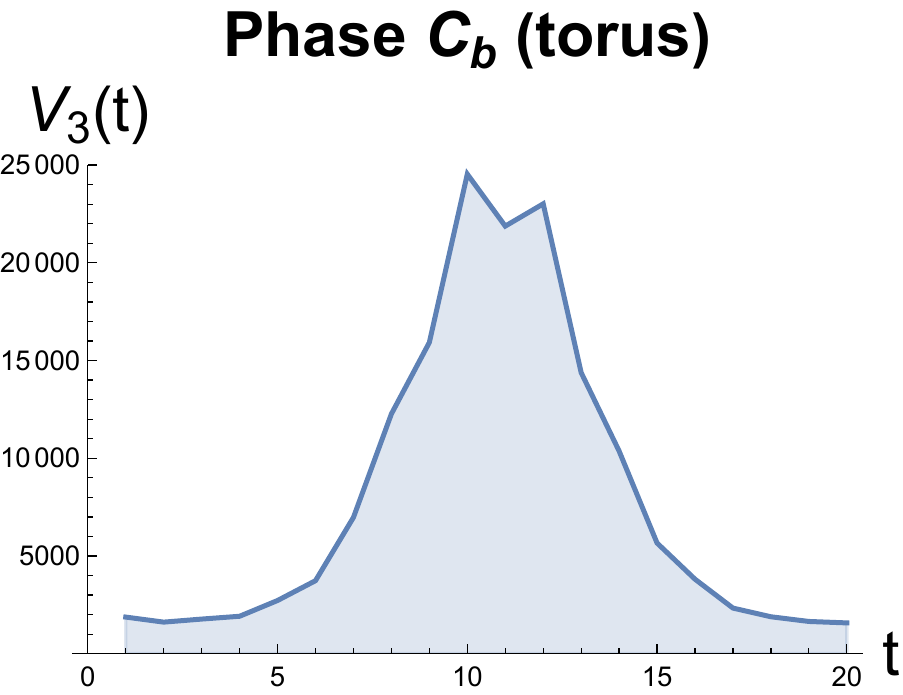}}
\frame{\includegraphics[width=0.23\textwidth]{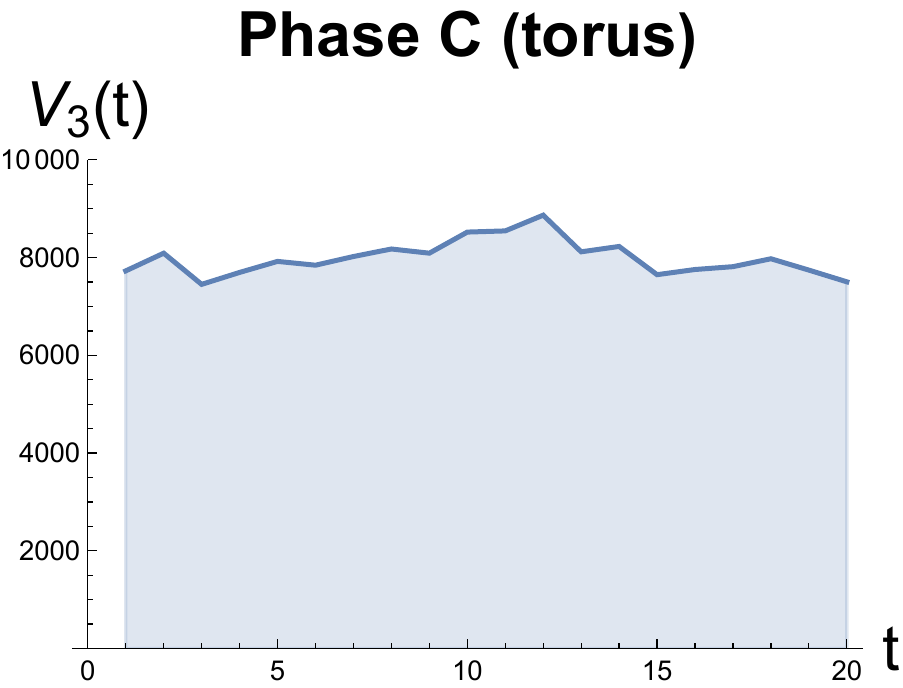}}\\
\caption{Spatial volume profiles of generic CDT configurations in different phases. Top: Spherical CDT: $A$, $B$, $C_b$, $C$; Bottom: Toroidal CDT:  $A$, $B$, $C_b$, $C$, respectively.}
\label{fig:volprofs}
\end{figure}

CDT was formulated in the beginning of the 21st century and became recognized by the quantum gravity community in the following years. The introduction of the foliation to the triangulation allowed for the addition of the asymmetry parameter between space and time, which was promoted to a new coupling constant $\Delta$ in the action (\ref{eq:ation_kappa}). This particular change had a huge impact on the properties of the CDT model, compared to DT, as due to the enforced causality constraint the ensemble of triangulations present in the partition function (defined by eq. (\ref{eq:partfun})) became significantly reduced. At the same time, the third coupling constant ($\Delta$) allowed for an extended view on the phase diagram of simplicial quantum gravity. There were only two phases in DT, one phase where a link of the generic triangulation gathered a significant number of simplices around itself, and its end vertices experienced a huge coordination number\footnote{The coordination number of a vertex is defined as the number of  four-simplices which share the vertex.}, comparable to the system size, thus the name "collapsed phase". The generic geometries of the other phase could be described by the so-called, branched polymers \cite{branchedpoli}, hence the name "branched polymer phase". The analogs of these phases \cite{phase_struct_cdt} are present in CDT\footnote{Phase $B$ is the collapsed phase and phase $A$ is the branched polymer phase}, however, the topological restriction related to the foliation resulted in the appearance of two new phases \cite{ccc,cb1,cb2}. This became apparent when new observables were used related to the newly introduced time foliation. The number of spatial tetrahedra at a given CDT foliation leaf (with integer lattice time $t$) can be computed and it is, by definition, proportional to the spatial three-volume at $t$, which defines the so-called, volume profile $V_3(t)$, shown in Fig. \ref{fig:volprofs}.

Apart from a "collapsed" volume profile of phase $B$ (where all three-volume is concentrated in one spatial "slice", i.e., the 3-dimensional  foliation leaf of integer $t$), and the heavily fluctuating volume profile of the "branched polymer" phase $A$ (independent number of tetrahedra in each spatial slice) there are new phases where the  volume profiles averaged over MC configurations follow a particular smooth function. The most interesting new phase is phase $C$ , where, in the case of the fixed spherical topology of spatial slices, the resulting average volume profile behaves as $cos^3(t)$, which corresponds to the (Euclidean) de Sitter solution of GR \cite{nonperturb_desitter}. Therefore phase $C$ is also called the de Sitter or the semi-classical phase and it is related to the IR limit of quantum gravity. The fourth phase, which is called the bifurcation phase ($C_b$), exhibits a smooth volume profile in case of large enough fixed total volumes (lattice sizes). The volume profile in phase $C_b$ is similar to the volume profile in  phase $C$ measured for the spherical spatial topology, however, it scales in a non-canonical  way when the lattice volume is increased. Furthermore, in phase, $C_b$ every second  spatial slice of integer lattice time coordinate contains a vertex with a macroscopically large coordination number, similar to "high-order" vertices encountered in phase $B$.\\

By analyzing  fluctuations of the spatial volume it was possible to derive an effective action\cite{impact_top} of CDT parametrized by the spatial volume, or alternatively by the scale factor. The effective action in the de Sitter phase ($C$) \cite{transfer_matrix} turned out to be consistent with the Hartle-Hawking minisuperspace model \cite{mini1,mini2,mini3}. This result is non-trivial, as in the case of CDT the scale factor is obtained after "integrating out" all other geometric degrees of freedom present in the lattice, while in the minisuperspace model, where spacetime isotropy and homogeneity are put in by hand, the scale factor is the only degree of freedom. Therefore, this feature of CDT is fully emergent. One could also show that the notion of effective dimension of spacetime first measured in the case of 2D CDT\cite{eff_dim_2d} and also in the case of Locally Causal Dynamical Triangulations (LCDT) \cite{spectraldim} was extended to higher dimensions. In the case of 4-dim CDT in phase $C$ it was measured to be consistent with the topological dimension four. This was not so obvious as the effective dimension measured in other phases of CDT (and earlier in DT) was different than four\cite{scaling_in_4d_grav}. Both the so-called, Hausdorff dimension \cite{cdt_desit_fi}, related to the scaling of an area and volume, and the spectral dimension \cite{spect_dim_uni_scale}, defined by a heat kernel of the Laplace operator, were measured. Additionally, the spectral dimension was shown to exhibit a non-trivial scale dependence changing from four in large scales (comparable to the size of the configuration) to approximately two in short scales and also in the presence of matter fields\cite{spectral} it can deviate from the classical values. The above phenomenon  of "dimensional reduction" was also confirmed in many other approaches to quantum gravity (e.g., in ST \cite{dim_red_st}, NCG \cite{dim_red_noncommgeom},  HLG \cite{dim_red_hlg}, AS  \cite{dim_red_1,dim_red_2,dim_red_3} and LQG \cite{dim_red_4}). \\

Most of the phase transitions present in the CDT model with spherical spatial topology were analyzed, and the $A-C$ phase transition was found to be first-order \cite{ac,pts_in_cdt,phase_struct_cdt}, while the $B-C_b$ and the $C-C_b$ turned out to be continuous  \cite{pts_in_cdt,cb1}. The existence of higher order (continuous)  phase transitions is an important result in view of the perspective existence of the UV fixed point of quantum gravity\footnote{As explained in Chapter \ref{chapter3}, such a fixed point should appear as a higher order transition point in CDT. }, however, the first study of the RG trajectories in CDT \cite{rg_flow1,rg_flow2} did not show convincing evidence for the existence of the UV fixed point. One of the issues was that a part of the phase diagram was out of reach due to computational difficulties, thus the analysis of some phase transitions was not possible. Also at that time, the available computational power was significantly smaller than presently. With the help of modern technology, much larger system sizes can be analyzed nowadays within available computational resources.\\ 

Most of the results presented above were obtained for the CDT model with the fixed spherical topology of the spatial slices. As the spatial topology choice is one of the free parameters of the model, in the past few years the main focus of the 4-dimensional CDT research was on models with  toroidal spatial topology. It was found that the phase diagram is almost invariant under the change of the topology, as all observed phases were present in both cases \cite{phase_structure_torus}. A huge difference between the spherical and the toroidal case is visible in the volume profile of phase $C$, see Fig. \ref{fig:volprofs}, and it is related to the potential term appearing in the effective action of CDT, which is different in the two cases \cite{impact_top}. In the spherical case, the potential term can be interpreted as coming from GR and it is consistent with the minisuperspace model, which also contains such a potential term for the scale factor. However, in the case of the toroidal CDT, one does not have a classical analog of the measured  potential, thus it can be interpreted as a quantum correction. Using the spatial topology of a three-torus allowed for an introduction of many new methods of analyzing the lattice-regularized quantum geometries, as it will be presented in Chapter \ref{chapter4}. It was also possible to investigate the region of the phase-diagram which was thought to be not available in the spherical CDT, see Chapter \ref{chapter3}. \\

In this thesis, we will present results related to the toroidal CDT: the study of the remaining phase transitions, including critical phenomena at the phase-transition lines. Then we will also discuss how to add scalar fields to the model of CDT, and either use them as semi-classical maps defining a coordinate system on the geometry or couple them to the geometry and analyze the effects of their back-reaction. But first, let us turn our attention to the numerical implementation of the CDT model, which is the topic of the next chapter.

%% file: Chapters/Chapter2.tex

\chapter{Numerical Simulations} 

\label{chapter2}  


\section{The Numerical Setup}

\textit{"The student should not lose any opportunity of exercising himself in numerical calculation and particularly in the use of logarithmic tables. His power of applying mathematics to questions of practical utility is in direct proportion to the facility which he possesses in computation."  - \textbf{Augustus De Morgan}}\\\\\\

In the case of the four-dimensional CDT, there is no analytical solution, however, certain numerical methods provide useful tools in the quest to find out more about the nature of the model. One of those tools is a Monte Carlo (MC) simulation \cite{tellerteller}. In an MC simulation one attempts to numerically approximate the path integral or rather, in the Euclidean formulation, the partition function of eq. (\ref{eq:partfun}), and estimate expectation values or correlators of various observables based on a sample of independent configurations generated with a probability proportional to the Boltzmann weight: $\exp(-S)$. There are various algorithms enabling to the realization of this goal. In this discussion, we will present the Metropolis Algorithm, as this is the one that is used in the case of four-dimensional CDT. One starts from any initial state of the model (in the CDT case any allowed triangulation with a given fixed topology\footnote{In practice one usually uses an initial configuration which is easy to be constructed "by hand".}), and applies a set of local changes (moves) transforming state $A$ to $B$. In order to ensure that the probability of generating a state converges to the required equilibrium probability $\propto \exp(-S)$, the probability of performing the move has to satisfy the detailed balance condition:

\begin{equation}
\mathcal{P}(A)\mathcal{W}(A \to B) = \mathcal{P}(B)\mathcal{W}(B \to A),     
\end{equation}
where $\mathcal{P}\propto e^{-S}$ is the probability distribution of a state and $\mathcal{W}$ is the transition probability from one state to another. Additionally, the moves have to be selected in such a way, that all possible states can be reached with a finite number of performed moves, in other words, the configuration space should be closed with respect to the selected moves. This condition provides ergodicity, which is crucial to ensure meaningful statistics. In the Metropolis algorithm the transition probability $\mathcal{W}$ is chosen to be  

\begin{equation}
    \mathcal{W}(A \to B) = min\{1, e^{-\Delta S}\},
\end{equation}
where $\Delta S = S(B)-S(A)$ is the change of the action by the move.\footnote{In the case of CDT the transition probability $\mathcal{W}$ depends also on a "geometric" factor related to the number of possible locations in a triangulation where the move and its inverse can be performed.} As already mentioned,  after the so-called {\it thermalization} period, the probability distribution of configurations generated by the Metropolis algorithm reaches an equilibrium defined by the  partition function (the action $S$). This is the point when one can start collecting a sample of configurations, which has to be large enough to ensure good statistics of the measured observables.\\

\begin{figure}[ht]
    \centering
    \includegraphics[width = 0.8\textwidth]{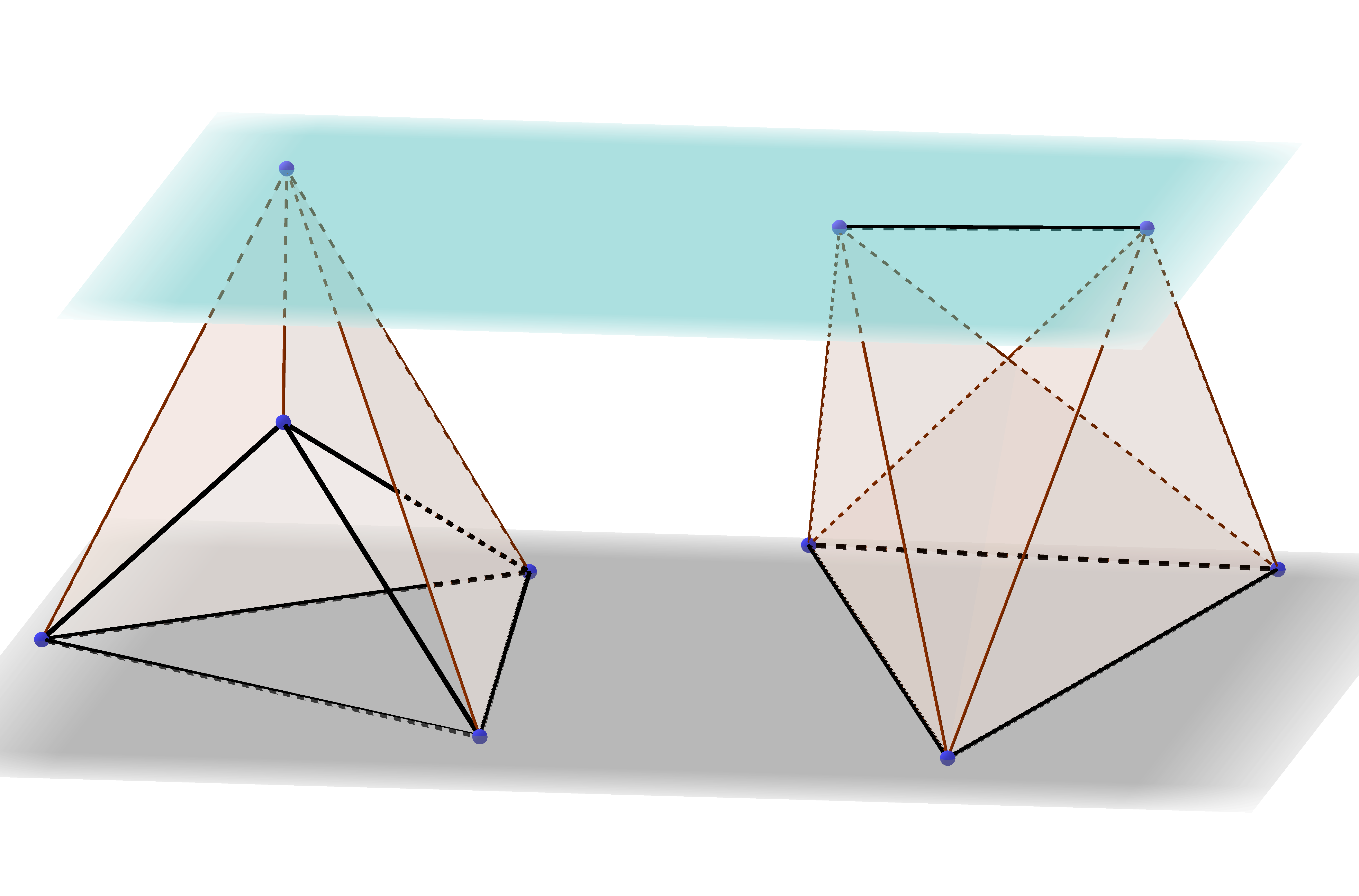}
    \caption{Two building blocks of the triangulation. The left simplex is $s_{41}$ and the right one is $s_{32}$. The other two types, $s_{14}$ and $s_{23}$, are mirrored-symmetric versions of them.}
    \label{fig:building-blocks}
\end{figure}

CDT is perfectly suitable for numerical simulations due to its relatively simple construction. The foliated space-times (MC states) are constructed by gluing the four-dimensional simplicial building blocks, presented in Fig. \ref{fig:building-blocks}, to each other, fulfilling some global and local constraints (discussed later in detail).

 Due to the nature of the triangulation, every simplex has exactly five neighbors, thus the local neighborhood of a hinge (i.e. a triangle in the four-dimensional CDT) can be simply discussed. The MC moves used in CDT are based on the so-called Pachner (or Alexander) moves \cite{pachner}, modified in such a way that, as shown later, the foliation structure with the fixed  topology of each spatial slice is conserved. Within our simulation code, we keep track of the vertices forming the 4-simplices and adjacency relations between the simplices. This information is enough to reconstruct the whole triangulation. Nevertheless, in order to optimize and speed up the code, we also keep track of some additional information, e.g., some specific types of sub-simplices or their coordination numbers.\footnote{The coordination number  measures how many $4$-dimensional simplices meet at a given vertex, link or triangle.} When we perform a measurement we usually have to calculate observables from the actual adjacency relations or other data that we store. As Fig. \ref{fig:building-blocks} shows, the graphical representation of simplices on a 2-dimensional figure is difficult. Due to this reason, we will present the idea of the moves of 4-dimensional CDT by showing how they impact the triangulation at the $t+\frac{1}{2}$ plane. Technically the $t+\frac{1}{2}$ plane describes the connectivity structure of the triangulation in a {\it slab}, defined by all simplices between spatial slices at (integer) lattice time $t$ and $t+1$. This treatment simplifies the discussion as it reduces the dimensionality of the problem by one, because on the $t+\frac{1}{2}$ plane a slab of the 4-dimensional triangulation is mapped to a 3-dimensional graph decorated by colors. The construction of the building blocks of this 3- dimensional graph is analogous to the method used in \cite{abab} in the case of three-dimensional CDT. Specifically, the color or, in other words, the type of the links (solid black/grey or dashed) is important, as only links of the same type can be connected. When referring to the links of such a graph, the words "color" or "type" will be used interchangeably.

\begin{figure}[ht]
    \centering
    \includegraphics[width = 0.8\textwidth]{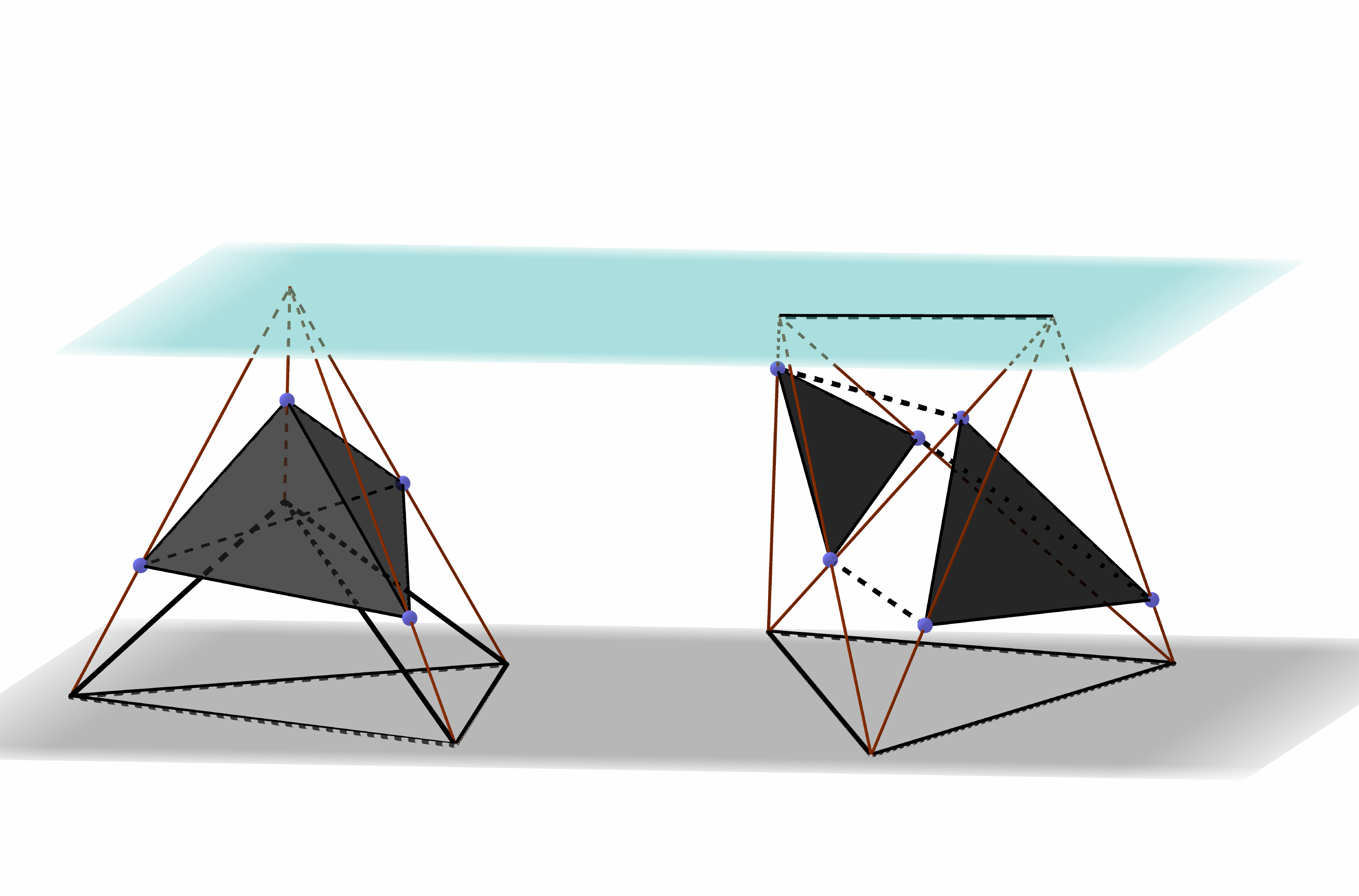}
    \caption{The figure shows the representations of $s_{41}$ (left) and $s_{32}$ (right) simplices in  the $t + \frac{1}{2}$ plane. The simplex $s_{41}$ is a single-colored tetrahedron, $s_{32}$ is a bi-colored prism with two triangular and three rectangular faces. The other two types of simplices $s_{14}$ and $s_{23}$ are mirror-reflections.}
    \label{fig:projection}
\end{figure}

In the $t+\frac{1}{2}$ plane, instead of the 4-simplices,  we now have 3-dimensional objects:  tetrahedra and prisms, see Fig. \ref{fig:projection}. In order to distinguish between the $s_{41}$ and the $s_{14}$ simplices we  attribute colors to the tetrahedra, such that a tetrahedron belonging to the $s_{41}$ simplex will have all-black (triangular) faces, and all triangles of a  tetrahedron belonging to the  $s_{14}$ simplex will be  grey. Even though each 4-simplex  has formally 5 neighbors, the tetrahedra have only four, which means that (for better clarity of the graphs) we omit the neighbors belonging to the previous/next slab. The prisms with black/grey triangular faces and transparent rectangular sides represent the $s_{32}/s_{23}$ simplices, respectively. Due to the topological constraints imposed on the CDT triangulations (the fixed spatial topology must be preserved  in all  layers interpolating between the spatial slices at $t$ and $t+1$)  the discretized geometry of the $t+\frac{1}{2}$ layer must be also connected in a specific way, meaning that a black triangle can be glued only to a black triangle, a grey triangle to a grey triangle, and a  transparent rectangle to a  transparent rectangle. It reflects the fact that the $s_{41}$ simplex can be adjacent only to  $s_{41}$ or $s_{32}$ simplices\footnote{Here we disregard the connections to the $s_{14}$ simplices of the previous slab.}, the $s_{32}$ simplex can be adjacent only to $s_{41}$, $s_{32}$ or $s_{23}$ simplices, etc.

\begin{figure}[h]
    \centering
    \includegraphics[width = 0.45\textwidth]{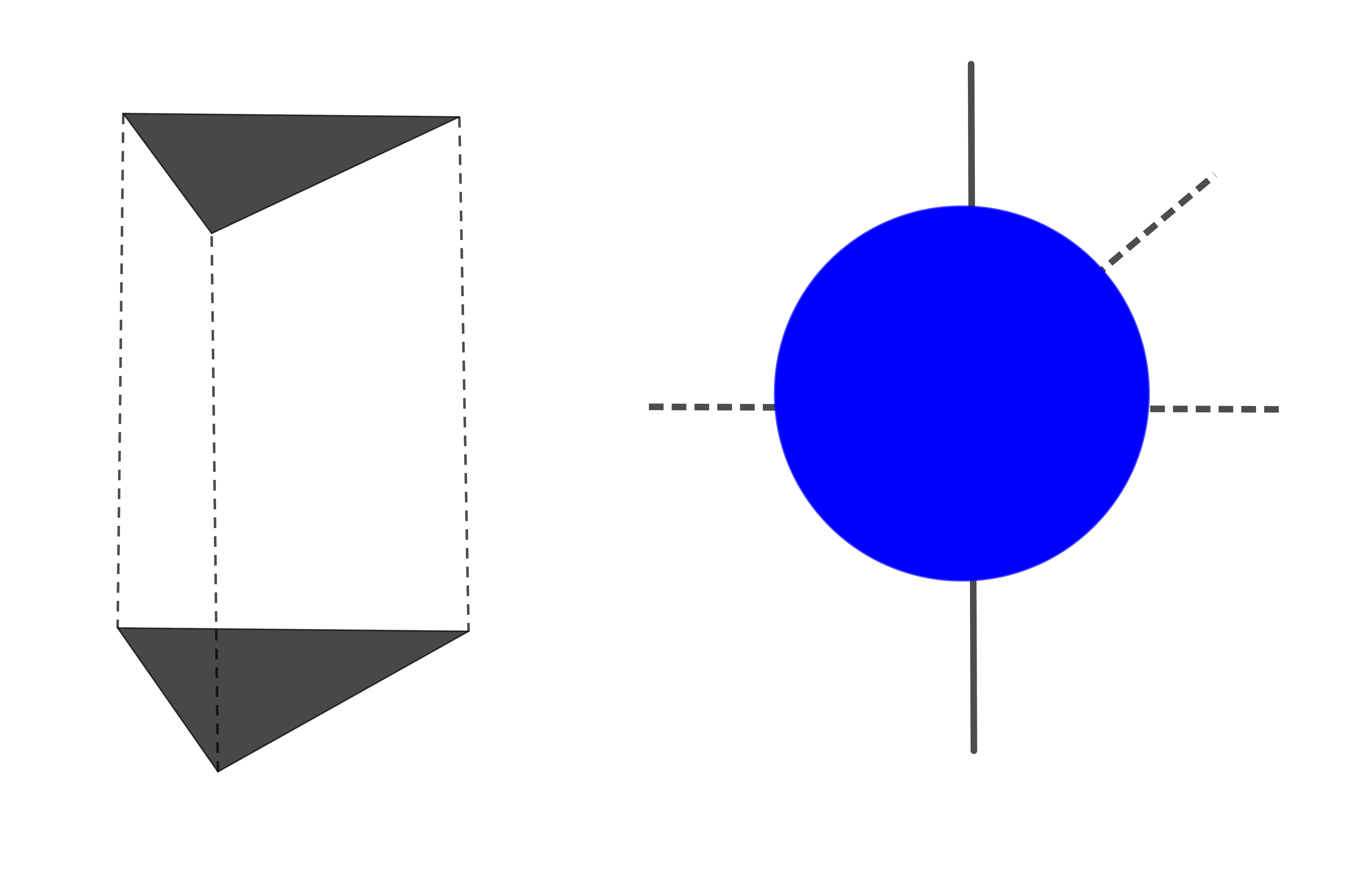}
    \includegraphics[width =0.45\textwidth]{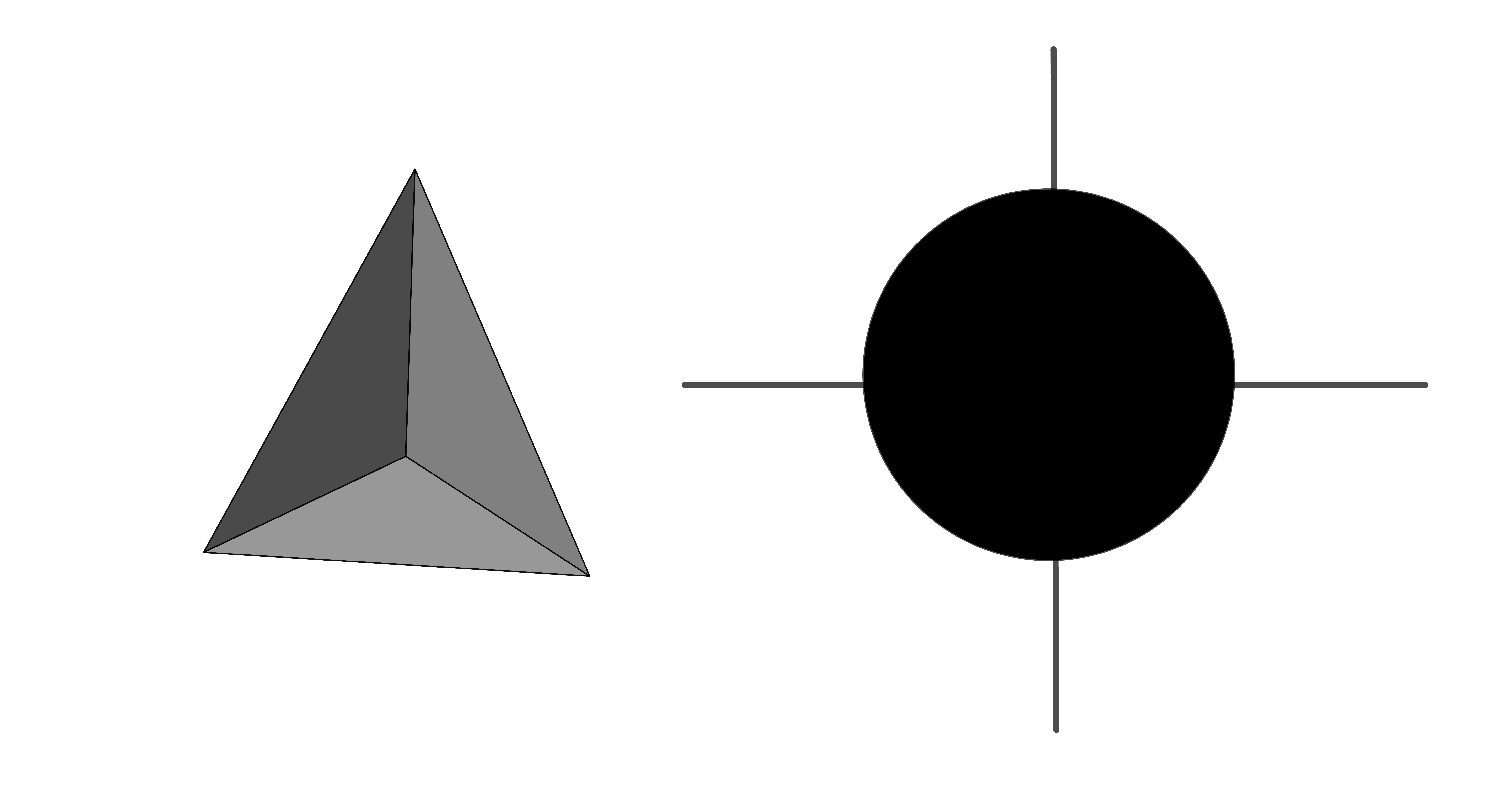}
    \caption{The figure presents the
    3-dimensional elements of the $t + 1/2$ plane. The prism with triangular bases and rectangular sides (left panel) comes from a $s_{32}$ simplex. In the graphical representation, it will be a blue dot with two solid black and three dashed legs. The tetrahedron  (right panel) comes from a $s_{41}$ simplex. In the graphical representation, it will be a black dot with four solid black legs (connections to neighboring slabs are omitted). Similarly, one has a red dot with two solid grey and three dashed legs, and a grey dot with four solid grey legs, coming from the mirror-reflected $s_{23}$ and $s_{14}$ simplices, respectively, which are not shown in the plot.}
    \label{fig:P_types}
\end{figure}

\begin{figure}[h!]
    \centering
    \includegraphics[width =0.8\textwidth]{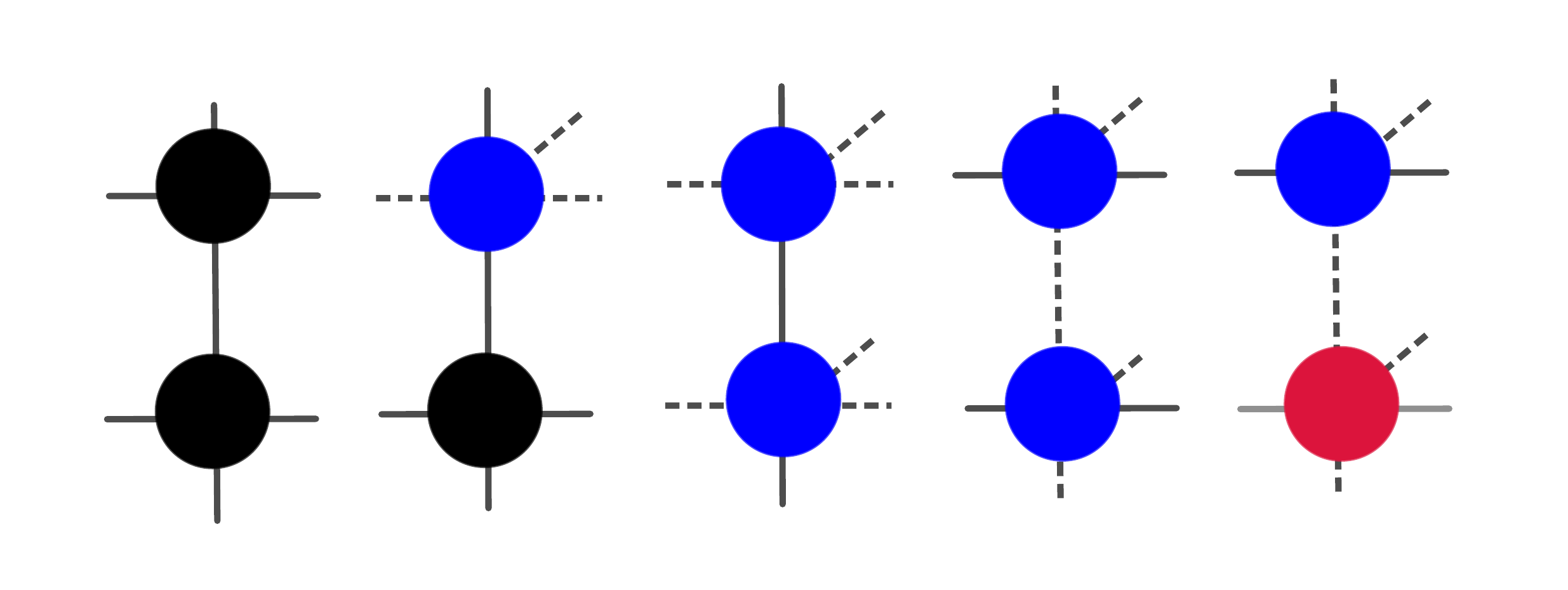}
    \caption{The figure presents  possible connections between various objects of the $t+\frac{1}{2}$ plane. Black dots (tetrahedra) can be connected to each other and to blue dots (prisms) via solid black legs (triangles). Similarly, blue dots can be connected to black and blue dots via solid black legs, but they can be also connected to blue and red dots via dashed legs (rectangles). The red dots have two solid grey legs, which can be connected to other red or grey dots, which are not shown in the figure.}
    \label{fig:connectivities}
\end{figure}

\begin{figure}[h]
    \centering
    \includegraphics[width =0.8\textwidth]{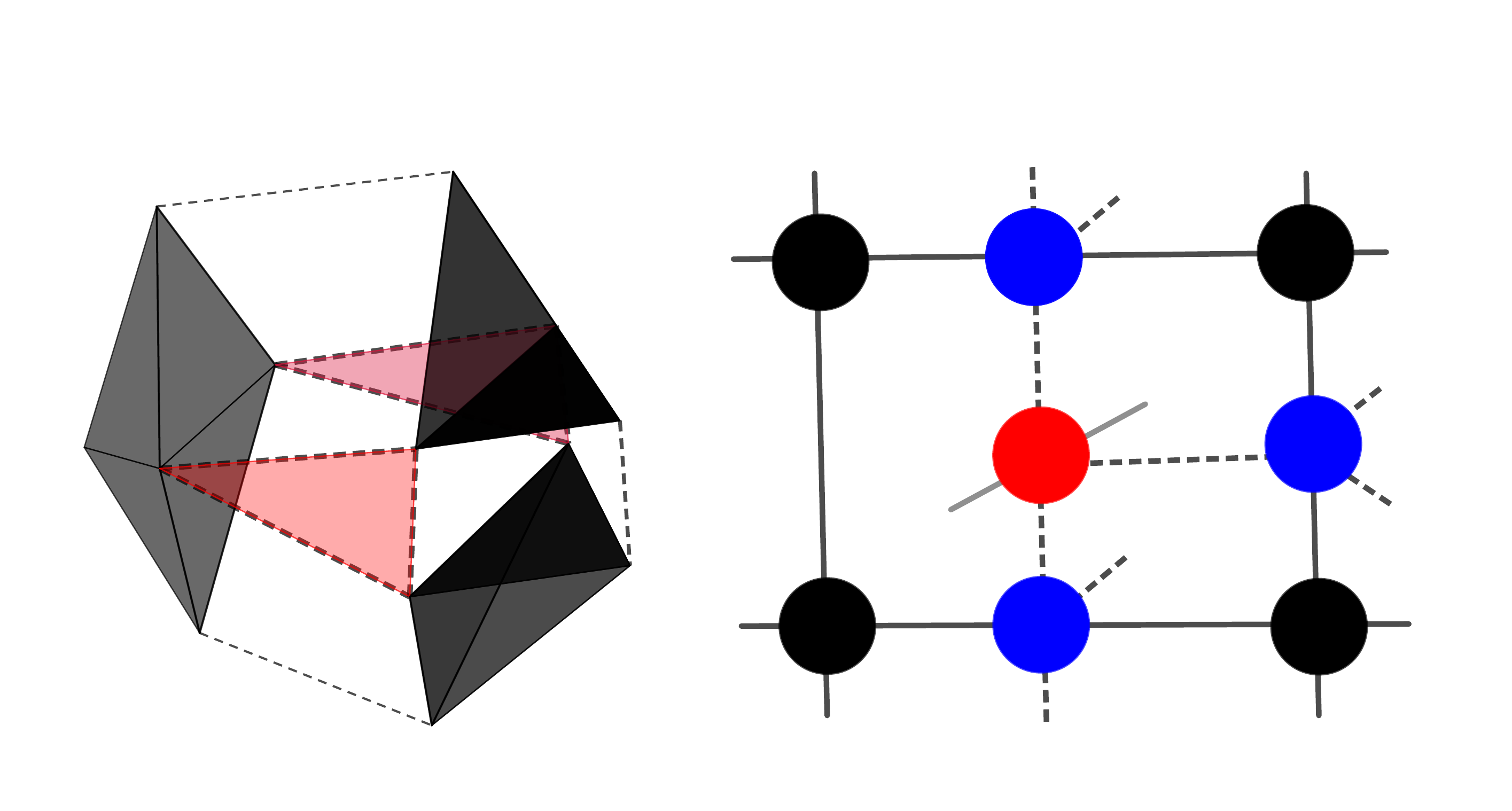}
    \caption{An example of a possible connection between four $s_{41}$, three $s_{32}$ and one $s_{23}$ simplices in the $t+\frac{1}{2} $ plane (left panel) and the corresponding graphical representation (right panel). A solid black loop in the graphical representation  is a spatial link in the  CDT triangulation. I deleted here a sentence with "red triangles" - you did not introduce such triangles in the description - it becomes a mess}
    \label{fig:bbbbr_example}
\end{figure}

In order to simplify notation we will represent the black/grey tetrahedra by the black/grey dots, and the prisms by the blue/red dots, such that a blue dot represents a prism with two black triangles (and three transparent rectangles) and a red dot is a prism with  two grey triangles (and also three transparent rectangles). In the 4-dimensional context, the black/grey dots will correspond to the $s_{41}$ / $s_{14}$ simplices in the slab, and the blue/red dots will correspond to the $s_{32}$ / $s_{23}$ simplices, respectively. The dots will be connected by "legs" of various types, representing the different types of connections (through colored triangles or rectangles) in the $t+\frac{1}{2} $ plane. Thus a solid black/grey  leg will represent a black/grey triangle, and the dashed leg will be a transparent rectangle, see Fig. \ref{fig:P_types}. In order to preserve the topological restrictions, only the legs of the same color/type can be connected. All possible  connections between colored dots are presented in Fig \ref{fig:connectivities} (up to mirror-reflections). 
As, by definition, the manifold constraints of the original triangulation are not violated, and the description of the triangulation in the $t+\frac{1}{2}$ plane is still a manifold (a three-dimensional one), it is in one-to-one correspondence with the transition tensor of the triangulation from slice $t$ to $t + 1$. An example (part of the) $t+\frac{1}{2}$ slice of a CDT triangulation and the corresponding graph with colored dots and various types of legs is presented 
in Fig.~\ref{fig:bbbbr_example}. \\

One should note that if, in the original CDT triangulation, two $s_{41}$ simplices are connected to the same vertex at $t+1$ then these simplices correspond necessarily to two adjacent tetrahedra in the $t+\frac{1}{2}$ plane, or in the graphical representation two black dots connected by a solid black line. The same is of course true for the mirror-reflected $s_{14}$ simplices and thus the grey dots connected by a solid grey line. Additionally, using the graphical representation one can recognize the links of the original CDT triangulation as closed loops in the colored dot graphs. Closed solid loops are spatial links (black on slice $t$ and grey on slice $t+1$), while closed dashed loops are time-like links of the original triangulation. Then, the coordination number of a link in the original triangulation is related to the number of dots along that loop. Another important feature of this graphical representation is, that the vertices of the original triangulation are represented as 3-dimensional objects defined by the surrounding colored dots and closed loops. As it was already mentioned, the above graphical representation contains only elements of the $t + \frac{1}{2}$ plane of a slab, therefore the true coordination number of spatial links will actually also depend on a similar graph in the adjacent slab.\\

As the CDT moves are local, i.e., they change only the interior of a small region in a CDT triangulation, the connection to the outside region of the triangulation is preserved, which, in the graphical representation, manifests itself by the fact that the type and number of  external legs remain unchanged when the move is performed. \\

Now, we are ready to discuss the moves with the graphical representation defined above. In the following discussion, if only black or black-and-blue dots are shown, then recoloring black to grey and blue to red will lead to the mirror-reflected version of the movie. 
 
\subsection{Move-2}

\begin{figure}[h]
    \centering
    \includegraphics[width =0.45\textwidth]{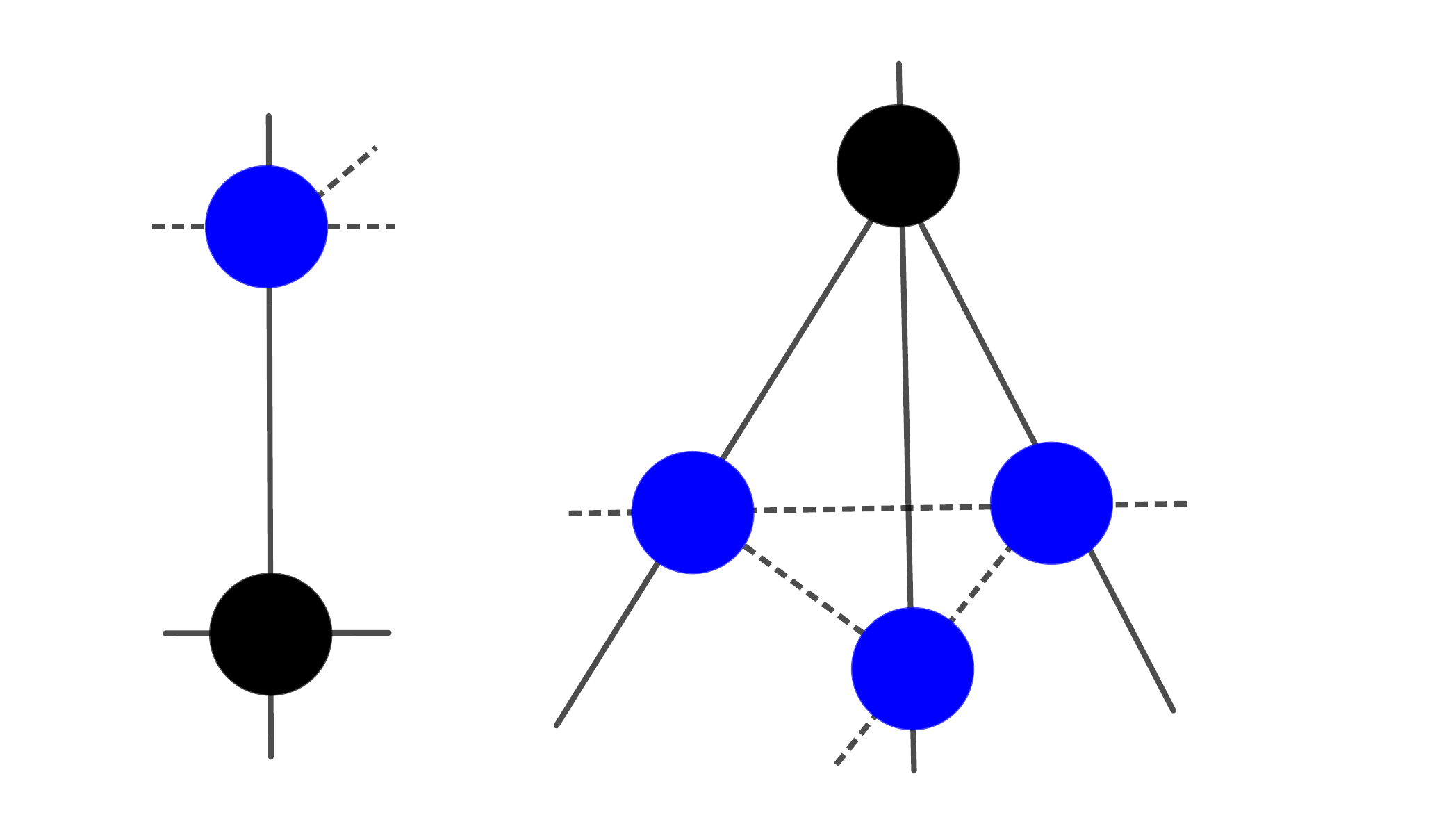}
   \includegraphics[width = 0.45\textwidth]{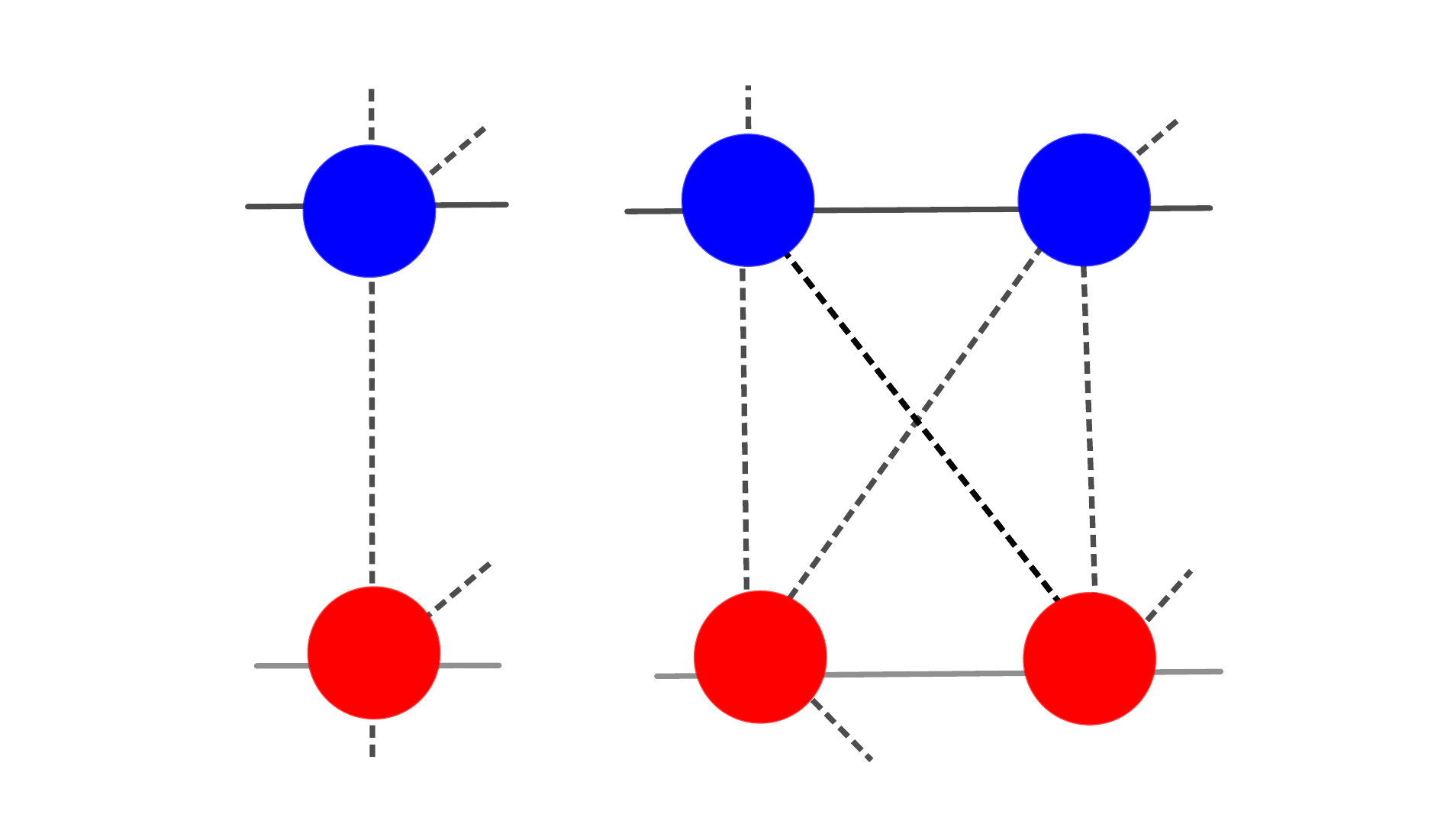}
    \caption{Move-2: version-1 (left) and version-2 (right). In the CDT triangulation, it replaces a (tetrahedral) interface between 4-simplices with a link, creating additional three 4-simplices.}
    \label{fig:m2}
\end{figure}

"Move-2" is a move that changes the interface between two (black-blue or blue-red) dots and increases the number of dots by two.  It exists in two versions. Version one can be done between a black and a blue dot. The move removes two dots and replaces them with four, see Fig. \ref{fig:m2}. After the move, the black dot will be connected to the external leg, which was earlier connected to the blue dot, and, at the same time, all of the original black dot's external legs will become the external legs of the three new blue dots. These blue dots are also connected via dashed legs. The second version of the movie can be done between a blue and a red dot. The move replaces the dashed line between the original blue and red dots with four dashed lines between the  blue and red dots. These new blue and red dots are connected to the external dashed legs of the original configuration.

\subsection{Move-3}

\begin{figure}[h]
    \centering
    \includegraphics[width = 0.45\textwidth]{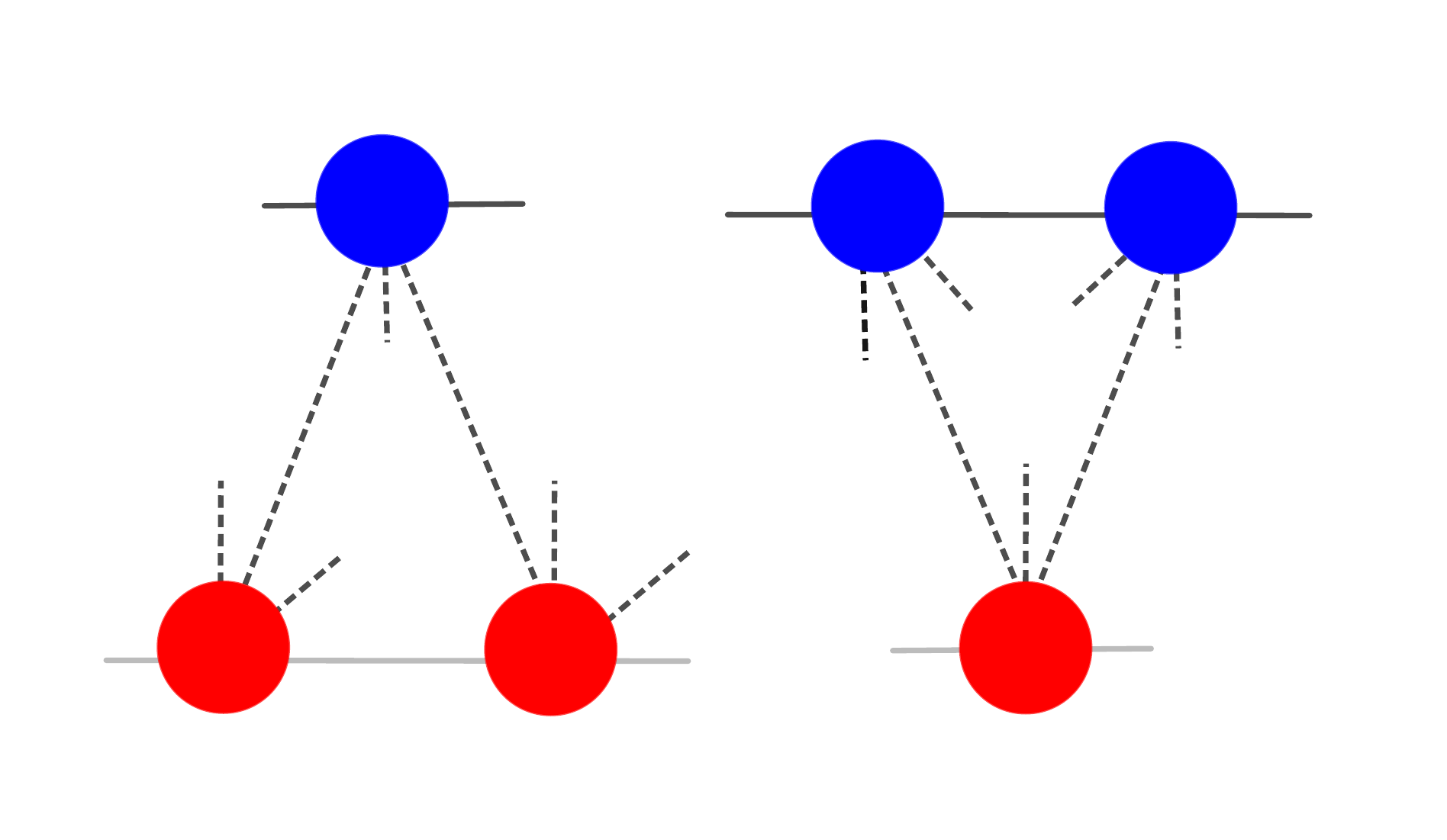}
    \includegraphics[width = 0.45\textwidth]{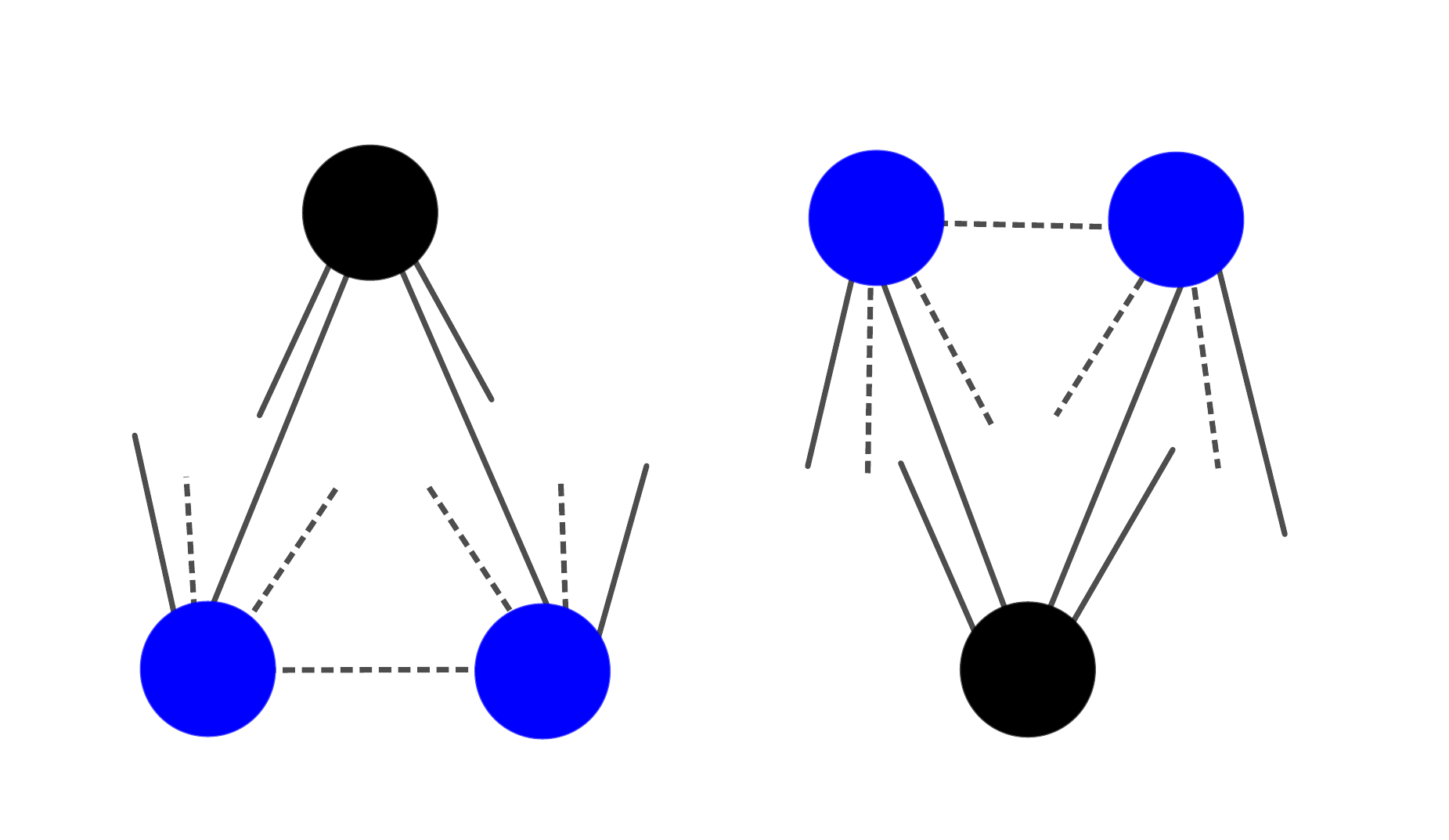}
    \caption{Move-3: version 1 (left) and version 2 (right). In the CDT triangulation, it replaces the triangular interface with a dual one. }
    \label{fig:m3}
\end{figure}

The next move is "move-3", shown in Fig. \ref{fig:m3}, which is an analog of the "flip" move used in the two-dimensional CDT. It also comes in two versions. In version one, it replaces one blue and two red dots with one red and two blue dots, which corresponds to replacing an $s_{12}$ triangle with an $s_{21}$ in the CDT triangulation. The second version removes two adjacent blue dots connected with the black dot and places them on the other side, i.e. connects them to two external legs of the original black dot. At the same time, the black dot gets connected to the two external legs originally connected to the blue dots. The move does not change any of the global numbers in the triangulation.

\subsection{Move-4}

\begin{figure}[h]
    \centering
    \includegraphics[width = 0.5\textwidth]{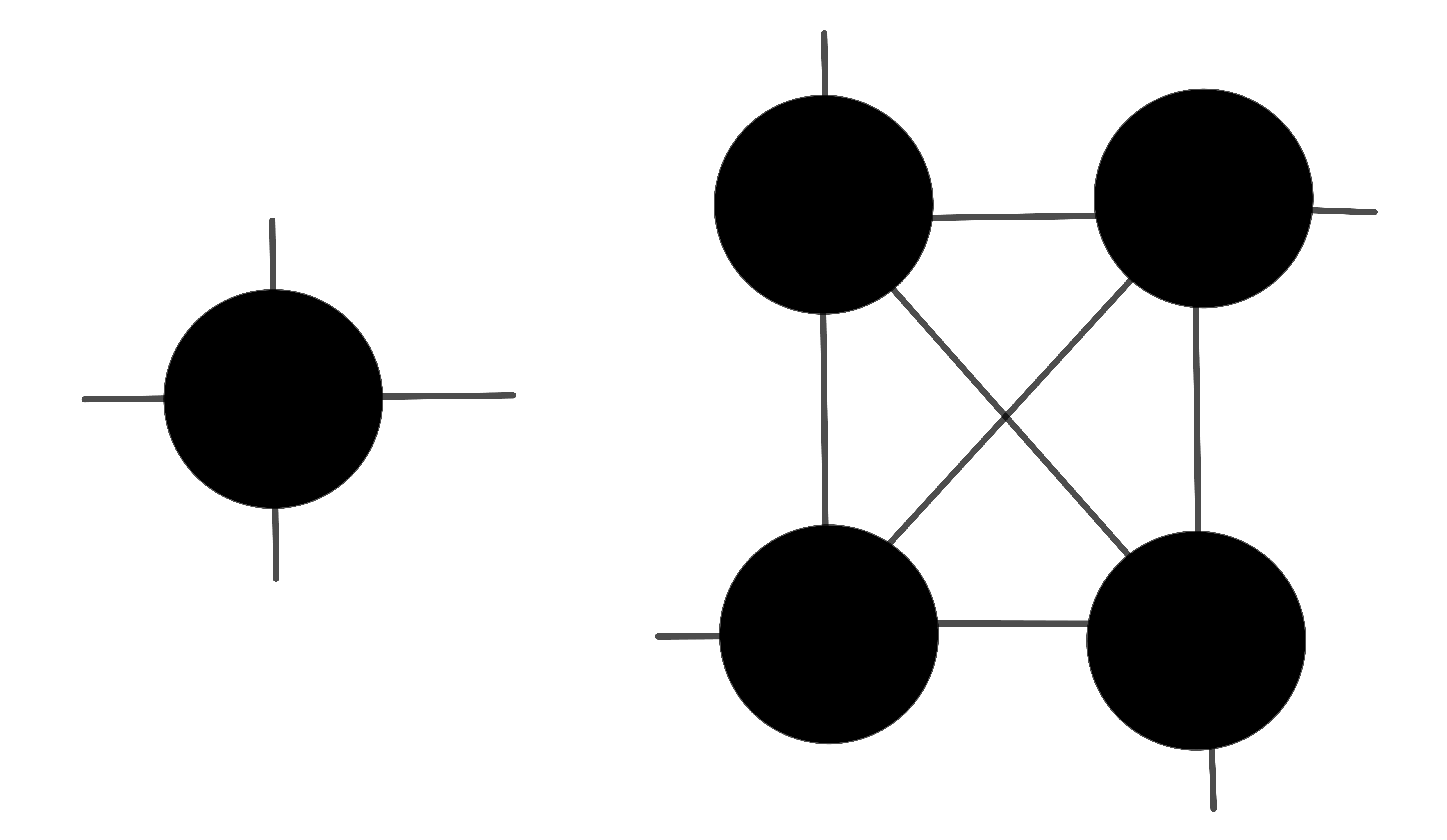}
    \caption{Move-4 replaces a black dot with four fully connected black dots, connecting each of them to the external link of the original configuration. In the CDT triangulation, it adds a vertex inside an $s_{41}$ simplex, replacing the simplex with four new $s_{41}$ simplices.}
    \label{fig:m4}
\end{figure}

The "move-4" is one of the simplest ones, and is shown in Fig. \ref{fig:m4}. Move-4 and its inverse are effectively a special case of a "split-merge" move. It removes a black dot and replaces it with a fully connected set of four black dots. The four dots are also connected to the external legs of the original configuration, one by one.

As every solid loop in the graphical representation corresponds to a spatial link, and as it is visible in the right panel of Fig. \ref{fig:m4} there are four such solid loops, thus in the real triangulation four new spatial links are created. As all the four black dots are adjacent to each other, it can happen only if they share a vertex, thus the move creates a vertex in the original triangulation, this vertex has coordination number  four.\footnote{In fact, the coordination number is eight, as there are additional $s_{14}$ simplices in the previous slab.} 

\subsection{Move-5}
\begin{figure}[h]
    \centering
    \includegraphics[width = 0.5\textwidth]{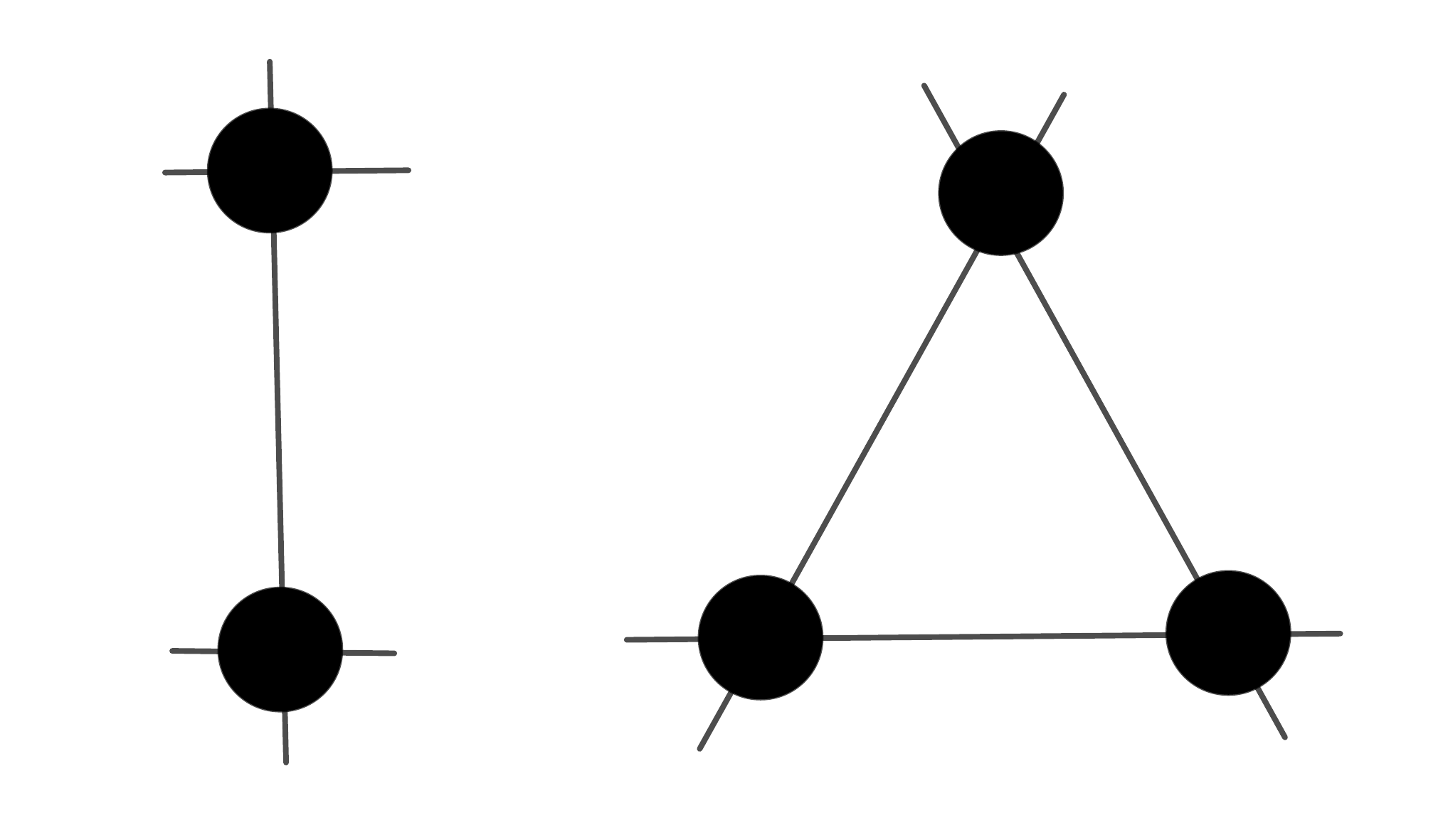}
    \caption{Move-5 replaces two adjacent black dots with three black dots. In the CDT triangulation, it creates a spatial link with coordination number three. The link is signaled by a solid black loop in the graphical representation.}
    \label{fig:m5}
\end{figure}

The last move is "move-5", shown in Fig. \ref{fig:m5}.

The move takes two adjacent black dots (tetrahedra) and replaces the triangular interface formed by the three common vertices with a link that connects the remaining two vertices. The move creates a link with coordination number three\footnote{In fact, the coordination number is six, as there are additional $s_{14}$ simplices in the previous slab.}, signaled by the solid black loop connecting the three black dots on the right panel in Fig. \ref{fig:m5}. The inverse move requires a link with coordination number three.\\ 

One should also note, that in the  CDT code, we use the full four-dimensional triangulation. In the graphical representation, it could be achieved by  adding a single solid external leg to each black/grey dot. This way  each black (grey) dot of the $t+\frac{1}{2}$ plane would be connected to a grey (black) dot of the previous (next) slab, defined by the plane at $t+\frac{1}{2}-1$ ($t+\frac{1}{2}+1$). Move-4 and move-5 are the only moves that are affected by the neighboring slabs, and the connected grey/black dots of the adjacent slabs would behave exactly the same way as the black/grey dots behave in the above graphical representation description. \\

So far we discussed the moves which are currently used in the MC simulations of the four-dimensional  CDT. In principle, one could try to define some new moves, but doing so is a hard task as they must be efficient numerically and their required components (a vertex/link/triangle with a given coordination number) must be easy to be tracked during the simulations, e.g., a vertex with a given coordination number is easy to track, but it is not the case for more complex structures. In appendix \ref{AppendixB} we discuss some proposals for new moves with the help of the  graphical representation defined above. 


%% file: Chapters/Chapter3.tex

\chapter{Empty Universes} 
\label{chapter3}  

\textit{This chapter gives a brief summary of the following articles: \cite{pub1, pub2, pub3} which are presented as publications in the last chapter}.\\\\

\section{About criticality at phase transitions}
\textit{"My memory for figures, otherwise tolerably accurate, always lets me down when I am counting beer glasses”} - \textbf{Ludwig Boltzmann}\\\\

Transitions between phases can typically be described by simple models. The general idea is that one has to find some (macroscopic) properties of a given physical system that characterize the phases, and track their changes by varying the coupling constants of a given theory. For finite size systems, as the ones observed in numerical MC simulations,  one cannot observe true phase transitions but only pseudo phase transitions, i.e. cross-overs, as finite size systems have finite thermodynamic potentials and also all derivatives of such potentials are finite.  Anyway, one can observe that order parameters, related to some derivatives of the thermodynamic potential, become more and more singular with increasing system size (lattice volume), and by taking the volume to infinity one encounters a true phase transition. There are several phase transitions that exhibit similar behavior and can be characterized the same way, thus they will belong to the same universality class. Phase transitions belonging to the same universality class will show the same type of finite volume scaling properties. It manifests itself by universal values of scaling exponents, which can be used to measure the order of a phase transition. The notion of an order of a phase transition was introduced by Ehrenfest, who characterized  phase transitions using derivatives of thermodynamic potentials (e.g. free energy, entropy, chemical potential... etc). If the $n^{th}$ order derivative diverges at the transition point then one has an $n^{th}$ order phase transition. This picture was refined by Landau by introducing the notion of local order parameters (OP) (in the context of the Ising model), and Ginzburg \cite{ginz} improved Landau's theory by adding fluctuations to the model. Since then the classification of phase transitions shifted towards distinguishing between two types of phase transitions: first-order, which has a divergent first-order derivative of the thermodynamic potential, and  higher-order (also called continuous), where the second- or higher-order derivative diverges. Furthermore, in the above classification, there is a relation between the order of a transition and the correlation length. For a first-order transition one typically has finite correlation lengths, while divergent correlation length signals a continuous phase transition. In the lattice approach, as those discussed in the thesis, finding a phase transition where the correlation length diverges is crucial as only then can the lattice spacing be taken to zero to reach the continuum limit, while keeping the physical quantities fixed. One should also note, that recent models of solid state physics revealed, that the Landau-Ginzburg characterization can also fail, when a phase cannot be characterized by a local order parameter, but rather by long-range entanglement, called topological order \cite{topord}. If the nature itself exhibits such phenomena where the traditional description of phase transitions fails then we cannot take it for granted that such a description works for a model of quantum gravity. Nevertheless, in this chapter we will stick to the Landau approach. As we will later see, most CDT transitions falls in this category, however some of them show atypical features having  relation to phase transitions involving the topology of the underlying manifold.\\

We will present the idea of critical exponents by taking an example of the Ising model \cite{Ising}. The Ising model is one of the simplest lattice models of spin chains with nearest-neighbor interactions. In one dimension it is literally a chain of spins, in two dimensions the spins are placed in vertices 
of a regular lattice. The Hamiltonian of the model is:

\begin{equation}
    \mathcal{H}_{I} = -J \sum_{i\leftrightarrow j} \sigma_i \sigma_j,
\end{equation}
where $J$ is a coupling constant, $\sigma_i = \pm 1$ is a spin and the sum is over nearest neighbors in the lattice. Including an external magnetic field $h$ one may write the partition function including the magnetization ($M = \sum_i \sigma_i$) as

\begin{equation}
    \mathcal{Z}(T,h) = \sum_{\{\sigma_i\}} e^{-\beta(\mathcal{H}_{I} -hM)},
\end{equation}
where $\beta = \frac{1}{k_b T}$ is the inverse temperature and the sum is over all possible spin configurations. An example order parameter is $\mathcal M$, the average magnetization:

\begin{equation}
    \langle \mathcal{M} \rangle = \frac{\partial \mathcal{Z}}{\partial(\beta h)} = \frac{1}{\mathcal{Z}}\sum_{\{\sigma_i\}} M e^{-\beta(\mathcal{H}_{I} -hM)}.
\end{equation}
The susceptibility ($\chi$) is the first-order derivative of the magnetization:

\begin{equation}
    \chi(T,h) = \frac{1}{V}\frac{\partial \langle \mathcal{M} \rangle }{\partial h},
\end{equation}
where $V$ is the volume of the system. The relation which follows from this is then:

\begin{equation}
    \frac{V\chi}{\beta}
    = \langle \mathcal{M}^2 \rangle - \langle \mathcal{M} \rangle^2,
\end{equation}
so the susceptibility $\chi$ is related to the magnetization variance. Taking the continuum limit, i.e., the lattice spacing $a\to 0$ and the lattice size $N \to \infty$  such that the physical volume  $V=a^d N $ (where $d$ is the  dimension of the system) remains constant, one can compute a two-point correlation function, where the susceptibility will depend on the spatial distance of two points in the following way

\begin{equation}
    k_bT\chi = \frac{1}{V}\int dx \int dx' [\langle m(x)m(x')\rangle - \langle m(x) \rangle \langle m(x') \rangle] = \int dx \langle m(x) m(0) \rangle_c, 
\end{equation}
where $\langle m(x) m(0) \rangle_c$ denotes the connected correlator $G_c$, which  typically decays exponentially with some characteristic correlation length $\xi$. In case of $|x| < \xi$ the susceptibility will behave as:

\begin{equation}
    k_B T\chi < g \xi^d,
\end{equation}
where $g$ is a constant, yielding the correlation length divergent in case of  the  divergent susceptibility. The correlation function can be measured with respect to the change of the temperature yielding

\begin{equation}\label{eq:scalrel}
    \xi(T, H = 0) \propto  |T-T_{crit}|^{-\nu},
\end{equation}
which means that the correlation length scales  with the critical exponent $\nu$ as $T$ approaches the critical temperature $T_{crit}$.\\

In the lattice MC measurements, the largest available correlation length is controlled by the lattice size $N$, i.e., $\xi(N) \sim V^{1/d} =  a N^{1/d}$. Using equation (\ref{eq:scalrel}) it follows that the (pseudo-) critical temperature, or in the general the (pseudo-) critical coupling constant, which triggers the phase transition, will show the following finite-size scaling dependence:
\begin{equation}
    T^{crit}(N) = T^{crit}(\infty)+ \text{const} \times {N}^{-\frac{1}{\gamma}},
\end{equation}
where $T^{crit}(\infty)=T^{crit}$ is the (true) critical temperature in the thermodynamical limit ($N\to \infty$), and $\gamma=\nu d$ is the critical scaling  exponent. The above scaling relation  was used in the studies presented in this chapter. One should note that for a higher-order transition one expects the scaling exponent $\gamma>1$, while for a first-order transition  one typically has $\gamma=1$. 

\section{Order parameters and the internal structure of the configurations}
\textit{"Einfach wie möglich aber nicht einfacher." / "Everything should be made as simple as possible, but not simpler. "} - \textbf{Albert Einstein}\\\\

\label{sec:OPs}

The idea behind Monte Carlo numerical simulations is quite simple. As discussed in the Chapter \ref{chapter2}, one can generate a set of  (almost)  statistically independent configurations using a Markov chain of "moves" applied randomly with a proper transition probability, and then use it to estimate expectation values or correlators of observables, such as order parameters related to  phase transitions. The Regge action of CDT, see eq. (\ref{eq:ation_kappa}), contains  a linear combination of the total number of vertices and simplices of various types, weighted by the bare coupling constants. When changing the couplings the (averaged) values of the above mentioned {\it global} numbers, and also other characteristics of the triangulations, change as well. Therefore, these observables  can be used to define the order parameters of CDT.\\

In the four-dimensional pure gravity CDT model we have three coupling constants, thus we can use them to parametrize the phase-diagram. As we will see in this section, the numerical MC simulations used in CDT reveal four distinct regions  (phases) in the CDT parameter space. In order to be able to perform the MC simulations we fix the (average) lattice volume $\bar{N}_{41}$. The volume fixing means that, throughout the MC simulation, the observed lattice volume ${N}_{41}$ will perform fluctuations around the fixed value $\bar N_{41}$, also restricting the values of related quantities, such as coordination numbers of various sub-simplices. It also corresponds to fixing the total spatial volume $\sum_t V_3(t)=\frac{1}{2} N_{41}$ (total number of spatial tetrahedra in slices with integer lattice time coordinate $t$). Using different values of $\bar{N}_{41}$ one is able to perform the finite volume scaling analysis, where the change in the order parameters (OPs) can be related to the change in the lattice size, as discussed in the previous section. This way one can track the approach to the thermodynamical limit. In order to enforce fluctuations of the lattice volume $N_{41}$ around $\bar{N}_{41}$ it is also necessary to tune the bare cosmological coupling constant $\kappa_4\to \kappa_4^{c}(\kappa_0, \Delta,\bar N_{41})$. This way one trades the  $\kappa_4$ coupling for the  $\bar{N}_{41}$ volume fixing. The fixing slices-off a two-dimensional hyper-surface $\kappa_4(\kappa_0,\Delta)$ from the full parameter space for fixed $\bar{N}_{41}$.\\

Already, before starting any deeper analysis, one can  look at the freedom of the {\it global} numbers characterizing a CDT triangulation and appearing in the bare Regge action (\ref{eq:ation_kappa}), i.e., $N_0$, $N_{41}$ and $N_{32}$. A single triangulation (a path in the path integral) is itself physically not meaningful, however if some features of a given triangulation repeat in the ensemble of generic triangulations observed in a given phase, then in such a case it makes sense to discuss these features of a particular triangulation, as they will also appear in the expectation values (averages) of the measured observables. As all global numbers ($N_0, N_{41}$ and $N_{32}$) are independent of each other\footnote{Although there are theoretical lower and upper limits, which in itself features an unsolved mathematical problem.}, a configuration with a given fixed $N_{41}$ can have small or high number of vertices or other (higher-dimensional) sub-simplices, which will result in a significantly different distribution of these numbers in different phases. One can also imagine that even if all global numbers $N_0, N_{41}$ and $N_{32}$ were constant, the local distribution of vertices and (sub-)simplices within a configuration can be not homogeneous. Even though every simplex has exactly 5 neighbors, every vertex has a different number of simplices connected to it, which gives rise to the possibility of non-trivial vertex coordination number distributions, where some vertices are shared by only a few simplices, but some other vertices will have a large coordination number. \\

The (ratios of) global numbers of a configuration are natural OPs as they are the simplest degrees of freedom in our geometric setup. Thus the first two OPs  can be defined as:
 
 \begin{equation}
\mathcal{O}_1 = \frac{N_0}{N_{41}} \quad , \quad    \mathcal{O}_2 = \frac{N_{32}}{N_{41}}.
\end{equation}
There are also some OPs which are not global in the sense that they are related to a local distribution of (sub-)simplices in a triangulation. For example, as we have a foliation, we can measure the distribution of vertices as a function of the lattice time coordinate $N_0(t)$. We can also measure similar distributions for the  4-simplices, but, as the above simplices  are four-dimensional objects, instead of talking about fixed $t$ we rather talk about the four-dimensional {\it slab} (part of the triangulation between $t$ and $t+1$), and denote the number of 4-simplices in the slab by $N_{41}(t)$ and $N_{32}(t)$, respectively. For example, the (three-dimensional) volume profile, introduced in Chapter~\ref{Chapter1}, is simply given by: $V_3(t)=\frac{1}{2} N_{41}(t)$. If the adjacent spatial slices contain similar number of tetrahedra, then $N_{41}(t)$ will be a flat function but if the volume profile has a non-trivial shape, then the difference between the adjacent slices will be larger. Therefore, one can define the third OP which quantifies this:

\begin{equation}
\mathcal{O}_3 = \frac{1}{N_{41}} \sum_{t}
(N_{41}(t)-N_{41}(t+1))^2.
\end{equation}
The shape function $\langle V_3(t) \rangle$ (the volume profile) could potentially be also used as an order parameter. For example, in the case of  spherical CDT the $\langle V_3(t) \rangle \approx \cos^3{(t)}$ \cite{sphere} and in the toroidal CDT it is $\langle V_3(t) \rangle = \bar{N}_{41}$ in the semi-classical phase~($C$), while it has a completely different shape in other phases. An example of a local OP is $\mathcal{O}_4$, defined by the highest vertex coordination number among the set of vertices: 
\begin{equation}
    \mathcal{O}_4 = \frac{1}{N_{41}} \argmax_v (coord(v)), 
\end{equation}
where $v$ is a set of all vertices in a  triangulation. One can as well measure the distribution of this quantity in the lattice time $t$.\\

Additional OPs can  also be useful. For example, one can measure the total number of $type_1$-type of simplices neighboring $type_2$-type of simplices in a triangulation, where $type$ refers to a general 4-simplex. The various types of these numbers are summarized in Table \ref{table:ABCD}.

\begin{center}
\begin{tabular}[h]{ a c c c c c }
\rowcolor{Gray}
\mc{1}{}  & \mc{1}{$s_{41}$} & \mc{1}{$s_{32}$} & \mc{1}{$s_{23}$} & \mc{1}{$s_{14}$} & \mc{1}{sums to} \\
 $s_{41}$ & $A_1$ & $C_1$ & 0 & $E$ & $\rightarrow 5\cdot N_{41}$ \\ 
 $s_{32}$ & $C_1$ & $B_{1a} + B_{1b}$ & $D$ & 0 & $\rightarrow 5\cdot N_{32}$ \\  
 $s_{23}$ & 0 & $D$ & $B_{2a} + B_{2b}$ & $C_2$ & $\rightarrow 5\cdot N_{23}$ \\
 $s_{14}$ & $E$ & 0 & $C_2$ & $A_2$ & $\rightarrow 5\cdot N_{14}$  
\end{tabular}
\captionof{table}{The table summarizes the numbers related to the adjacency relations of 4-simplices. All rows and columns sum up to the global numbers $N_{41}$ or $N_{32}$.}
\label{table:ABCD}
\end{center}
The rows and columns of Table \ref{table:ABCD} denote the adjacent $type_1$ and $type_2$ simplices, e.g.,  $A_1$ is the total number of common faces (tetrahedra) between two $s_{41}$ simplices in a given triangulation, while $C_1$ counts the total number of tetrahedra connecting the $s_{41}$ and $s_{32}$ simplices. The parameter $B_1$ (and $B_2$), which  measures the self connectivity between the $s_{32}$ (or respectively $s_{23}$) simplices, can additionally be split into two sub-categories, depending on the type of a connection between the sub-simplices.\footnote{See discussion in Chapter \ref{chapter2}.} Subscript $a$ denotes the connectivity via a spatial tetrahedron ($s_{31}$) and subscript $b$ via a time-like tetrahedron ($s_{22}$). Even though Table \ref{table:ABCD} contains in general $10$  different additional parameters characterizing a CDT triangulation, one can show that only some of these parameters are  independent, but surprisingly not all can be expressed via the \textit{global} numbers. Taking also into account all different types of sub-simplices in a triangulation (e.g. vertices, space-like links, time-like links, spatial triangles, time-like triangles,..., etc.) the topological constraints of the CDT manifolds restrict the total number of independent parameters (including the elements of Table \ref{table:ABCD}.) to 8. The derivation of the relations is presented in Appendix \ref{AppendixA}.\\

During a Monte Carlo simulation, the topology of the triangulations, i.e., their Euler characteristic $\chi$, is fixed and to perform a simulation one also fixes the coupling constants $\Delta$ and $\kappa_0$, and tunes $\kappa_4$ to the critical value corresponding to a given lattice volume $\bar{N}_{41}$. There are then two independent global parameters that can change freely\footnote{Strictly speaking, $N_{41}$ also changes as it fluctuates around the fixed $\bar{N}_{41}$.}:  the total number of vertices $N_{0}$ and the total number of $s_{32}$ plus $s_{23}$  simplices $N_{32}$. Apart from the above mentioned global parameters, there are still three independent  parameters left, one can choose, e.g., $C_1, C_2$ and $D$. Statistically $\langle{C}_1\rangle \approx \langle{C}_2 \rangle$, therefore one can effectively increase the number of order parameters by two, defining:

\begin{equation}
\mathcal{O}_5 = \frac{C_1+C_2}{N_{41}}= \frac{C}{N_{41}},
\end{equation}
and 
\begin{equation}
\mathcal{O}_6 = \frac{D}{N_{41}}.    
\end{equation}

In next section we will show how to use the OPs to analyze the phase-diagram of the CDT model.

\section{Phase transitions}

\textit{"God does not play dice with the Universe!"} - \textbf{Albert Einstein}\\\\

\begin{figure}[h]
    \centering
    \includegraphics[width = 0.8\textwidth]{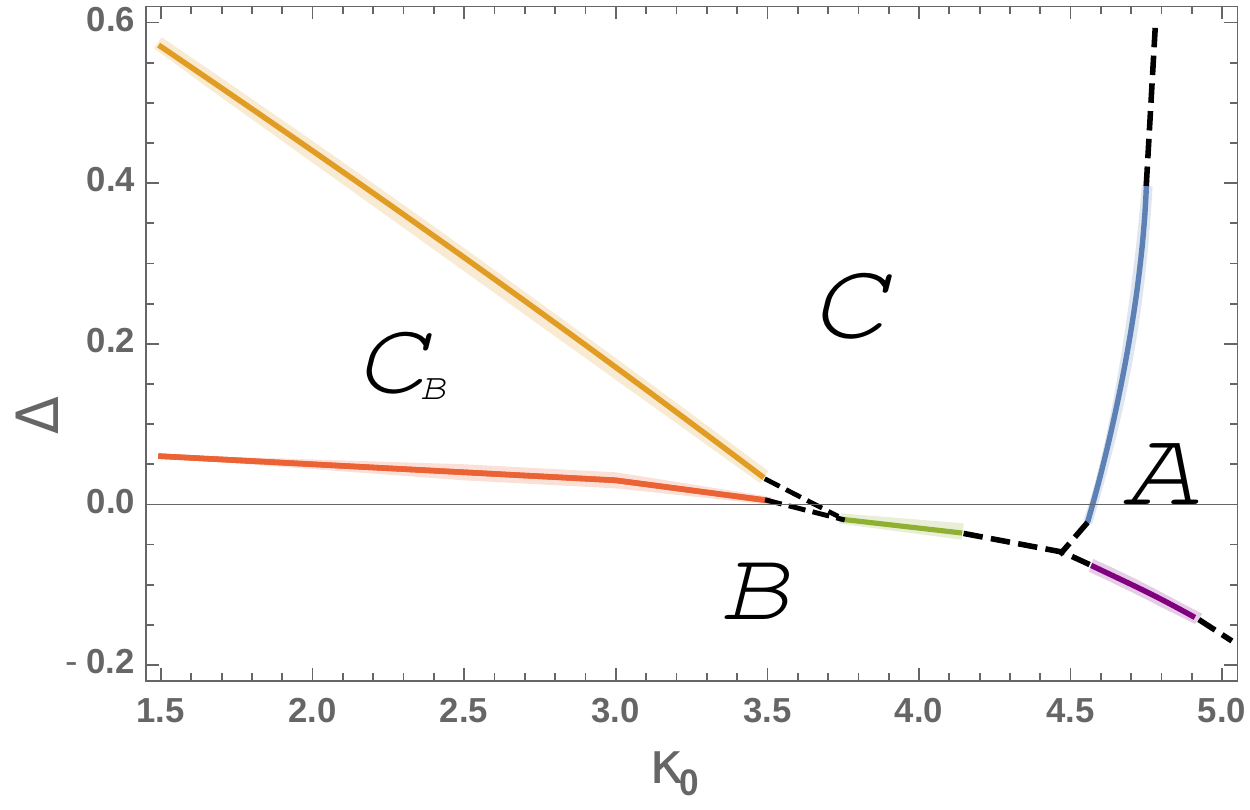}
    \caption{The phase-diagram of CDT, which shows four different phases: $A$ (branched polymer), $B$ (collapsed), $C$ (de Sitter) and $C_b$ (bifurcation).}
    \label{fig:phase_diag}
\end{figure}

Albert Einstein once criticized quantum mechanics and he said: "God does not play dice with the Universe!", maybe Gods don't but we do within our numerical simulations. As it was discussed in Chapters \ref{Chapter1} and \ref{chapter2}, CDT aims to study the lattice regularized path integral of  quantum gravity using numerical MC methods. In the simplest case one deals with triangulated empty "universes", i.e., pure gravity models, without additional matter fields. The properties of CDT emerge as a result of interplay between the bare Regge action:
\begin{equation}
    S_{R} = -(\kappa_0+6\Delta) N_0 + \kappa_4 (N_{41}+N_{32}) + \Delta N_{41}, \\
\end{equation}
and the entropy of states, i.e., the number of triangulations with the same value of the bare action in the partition function (\ref{eq:partfun}). Due to this entropic nature there are several phases which can be visualized in the two-dimensional parameter space\footnote{As explained above, the third coupling constant $\kappa_4$ is tuned to $\kappa^c_4(\kappa_0, \Delta, \bar N_{41})$ corresponding to the fixed lattice volume $\bar{N}_{41}$ of a MC simulation.} $(\kappa_0,\Delta)$. As we have a two-dimensional coupling-constant space $(\kappa_0,\Delta)$, sometimes we will refer to the coupling constants as coordinates in the phase-diagram. The four phases of CDT are presented in Fig. \ref{fig:phase_diag}.\\

Even though the  CDT model is simple in its construction, the resulting complexity arises in the variety of possible configurations. For very large   (inverse) bare gravitational coupling  $\kappa_0$ one recovers  phase $A$, which is characterised by a vanishing kinetic term in the effective action of CDT, parametrized by the spatial volume $V_3(t)$ (or alternatively by the scale factor) \cite{transfer_matrix}. The internal dynamics between the simplices results in an emerging geometry with a branched-polymer structure. For low enough asymmetry parameter $\Delta$ phase $B$ can be observed. It is characterized by the vanishing time-extent of the generic geometric configurations. All  spatial tetrahedra ($3-$volume) gather in one spatial slice, and each of the two adjacent  slices features a vertex with an almost full connectivity to the 4-simplices containing these tetrahedra. The occurrence of this phase is understandable in the context of the, so-called, balls-in-boxes model \cite{balls_in_boxes, condens_model}. The most interesting region of the phase-diagram is phase $C$, also called the de Sitter\footnote{Technically the name "de Sitter" should be used only for the spherical CDT case, as the toroidal CDT volume profile is constant and does not resemble any de Sitter-like solution.} \cite{nonperturb_desitter} or the semi-classical phase, which can be mostly observed for positive $\Delta$ and medium range of $\kappa_0$. In the case of toroidal CDT, the spatial volume profile $V_3(t)$ of generic phase $C$ triangulations is constant while in the case of the spherical CDT a de Sitter-like blob with the shape $V_3(t)\approx cos^3(t)$ forms. Last but not least, the remaining phase is the, so-called, bifurcation phase or shortly phase $C_b$. The phase is characterized by the appearance of vertices of high coordination number in every second spatial slice and the formation of a blob (different from that of phase $C$) in the volume profile both in the spherical and the toroidal CDT. As the de Sitter phase is  physically the most interesting one, the phase transitions surrounding this region were studied the most, especially as the perspective UV fixed point of quantum gravity should lie at the border of this region. It was found, that the lattice spacing decreases with increasing $\kappa_0$ and slightly decreases with decreasing $\Delta$ \cite{towardcontinum}, thus the part of the phase-diagram nearby the $C-B$ phase transition is of great interest, as the two "triple" points where the phase transition lines meet are natural candidates to be the UVFP of the theory. Due to this, it is very important to analyze the scaling exponents related to the phase transitions around the triple points. This is the reason why in this section we will present results related to the three-phase transitions: $A-B$, $C-B$, and $C_b - B$. If any of them turns out to be higher-order then it will support the possibility of the existence of the UVFP. However it is also known, that first-order phase transition lines may  end at a higher-order point (e.g., in the phase diagram of water).\\

\newpage

\begin{figure}[h]
    \centering
    \includegraphics[width = 0.45\textwidth]{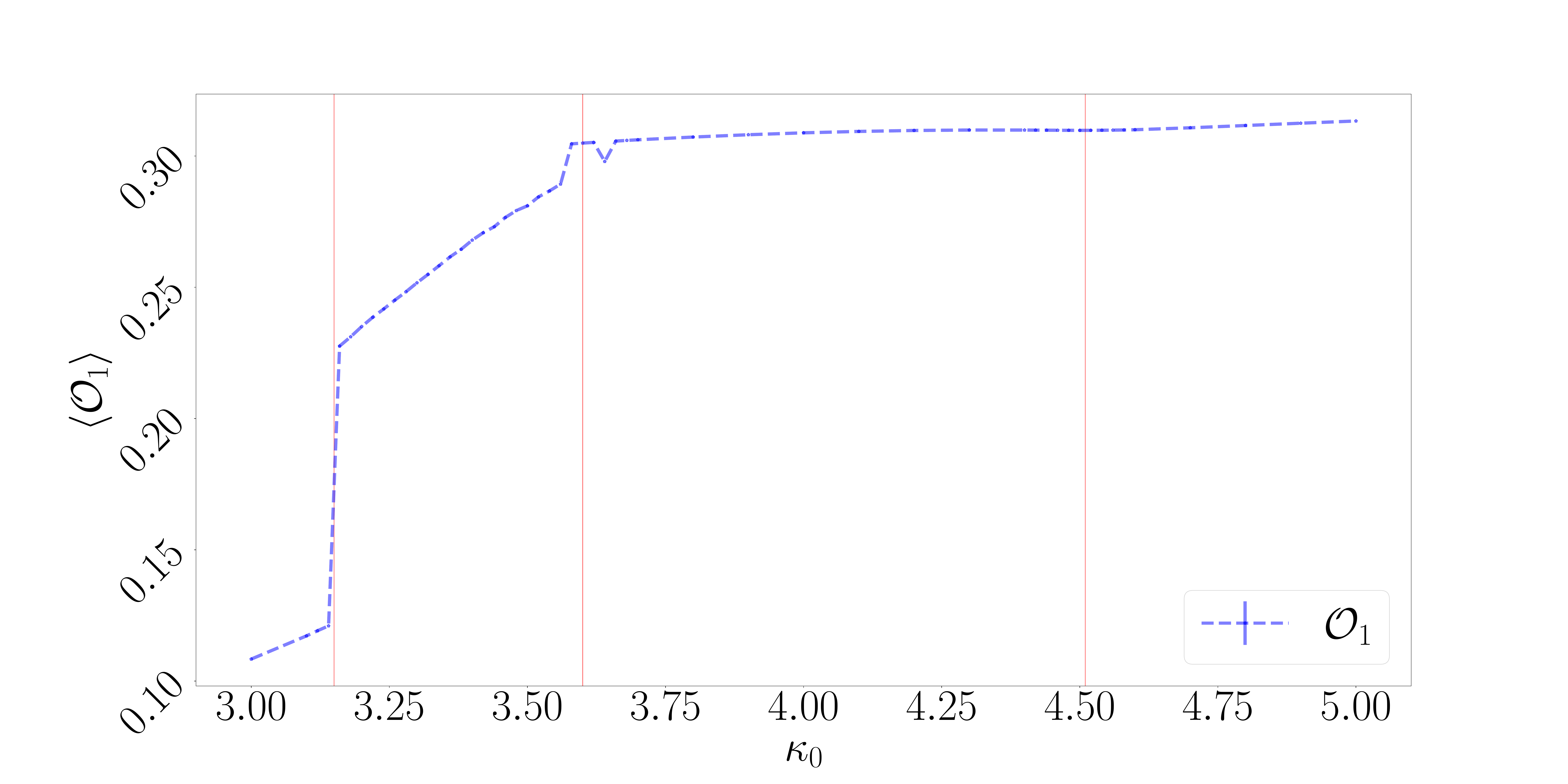}
    \includegraphics[width = 0.45\textwidth]{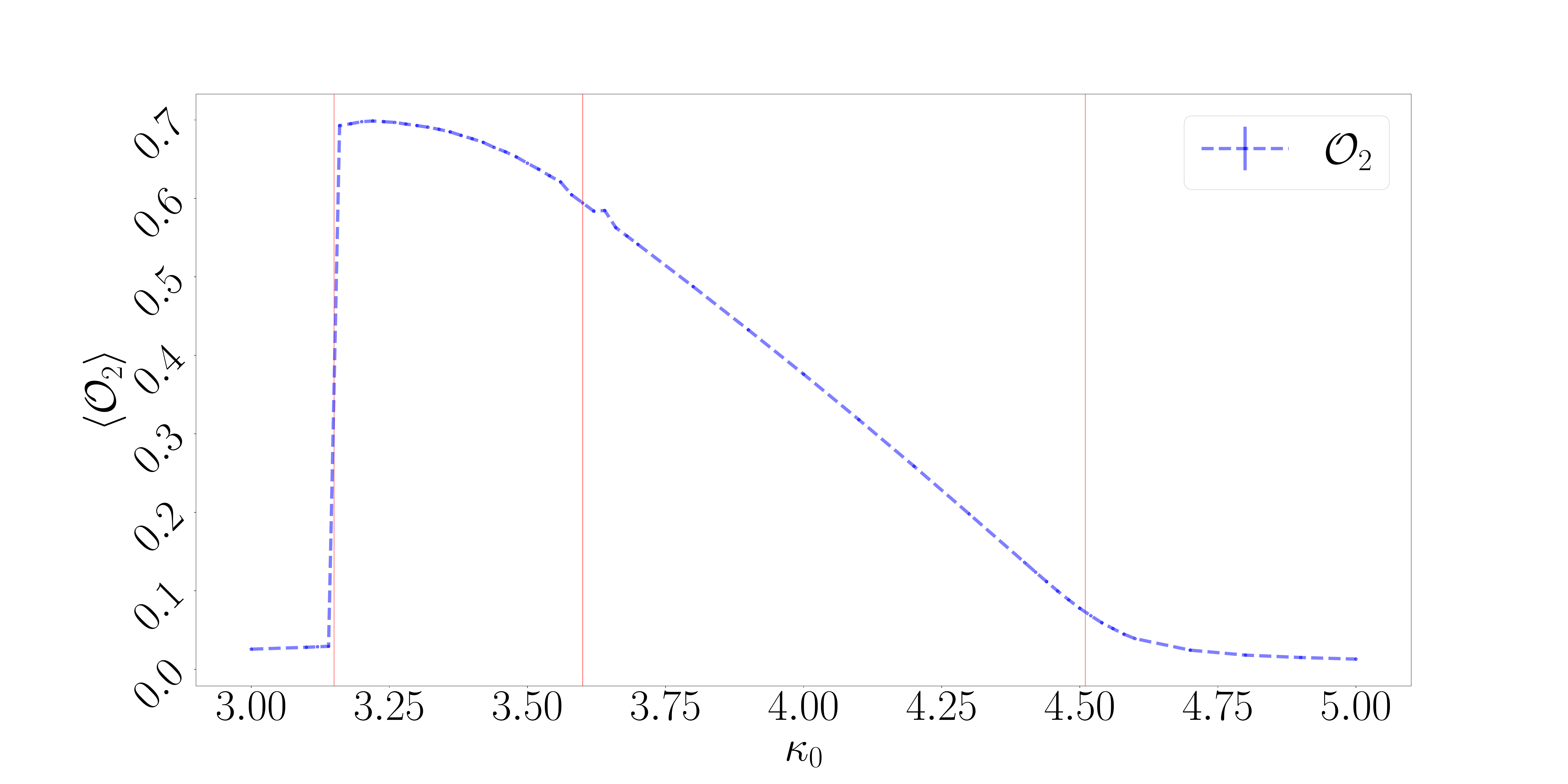}
    \includegraphics[width = 0.45\textwidth]{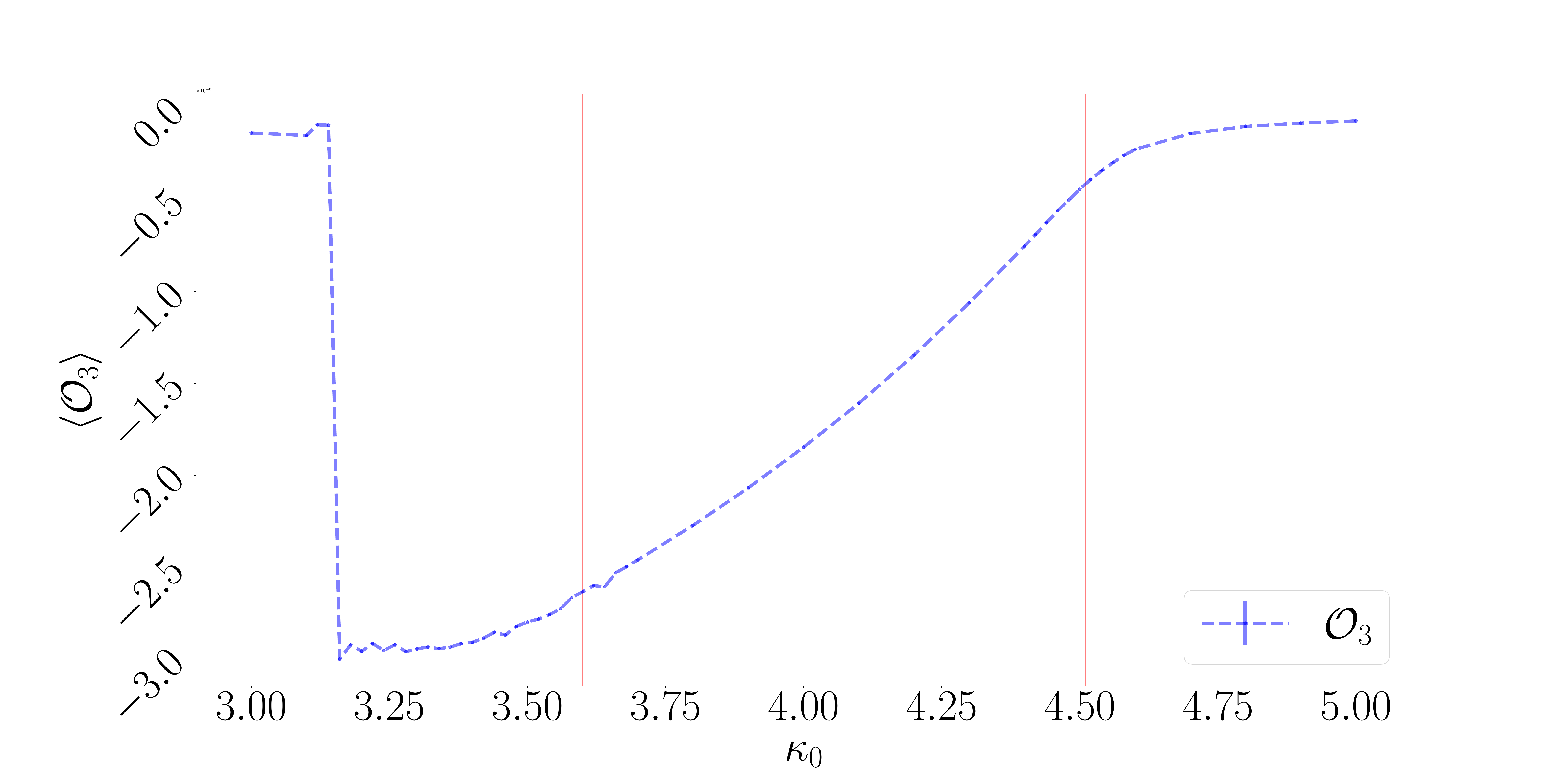}
    \includegraphics[width = 0.45\textwidth]{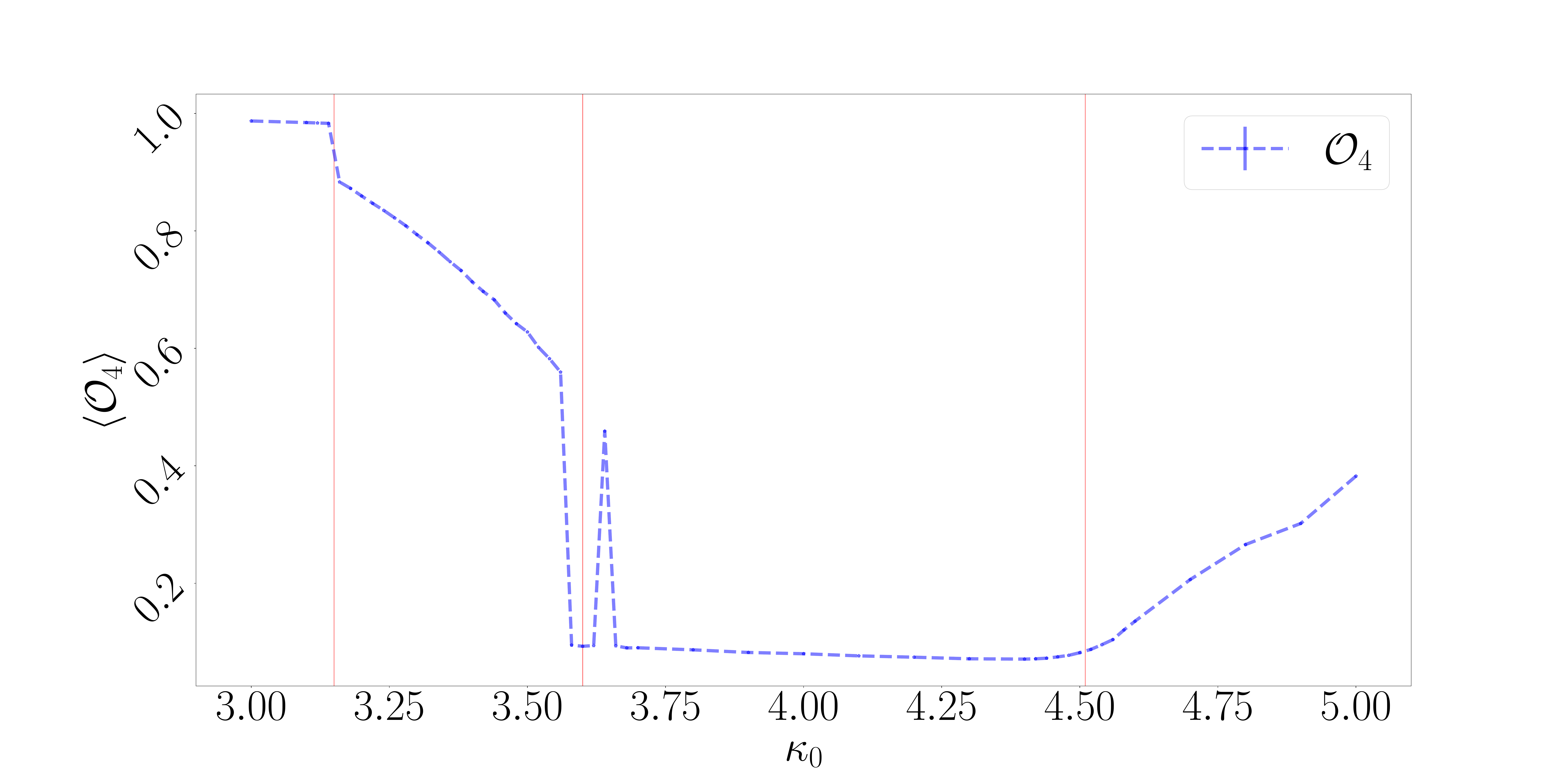}
    \caption{Example of the measured OPs from $\mathcal{O}_1$ (top left) to $\mathcal{O} 4$ (bottom rigt). The most left region on the plots is  phase $B$ (collapsed), next to it is phase $C_b$ (bifurcation), then phase $C$ (de Sitter), and next to it phase $A$ (branched polymer).}
    \label{fig:ops}
\end{figure}

\begin{figure}[h]
    \centering
    \includegraphics[width = 0.45\textwidth]{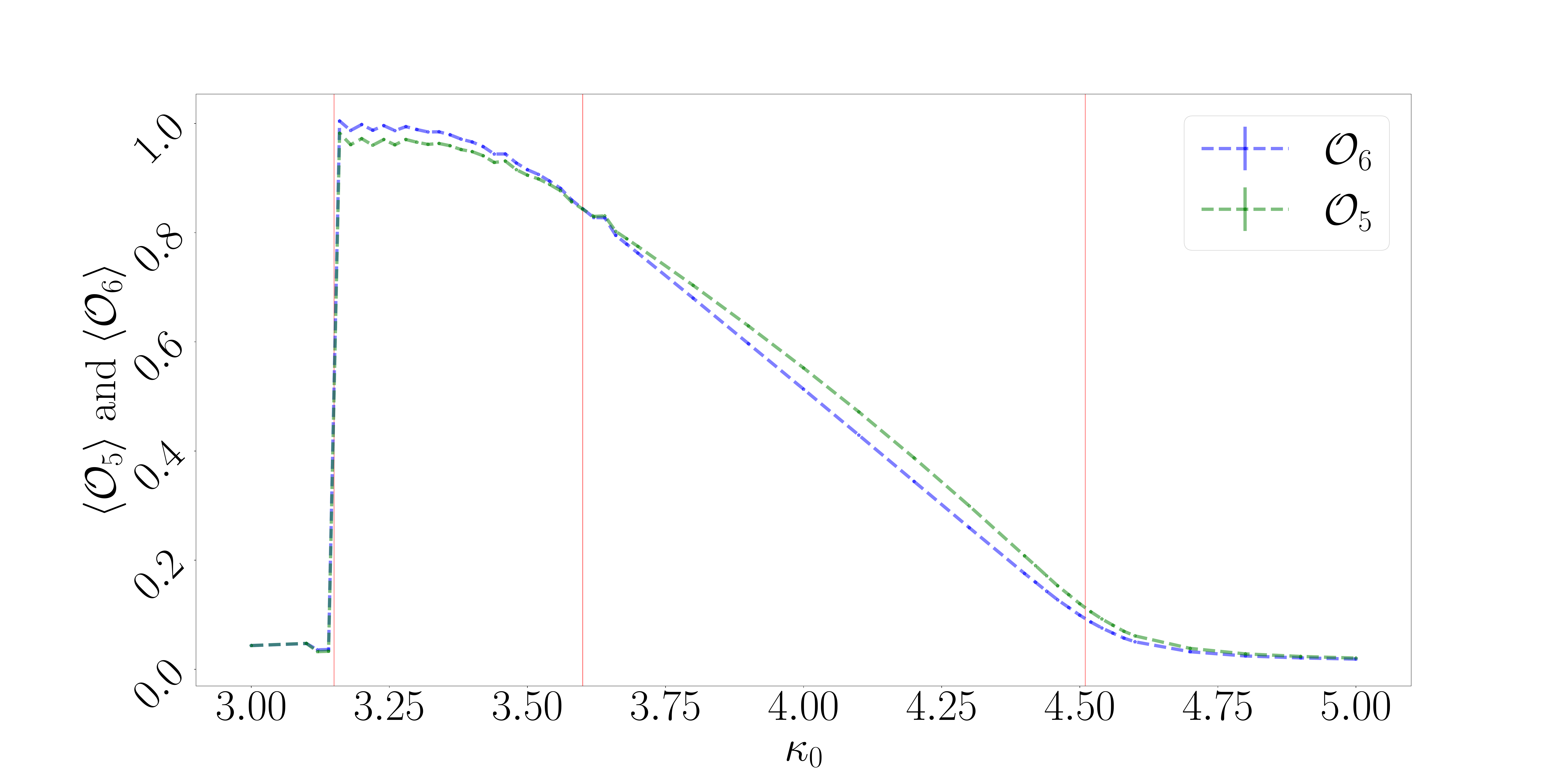}
    \includegraphics[width = 0.45\textwidth]{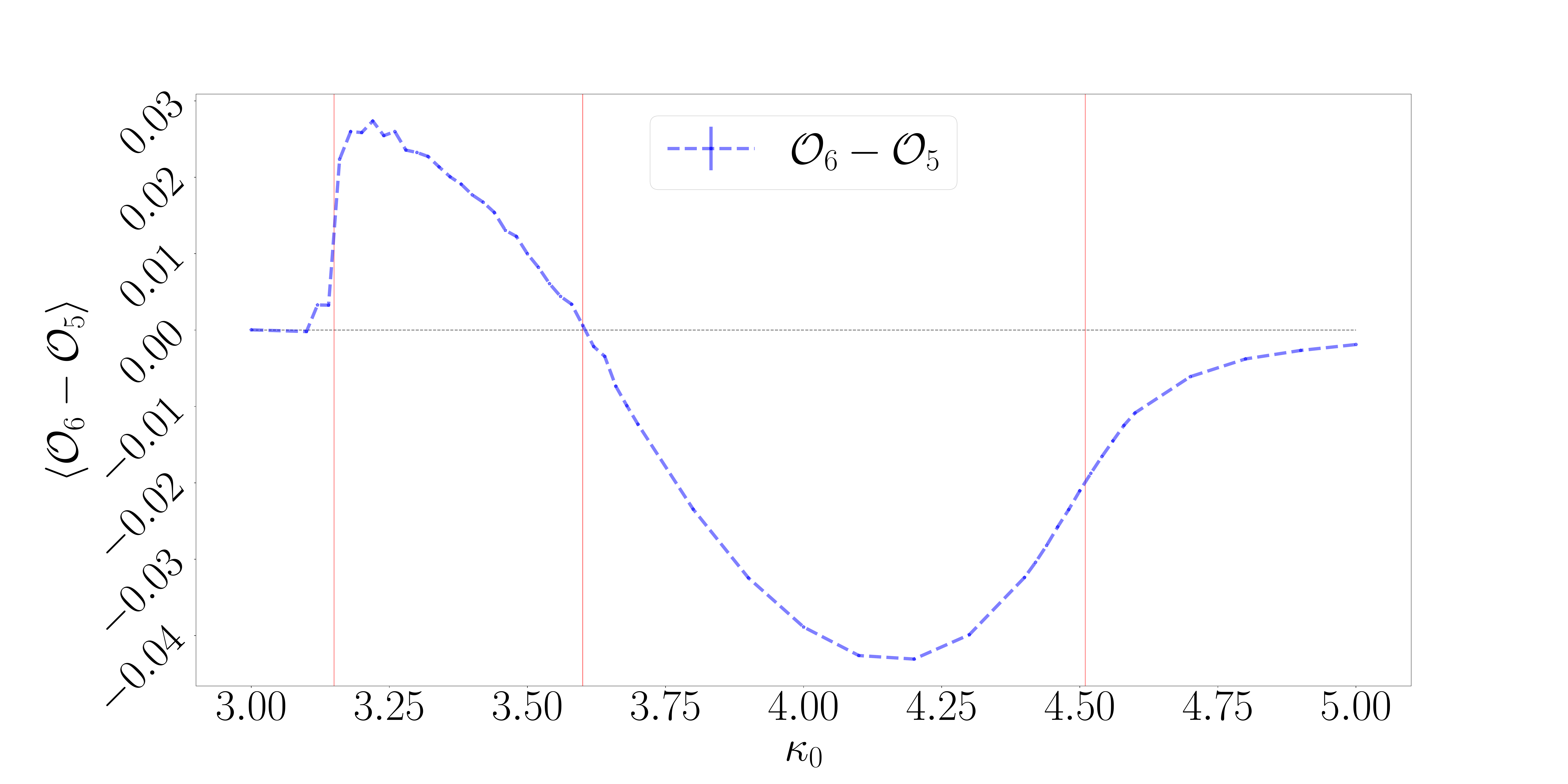}
    \caption{The new OPs $\mathcal{O}_5$ and $\mathcal{O}_6$ are shown in the left plot. Their behavior is similar to $\mathcal{O}_2$, but the crossing point additionally signals the $C_b - C$ phase-transition. The difference $\mathcal{O}_6-\mathcal{O}_5 $ is presented in the right plot, where the dashed horizontal line is at value zero.}
    \label{fig:op56}
\end{figure}

The typical way to find a phase transition is to fix one coupling constant, which will be either $\kappa_0$ or $\Delta$ in the case of 4-dimensional CDT, and then start a set of MC simulations for various values of the other coupling constant. To show the behavior of the order parameters, defined in Section, \ref{sec:OPs}., we present Fig. \ref{fig:ops} and Fig. \ref{fig:op56}, where the OPs were measured in CDT with toroidal spatial topology for fixed $\Delta= 0.02$, total lattice volume  $\bar{N}_{41} = 160k$ and length of the (periodic) lattice time coordinate (number of spatial slices) $T = 4$. 
\begin{figure}
    \centering
    \includegraphics[width = 0.8\textwidth]{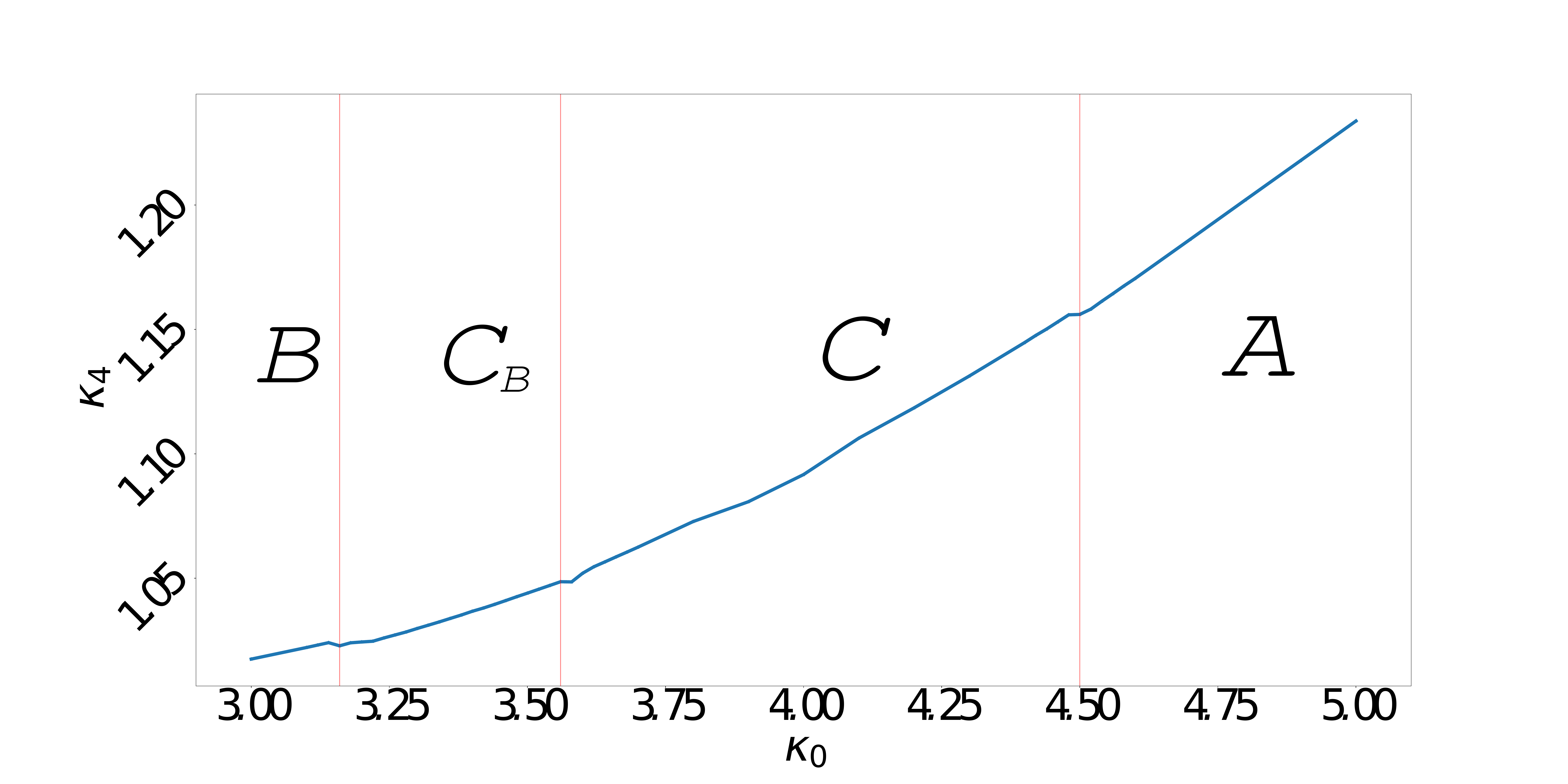}
    \caption{Values of $\kappa_4$ in the function of $\kappa_0$. Slight discontinuities in the function $\kappa_4(\kappa_0)$  signal the phase transitions, which is related to the change in entropy of the configurations on the two sides of the phase transitions. Between the vertical lines the corresponding phases are shown.}
    \label{fig:k4k0}
\end{figure}
One of the parameters that strongly depend on the volume is the bare cosmological coupling constant, that has to be tuned for each $\bar{N}_{41}$, however, its value also depends on the selected average volume. $\kappa_4(\kappa_0,\Delta)$ is a function of the other coupling constants, thus fixing one of it one may find how it changes in the function of the other (see Fig. \ref{fig:k4k0}). 

Red vertical lines visible in figures Fig. \ref{fig:k4k0} - \ref{fig:ops} show locations where the behavior of (at least some) OPs changes, thus they signal the phase transitions. In Fig. \ref{fig:k4k0} between the vertical red lines each corresponding phase is written. The slight discontinuities in the function $\kappa_4(\kappa_0,\Delta = fix)$ signal the locations of the phase transitions, which originate from the different entropy on the two sides.\\

\begin{figure}[h]
    \centering
    \includegraphics[width = 0.8\textwidth]{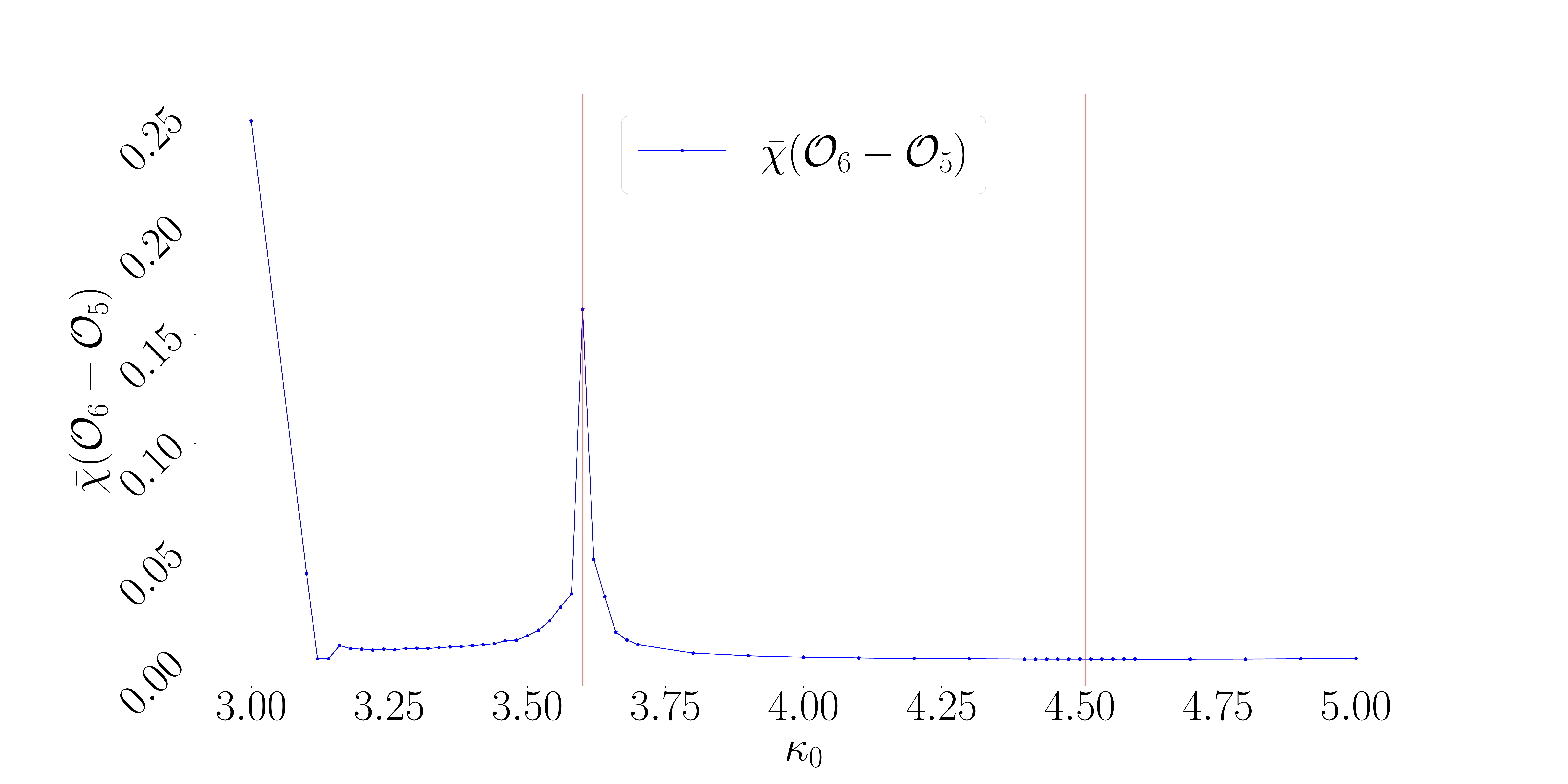}
    \caption{The figure shows the (normalized) susceptibility ${\bar{\chi}}( \mathcal{O}_6-\mathcal{O}_5 ) / \langle \mathcal{O}_6-\mathcal{O}_5 \rangle $. The peak of the susceptibility, i.e., the   variance of $( \mathcal{O}_6-\mathcal{O}_5)$, at the  $C_b-C$  transition is a clear signal of the phase transition.}
    \label{fig:varop56}
\end{figure}

Not every OP signals all of the phase transitions, but using different OPs one can find them. For example, $\mathcal{O}_1$, $\mathcal{O}_2$ and $\mathcal{O}_3$ distinguish seemingly three different regions, while $\mathcal{O}_4$ seems to be sensitive to all four phases. The behavior of the two new OPs, $\mathcal{O}_5$ and $\mathcal{O}_6$, is similar to $\mathcal{O}_2$, however, their crossing point additionally signals the $C_b - C$  phase-transition, as presented in Fig.  \ref{fig:op56}. This becomes apparent when one looks at the  susceptibility of $(\mathcal{O}_6-\mathcal{O}_5)$, as shown in Fig.  \ref{fig:varop56}. In the figure we plot the susceptibility ${\chi}(\mathcal{O}_6-\mathcal{O}_5)$, i.e., the variance of  $(\mathcal{O}_6 -  \mathcal{O}_5)$, normalized by its expectation value $\langle\mathcal{O}_6-\mathcal{O}_5\rangle$, which shows a clear peak at the $C_b - C$ transition point.
Fig. \ref{fig:op56} shows the two new OPs, and their difference, which is close to zero in phase $A$ and $B$, positive in phase $C_b$ and  negative in phase $C$, and thus is useful in recognizing all four phases of CDT. The above observations show, that it is not enough to look at one OP but rather a set of OPs should be used while analyzing phase transitions. \\

Similar plots can be drawn if one measures the OPs in the function of the coupling constant $\Delta$ for fixed $\kappa_0$, or even when choosing  both $(\kappa_0, \Delta)$ values on some straight but not vertical nor horizontal line in the CDT phase-diagram. 

\newpage

\subsection{Finite volume scaling analysis} 

Even though, as explained in Section \ref{sec:OPs}.,  CDT phase transition signals observed for any fixed lattice volume $\bar{N}_{41}$ are not real phase transitions, as for any finite volume the free energy is finite and formally one just observes a cross-over, nevertheless using finite-size scaling analysis one can investigate the (real) phase transitions and draw conclusions about critical exponents in the thermodynamical limit. By extrapolating the scaling relations to $\bar N_{41}\to \infty$ one can as well find the (infinite volume) critical values of the coupling constants $\kappa_0^\infty$ and $ \Delta^\infty$ and of the order parameters $OP^\infty$. Thus a typical finite-size scaling relation of a coupling $\mathcal{C}$ corresponding to the transition point will be described by a function

\begin{equation}
\label{eq:scaling}
    \mathcal{C}^{crit}({\bar{N}_{41}}) = \mathcal{C}^\infty - \alpha \bar{N}_{41}^{-\frac{1}{\gamma}},
\end{equation}
where $\gamma$ is the critical exponent, whose value may be used to distinguish between the first-order ($\gamma=1$) and the higher-order ($\gamma>1$) phase transition.
\\

At the thermodynamical limit of the higher-order phase transition one can perspectively find a UVFP, however doing so in lattice simulations is not an easy task as it was shown in  \cite{rg_flow1, rg_flow2}. Previous findings did not give convincing evidence for the existence of the UVFP of CDT, however, since these measurements, there were many improvements both in the CDT code (making the MC simulations more efficient) and computer technology. Therefore now, using the new possibilities, one can try to re-investigate this issue in more detail.

\subsection{$A - B$ phase-transition}
\label{sec:AB_trans}

\textit{This subsection is based on the publication \cite{pub3}.}\\\\

Since the appearance of the path integral formalism of Feynman, we know that not only those paths should be taken into account which can be imagined classically but also non-classical ones too. In the path integral formalism,  for example, a point particle takes all possible paths when traveling from point A to B including also classically forbidden paths when it tunnels through potential walls or simply goes outside of the light cone. The contribution of most such paths however cancels out and the classical trajectory can be computed as an average of all paths. A generic triangulation of phase B is characterized by the following pattern: there is a vertex with almost maximal coordination number in a spatial slice with time coordinate  $t-1$, almost all spatial tetrahedra (3-volume) are gathered in slice $t$, and again there is a vertex with almost maximal coordination number in slice $t+1$. All spatial slices with time coordinate different than $t$ has spatial volume close to the minimal allowed cutoff. As the configurations of phase, A are characteristically different, i.e., they can be characterized by branched polymers,  there is a difference in entropy of the configurations between the two phases, which results in a phase-transition between them. 

\begin{figure}[h]
    \centering
    \includegraphics[width = 0.7\textwidth]{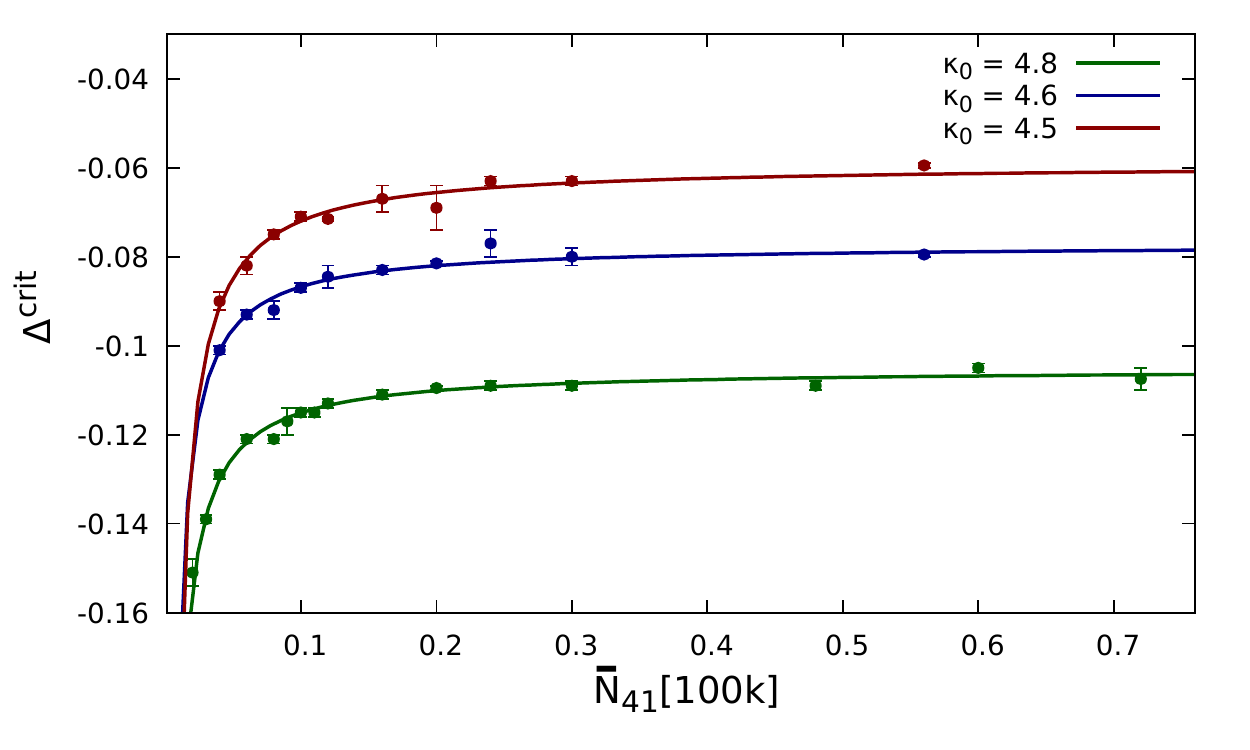}
    \caption{(Pseudo-) critical values of $\Delta^{crit}(\bar{N}_{41})$ measured for $\kappa_0 = 4.8$ (green), $\kappa_0 = 4.6$ (blue),
and $\kappa_0 = 4.5$ (red) together with the fits of eq. (\ref{eq:scaling}). The solid curves were fitted with the critical exponent fixed to $\gamma = 1$ for all three data sets.}
    \label{fig:AB_Dcrit}
\end{figure}

\begin{figure}[h]
    \centering
    \includegraphics[width = 0.8\textwidth]{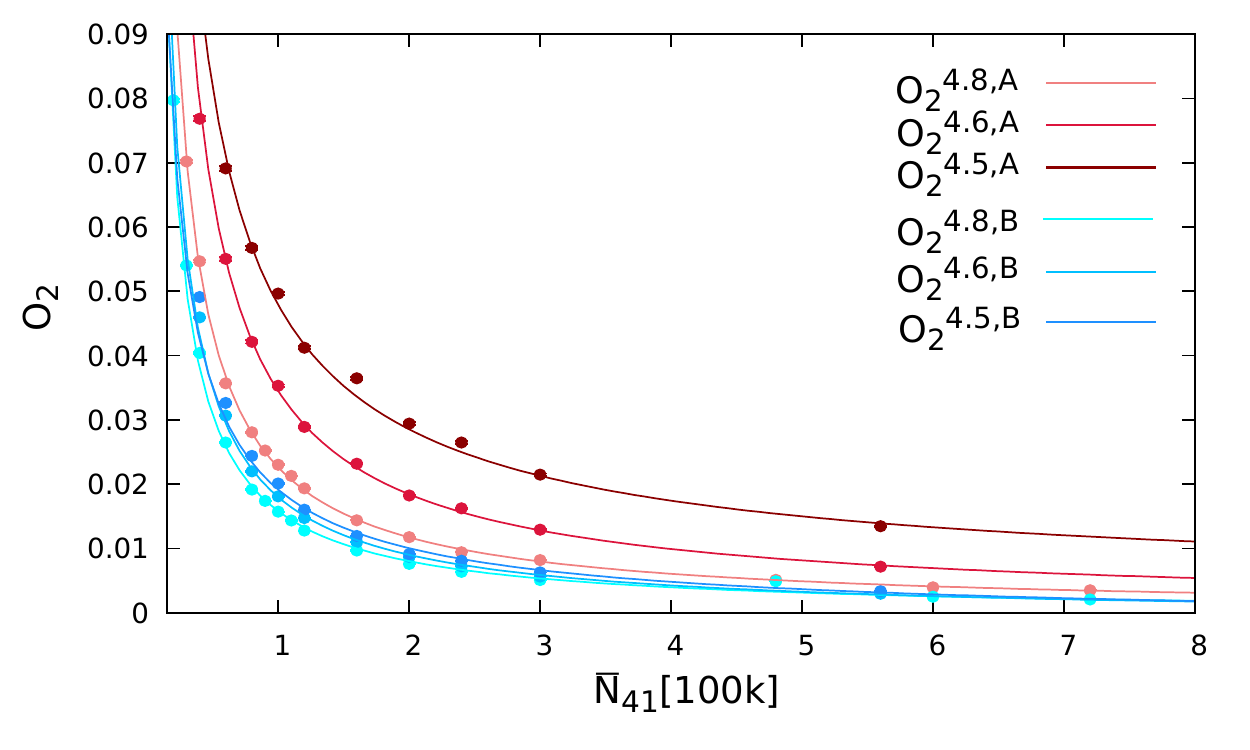}
    \caption{The running of $\mathcal{O}_2$ for $\kappa_0 = 4.8$, $\kappa_0 = 4.6$ and $\kappa_0 = 4.5$.  Blue colors correspond to data measured in phase $B$ and red in phase $A$ closest to the phase transition point, and the darker the color the lower the corresponding $\kappa_0$ coupling, i.e., the closer to the $A\!-\!B\!-\!C$ triple point. The error bars are smaller than the size of the data-points. The solid curves correspond to the fits of a relation similar to eq. (\ref{eq:scaling}) with the critical exponent fixed to be $\gamma = 1$ for all  data sets. }
    \label{fig:AB_o2}
\end{figure}

The fist-order nature of the A-B transition is obvious when one looks at the finite-size scaling of the critical coupling $\Delta^{crit}$ in the function of the lattice volume $\bar{N}_{41}$ presented in Fig. \ref{fig:AB_Dcrit}. Using eq. (\ref{eq:scaling}) one can fit the critical exponent $\gamma$.  The best fits resulted with critical exponent values $\gamma_{4.5} = 1.151 \pm 0.379$, $\gamma_{4.6} = 1.029 \pm 0.178$ and $\gamma_{4.8} = 1.088 \pm 0.101$ for three independent series of measurements with fixed $\kappa_0= 4.5, 4.6$ and $4.8$, respectively.  All three scaling exponents are in agreement with $\gamma = 1$ characteristic for the first-order transition. The same conclusion can be drawn if one looks at finite size scaling of $\mathcal{O}_2$. The fits of such scaling relations are presented in Fig. \ref{fig:AB_o2}. The value of $\mathcal{O}_2$ is very small in both phases. In the infinite volume limit it approaches zero in phase $B$ and for higher $\kappa_0$ also in phase $A$.

The conclusion is that phases $A$ and $B$ are thought to be non-physical in the sense of an emergent semi-classical geometry. It is well explained by the fact of the decreasing  connectivity between two adjacent spatial geometries, represented by the vanishing $\mathcal{O}_2$ parameter, although this phenomenon may be possibly related to the, so-called, "asymptotic silence" \cite{asymp_sil}.

\subsection{$C_b - B$ phase-transition} 
\label{sec:cb_trans}

\textit{This subsection is based on the publication \cite{pub1}.}\\\\

The $B$ and $C_b$ phases are not that different from each other, as in both phases there are vertices which connect to almost every tetrahedron on the adjacent spatial slices, and such high connectivity structure makes these phases to be effectively infinite-dimensional. Both spectral and Hausdorff dimensions differ from the topological value  4, thus these phases do not describe a four-dimensional Universe. Even though  their seemingly non-physical nature the phase transition becomes important to be studied as its endpoint leads to a candidate of the UVFP of the theory, to the $B-C-C_b$ triple-point.\\

\begin{figure}[h]
\centering
\includegraphics[width=0.4\linewidth]{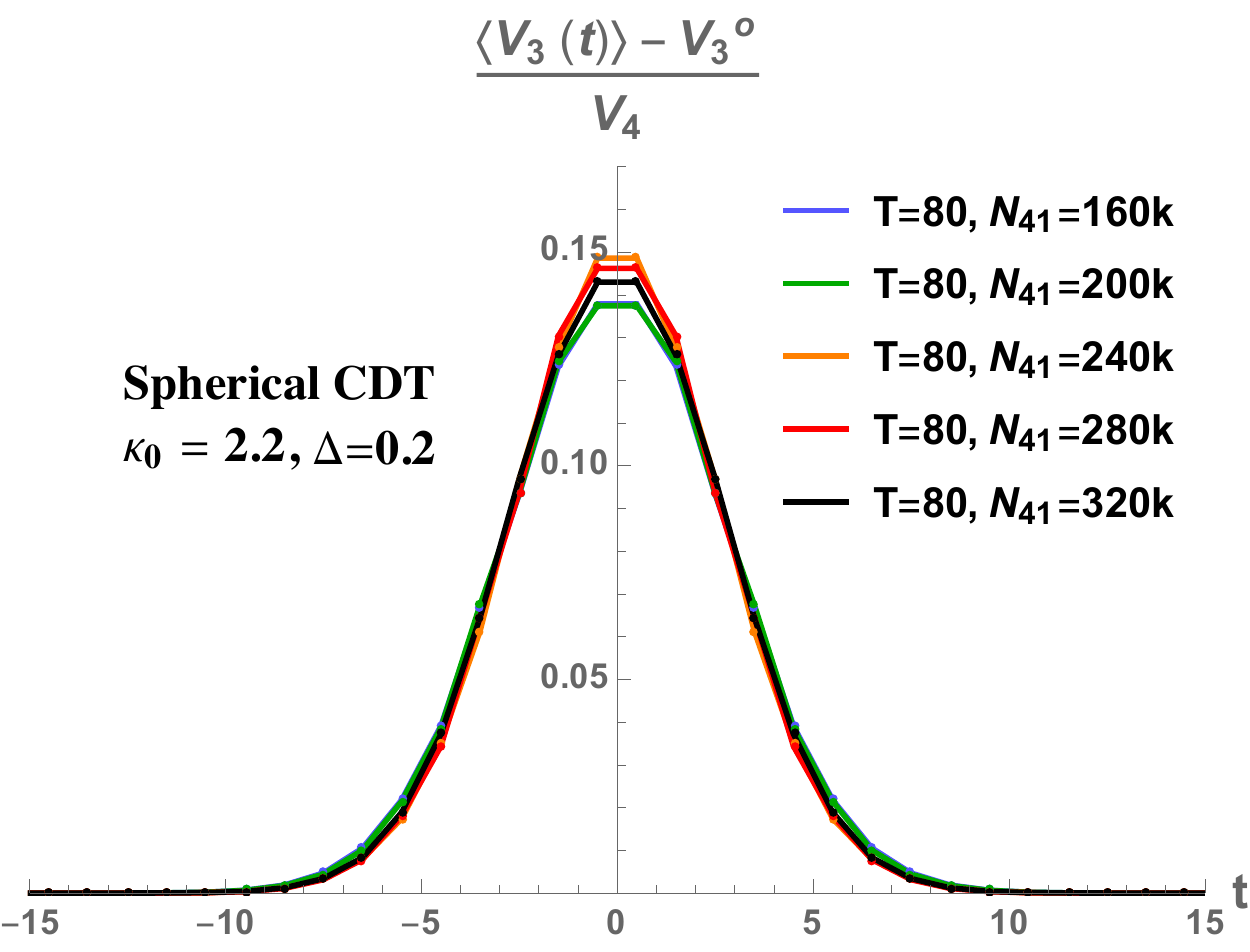}
\includegraphics[width=0.4\linewidth]{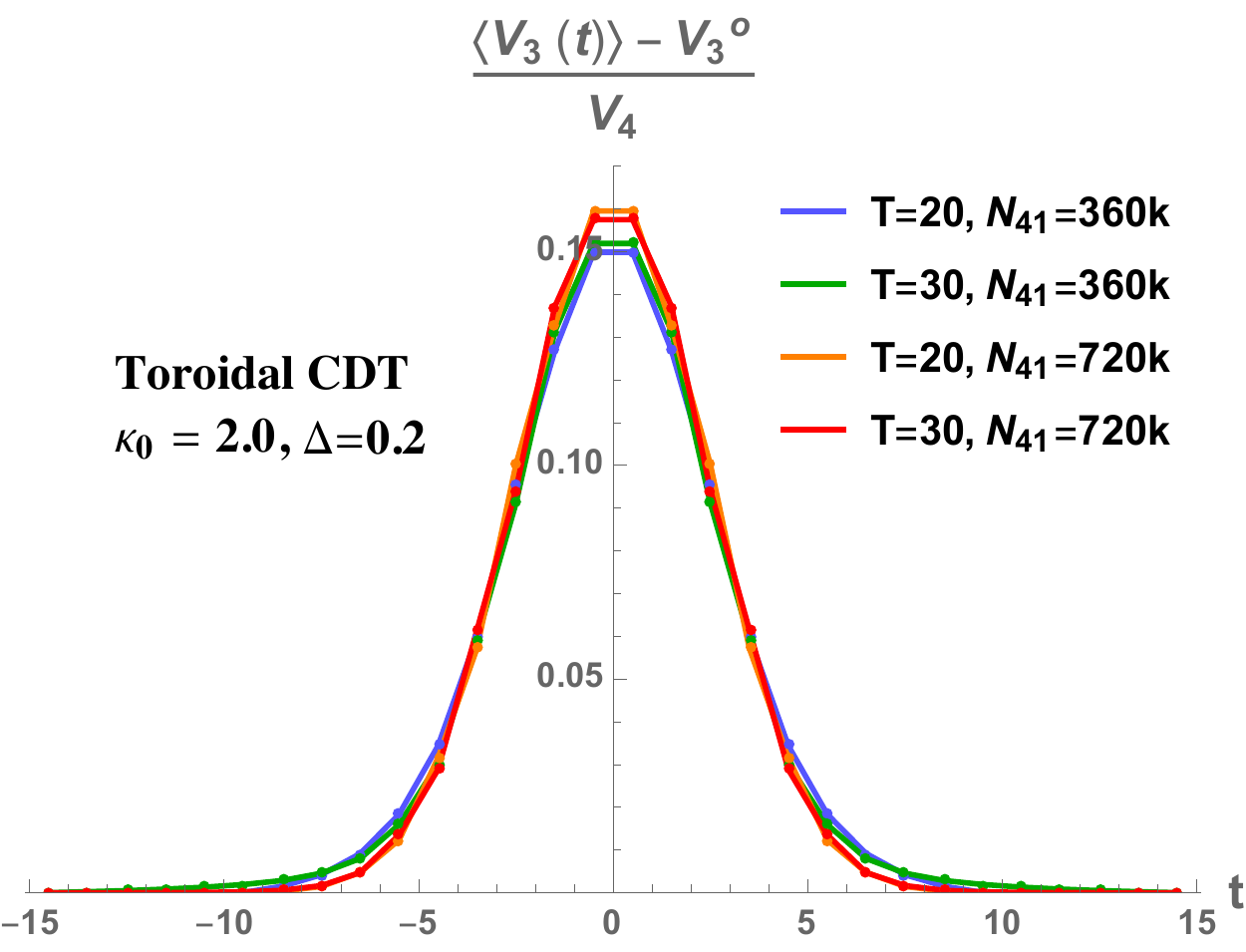}
\caption{\small The (re-scaled) average spatial volume profiles $\langle V_3(t)\rangle $ observed in the bifurcation phase $C_b$ in the spherical (left plot) and the toroidal (right plot) CDT. In both plots the spatial volume profiles were presented with respect to the center of the volume, set at $t=0$, and shifted  by a (constant  $V_3^0$) volume measured in the {\it stalk} range ($|t|>\sim 10$),  $V_3^0$ being different for each volume profile (in general  $V_3^0$ is bigger in the toroidal CDT where discretization effects are larger). Data measured for various total $\bar N_{41}$ lattice volumes and different $T$ were rescaled by $V_4=\sum_t(\langle V_3(t)\rangle - V_3^0$), i.e., in agreement with the Hausdorff dimension $d_H=\infty$.}
\label{fig:cb_prof}
\end{figure}

The volume profile of the bifurcation phase is presented in Fig. \ref{fig:cb_prof}. It looks the same in the spherical and toroidal version of CDT.

\begin{figure}[h]
\centering
\includegraphics[width=0.8\linewidth]{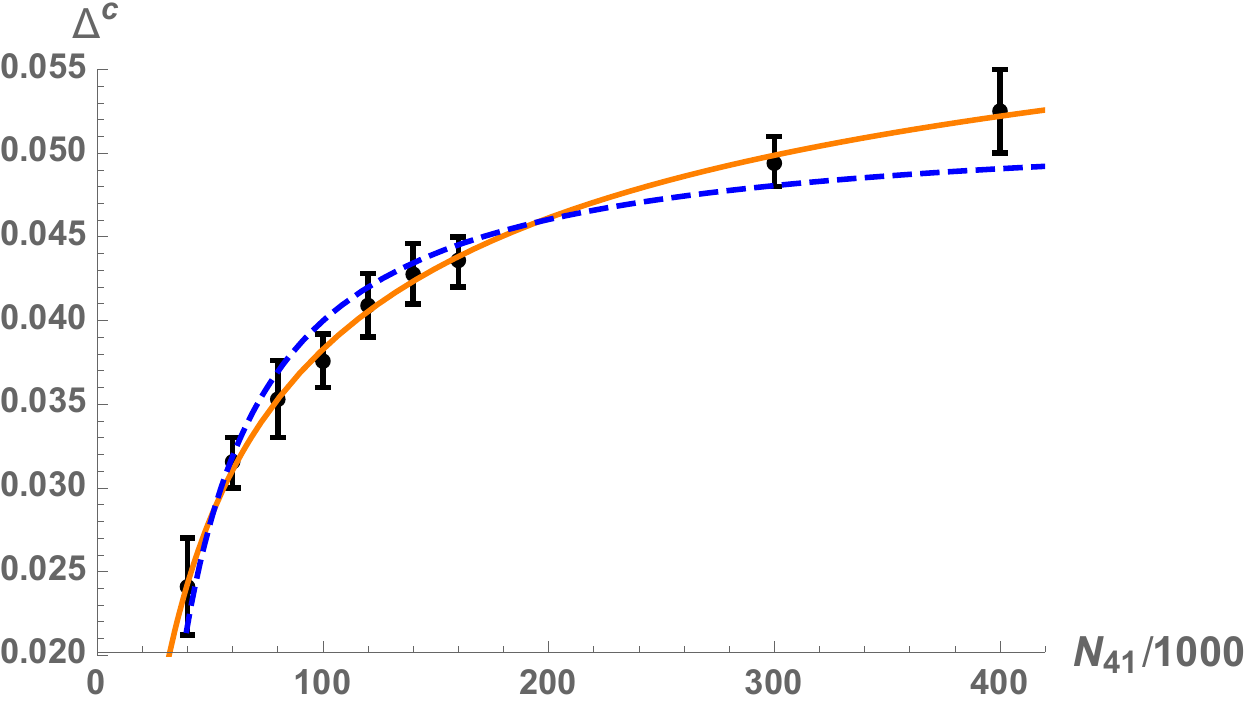}
\caption{\small Lattice volume dependence of the pseudo-critical $\Delta^c(N_{41})$ values in CDT with toroidal spatial topology measured for fixed $\kappa_0=2.2$ together with the fit of the finite-size scaling relation (\ref{eq:scaling}) with critical exponent $\gamma=2.51 \pm 0.03$ (orange solid line) and the same fit with a forced value of  $\gamma=1$ (blue dashed line). }
\label{fig:cb_scaling}
\end{figure}

The $C_b-B$  transition was analyzed for fixed $\kappa_0= 2.0$. Starting in phase $C_b$ and decreasing $\Delta$ close to $\Delta^{crit }\approx 0$ one finds the transition to phase $B$. The main difference between the two phases is the time extent of phase $C_b$, where the volume profile $V_3(t)$ resembles that of the spherical CDT in phase $C$, while in phase $B$ it is mostly collapsed to a single spatial slice. Although, as already mentioned, there are also many similarities between the two phases, for example, the Hausdorff dimension of generic geometries is very large or even infinite in the large volume limit. Probably due to these similarities, the $C_b-B$ phase transition was found to be the higher-order transition. Not only the fits to the finite size scaling relation of eq. (\ref{eq:scaling}) yielded a solution that was in disagreement with $\gamma = 1$ (as shown in Fig. \ref{fig:cb_scaling}) but also the order parameters showed a smooth transition between the two phases.
Furthermore, the Binder cumulants tend to vanish with increasing lattice volume, which is also characteristic of a higher-order transition \cite{bind}.\\ 

Summing up, in publication \cite{pub1}, the $C_b-B$ transition was shown to be a higher-order phase transition in CDT with the toroidal spatial topology. This is an important result in the quest for the UVFP of CDT. Due to the strong hysteresis observed in the toroidal CDT, the $C_b-C$ transition bordering the semi-classical phase was classified to be a first-order transition \cite{phase_structure_torus}.\footnote{In contrast to this, in the spherical CDT model the $C_b-C$ transition was shown to be a higher order transition \cite{cb1}.} The finding that the $C_b-B$  transition is higher-order provides a hope that its endpoint (i.e., the $B-C-C_b$ triple point) is also higher-order, yielding it a possible candidate  for the UVFP of the theory.\\

\subsection{$C - B$ phase-transition} 
\label{sec:CB_trans}

\textit{This subsection is based on the publications \cite{pub2} and \cite{pub3}.}\\\\

\begin{figure}[h]
    \centering
    \includegraphics[width = 0.7\textwidth]{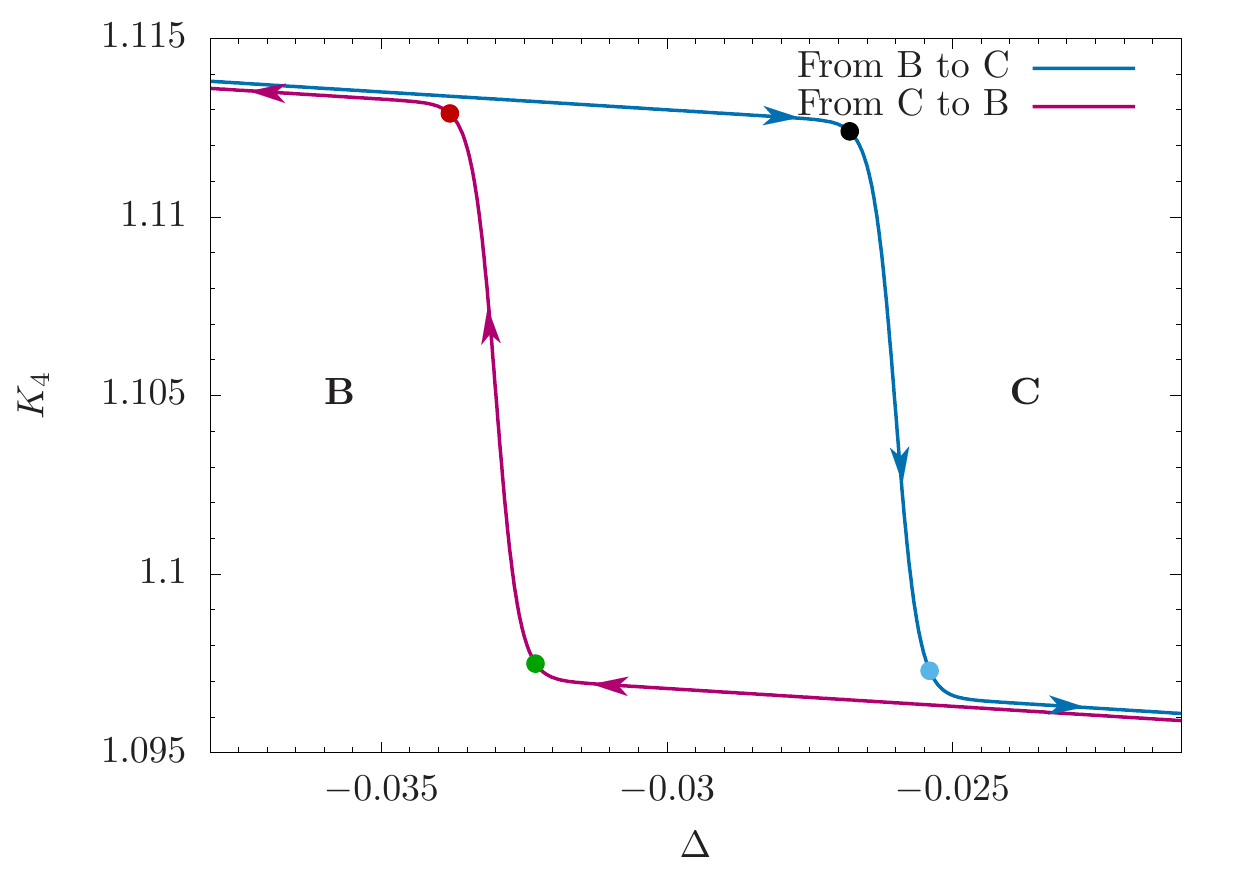}
    \caption{The plot illustrates the hysteresis measured during simulations for the lattice volume $\bar N_{41}=160\mathrm{k}$. The green and blue dots correspond to the location of the phase C side of the phase transition,  while the red and black dots correspond to the location of the phase B side of the phase transition.}
    \label{fig:hist}
\end{figure}

Fixing the value of $\kappa_0$ in the range [$3.5 : 4.5$] and changing $\Delta$ one can cross the $C-B$ phase transition. In the case of toroidal spatial topology the volume profile $V_3(t)$ of phase $C$ is almost constant, and thus invariant under the translation in time, but crossing to phase $B$ this symmetry of the generic configurations is broken to a "collapsed" volume profile. Phase $C$ is also characterized by quite homogeneous and isotropic  geometry in sufficiently large scales, but as one traverses to phase $B$ one can immediately observe that vertices with very high coordination numbers appear, which breaks the above homogeneity and isotropy. As a result, one observes a strong hysteresis  around the phase transition line, as presented
in Fig. \ref{fig:hist}.

\begin{figure}[ht!]
\centering
\includegraphics[width = 0.45\textwidth]{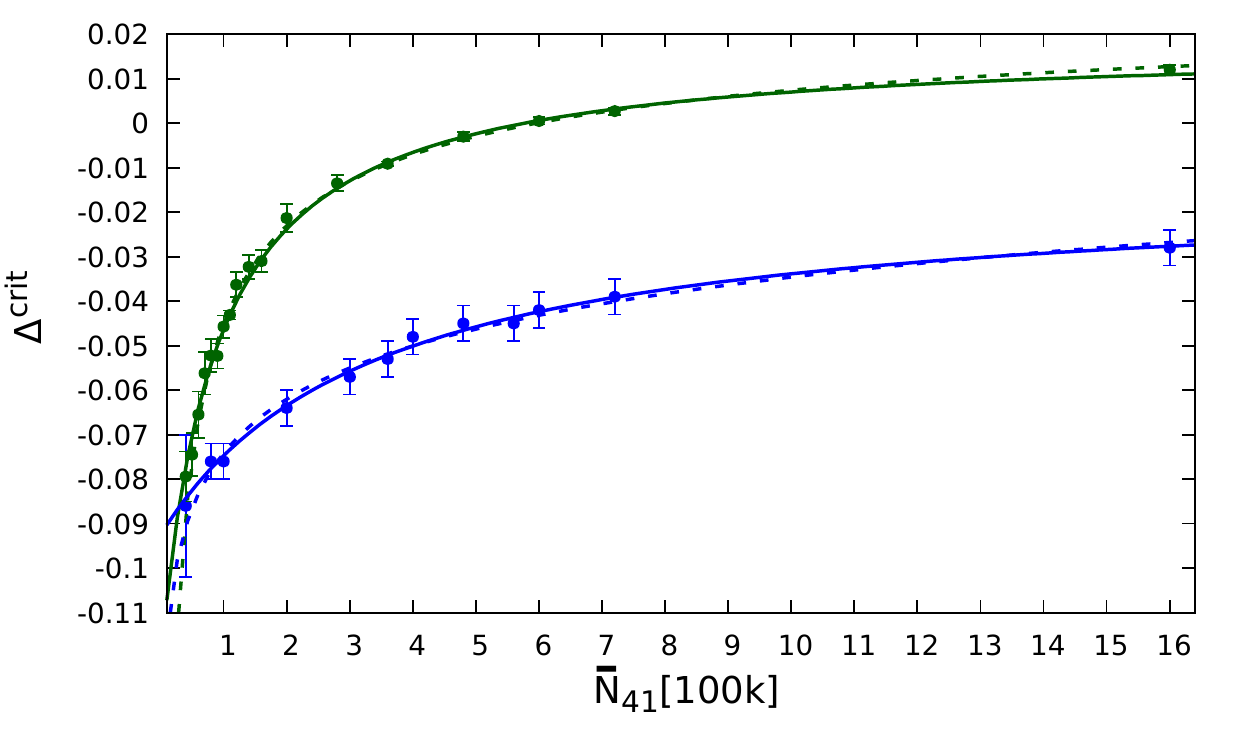}
\includegraphics[width = 0.45\textwidth]{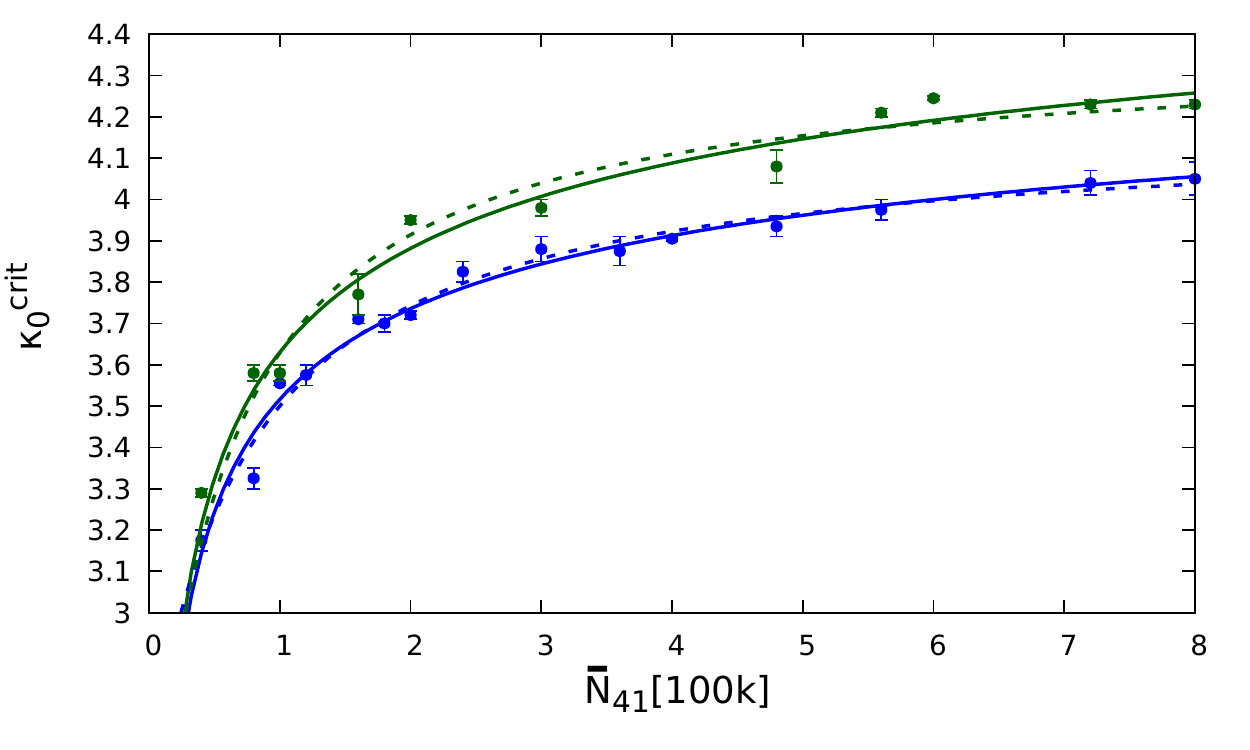}
\caption{Finite-volume scaling of the coupling constants $\Delta^{crit}$ (left panel) and $\kappa_0^{crit}$ (right panel) in the $C-B$ transition. The dashed and solid curves represent fits eq. (\ref{eq:scaling}) and eq. (\ref{eq:scaling_m1}), respectively. In the left panel (vertical measurements), green data points are for fixed $\kappa_0 = 4.0$ and blue are for $\kappa_0 = 4.2$. In the right panel (horizontal measurements), green data points are for fixed $\Delta = -0.02$ and  blue are for $\Delta = 0$. }
\label{fig:scaling_couplings}
\end{figure}

In order to encounter the $C-B$ transition, instead of fixing $\kappa_0$ in range [$3.5 : 4.5$] and changing $\Delta$, one can as well fix $\Delta$ in range [$-0.04:0.00$] and change $\kappa_0$. Therefore, the phase transition study, including the finite-volume scaling analysis, was performed for two different fixed $\kappa_0$ values ($\kappa_0 = 4.0$ and $4.2$) and for two different fixed $\Delta$ values ($\Delta = 0$ and $-0.02$). The  transition was determined to be the first-order  phase transition, but a rather atypical one. Although the values of  order parameters measured at both  sides of the hysteresis region do not converge to a common value, which itself signals a first-order transition, the size of the hysteresis region    shrinks when the lattice volume $\bar N_{41}$ is increased, which  can signal a higher-order transition in the thermodynamical limit. In order to resolve this inconsistency, in publication \cite{pub2} we measured the critical exponent resulting from the scaling relation of eq. (\ref{eq:scaling}), which turned out to be $\gamma = 1.62 \pm 0.25$, suggesting a higher-order transition. Nevertheless, in publication \cite{pub3} we repeated the finite-size scaling analysis using  much bigger data statistics and also  additional locations in the phase diagram. We also used a slightly modified finite-size scaling relation in the form:

\begin{equation}
\label{eq:scaling_m1}
    \mathcal{C}^{crit}({\bar{N}_{41}}) = \mathcal{C}^\infty - \alpha (\bar{N}_{41}-{c})^{-\frac{1}{\gamma}},
\end{equation}

where ${c}$ is a discretization correction. We found that the critical exponents are consistent with $\gamma=1$, see Fig. \ref{fig:scaling_couplings}, which signals the first-order transition. We, therefore, concluded that the $C-B$ phase transition is a first-order transition in the case of toroidal CDT.

\subsection{Summary} 

The three-phase transitions discussed in this chapter were the $A-B$, $C-B$, and the $C_b - B$ transitions, out of which only the last one turned out to be a higher-order (continuous) phase transition in CDT with the toroidal spatial topology. The $A-B$ and $C-B$ phase transitions were not yet analyzed in the case of the spherical CDT. What is more, the existence of the direct $C-B$ transition was not even known before the article \cite{phase_structure_torus} was published. The reason was that the region of the phase diagram analyzed in detail in this thesis was at that time thought to be unreachable via MC  simulations. In theory, the phase diagram of the CDT model does not have to be the same for different spatial topologies, thus a future analysis may potentially find that the $C-B$ transition is continuous in the spherical CDT.  An argument for this phase transition being continuous in the spherical CDT is related to the \textit{effective} topology of the phases. In the case when the topology of the spatial slices is chosen to be $S^3$ and the time direction is compactified to $S^1$, which is the case in the MC simulations, the full topology of the triangulations is $S^3 \times S^1$. However, in the semi-classical phase $C$, the emergent (Euclidean) de Sitter-like geometry, i.e.,  the four-sphere, transforms the {\it effective} topology to $S^4$. To clarify the notion of the {\it effective} topology, let us explain that, by definition, the imposed $S^3 \times S^1$ topology of the triangulations is not changed in the MC simulations, thus the two furthest (time-like) points (poles) of the  four- sphere are  connected by a thin \textit{stalk} of cutoff size, which can be treated as a lattice artifact. The {\it stalk} is necessary to preserve the imposed topological conditions numerically, but if only it was allowed by the MC algorithm it would completely disappear, yielding the change of  topology from  $S^3 \times S^1$ to $S^4$. It illustrates the fact that not only the space-time effective dimensionality but also  the effective topology are emergent concepts on the quantum level. Thus our conjecture, formulated in \cite{pub3}, is the following: \\

\noindent \textit{phase transitions which involve a change in effective topology will be first-order transitions}.\\

\noindent The argument is that if there are two adjacent phases separated by a phase transition, and these phases have different genera then the phase transition cannot happen smoothly resulting in the first-order  transition.\\ 

In the case of CDT with toroidal spatial slices, phase $C_b$ has, similarly to phase $C$ in the spherical CDT case, an effective topology $S^4$. At the same time, the semi-classical phase $C$ has the toroidal effective topology $T^4$. As a result of the phase transition the effective topology changes, yielding the $C-B$  transition first-order. Contrary to that, as explained above, in CDT with spherical spatial slices the effective topology of the semi-classical phase is $S^4$, so it does not change under the phase transition, potentially making the $C-B$ transition higher-order. \\  

So far all analysis was done in the case of empty Universes (pure gravity models). In the next chapter, we will discuss how the CDT model changes in the presence of matter fields.

%% file: Chapters/Chapter4.tex

\chapter{Universes with matter fields} 
\label{chapter4}  
\textit{This chapter gives a brief summary of the following articles: \cite{pub4,pub5,pub6}}.\\\\

\section{Scalar fields as coordinates}
\textit{"If people do not believe that mathematics is simple, it is only because they do not realize how complicated life is."} - \textbf{Neumann János}\\\\ 

\textit{This section is based on the publications \cite{pub4,pub5}.}\\\\

Life (the actual physical phenomenon) is indeed complicated, much more complicated than any  model designed in order to describe it. However, usually  simple models are the only ones that can be solved. Also in most physical theories vacuum solutions are the simplest, easiest and first ones to be found. The same is true in GR as most of known solutions of Einstein's field equations are vacuum solutions. It is also the case of CDT, where the first  twenty years of studies  were dominated by pure gravity models  (empty Universes) discussed above. As it was mentioned in Chapters \ref{Chapter1} and \ref{chapter2}, CDT is formulated in  a coordinate-free way, except from the time direction where one has a natural global proper-time coordinate $t$, consistent with the introduced foliation. It would be therefore beneficial to introduce some notion of coordinates, making contact with other gravitational research.\\

The simplest extension of the CDT model is the addition of massless scalar fields. As will be shown, such scalar fields can also play role of "clocks" and "rods", enabling one to define a coordinate system in the triangulated manifold, being an analogue of the harmonic (de Donder) gauge fixing in GR. As we will discuss below, in order to apply this method of defining coordinates in CDT one also has to make a proper choice of the target space of the scalar fields. As an example, take a (smooth) Riemannian manifold $\mathcal{M}$ equipped with a metric tensor $g_{\mu\nu}$ and another Riemannian manifold $\mathcal{N}$ with a trivial flat metric $h_{\alpha\beta}$. A harmonic map $\mathcal{M} \to \mathcal{N}$ can be defined with the help of a four-component scalar field $\phi^\alpha$, with $\alpha = 1, 2, 3, 4$. In case of our setup, if $\mathcal{M}$ has a topology of the four-torus $T^4$, then we also choose $\mathcal{N}$ to have the same toroidal topology, and each component $\phi^\alpha(x)$ is a map $\mathcal{M} \to S^1$, which minimizes the action

\begin{equation}
\label{eq:scalar_action}
S_M[\phi]	= \frac{1}{2} \int \mathrm{d}^{4}x \sqrt{g(x)} \;g^{\mu \nu} (x) \; h_{\alpha \beta}(\phi^\gamma(x)) \;\partial_\mu \phi^\alpha(x) \partial_\nu \phi^\beta(x).
\end{equation}

Due to our   choice of  the trivial metric $h_{\alpha \beta}$ on $\mathcal{N}$, the four-component scalar field can be decomposed into four independent components, later denoted $\phi^x$, $\phi^y$, $\phi^z$, and $\phi^t$. Due to this, it is enough to discuss the case of a single component (let's call it  $\phi$). The Euler-Lagrange equations for the field resulting from eq. (\ref{eq:scalar_action}) give rise to the Laplace equation:

\begin{equation}
\Delta_x \phi (x) =0, \quad 
\Delta_x = \frac{1}{\sqrt{g(x)}} \;\frac{\partial}{\partial x^\mu} 
\Big(\sqrt{g(x)} \,
g^{\mu\nu}(x)\Big) \frac{\partial}{\partial x^\nu},
\quad \phi(x) \in S^1.
\label{eq:laplace}
\end{equation}

In the case when $\mathcal{M}$ is closed, if we chose the target space of the scalar field $\phi$ to be $\mathbb{R}$ then the constant zero-mode of the Laplacian would be the only solution to the equation $\Delta_x \phi(x) = 0$. If instead, as we do, one chooses a nontrivial target space of the field to be $S^1$ (with circumference $\delta$) then one can obtain a nontrivial solution for the scalar field. Technically, the  condition $\phi(x)\in S^1$ can be obtained by considering a scalar field with the target space $\mathbb R$ and identifying 

\begin{equation}
\phi(x) \equiv \phi(x) + n \, \delta,\quad n \in \mathbb{Z}.
\end{equation}
The situation of interest is when we have the toroidal manifold $\mathcal M$ which can be thought of as an elementary cell  periodically repeating in all four directions. In such a case one can define four non-equivalent boundaries of the elementary cell,  i.e. 3-dimensional connected hypersurfaces $H(\alpha), \alpha = \{x,y,z,t\}$. Let us consider  the case when each component of the field $\phi^\alpha \in S^1$ winds around the circle once as we go around any non-contractible loop in  $\mathcal M$ that crosses a boundary in direction $\alpha$. In that case the field $\phi$ is a continuous function  except when one crosses the hypersurface $H(\alpha)$, where a jump of the field with amplitude $\delta$ happens, and the Laplace equation (\ref{eq:laplace}) acquires a nontrivial boundary term leading to a non-trivial solution for the field $\phi$. A corresponding function that is continuous despite the jump, will be a map

\begin{equation}
\phi \to \psi = \frac{\delta}{2\pi}\;e^{2\pi i \phi/\delta},
\label{eq:psi}
\end{equation}
which maps the scalar field $\phi$ to a circle in the complex plane. The interesting point is that, for a given direction $\alpha$, the  map $\psi$ does not depend on the exact choice of the boundary $H(\alpha)$ of the elementary cell.\footnote{Formally it depends only trivially, i.e., a continuous deformation  (a "shift") of the boundary $H(\alpha)$ will only cause a shift of the phase in the complex function $\psi$ by some constant.} \\

In CDT we  consider a discretization of the action (\ref{eq:scalar_action}) and the corresponding Laplace equation (\ref{eq:laplace}), where the field is localized in the center of simplices. We therefore consider a (discretized) Laplacian defined on the dual lattice, i.e., the graph whose  vertices represent the 4-simplices of the original CDT triangulation, and links represent the common interfaces between the 4-simplices in the triangulation. The Laplacian on the dual lattice can be defined via the adjacency matrix $A_{ij}$:

\begin{equation}
A_{ij} =
\begin{cases}
	1 & \textrm{if (the link } i \leftrightarrow j) \in \textrm{dual lattice},\\
	0 & \textrm{otherwise},
\end{cases}
\end{equation}
where $\leftrightarrow$ refers to the adjacency relation of simplex $i$ and $j$. The discrete Laplacian can be defined as:

\begin{equation}
L =  D - A,
\end{equation}
where $A$ is the adjacency matrix and $D$ is a diagonal matrix with $i$-th diagonal element containing the number of neighbors of a simplex labelled $i$. As, in the four-dimensional CDT, each simplex in the  triangulation has exactly 5 neighbors, the dual lattice of any triangulation is a five-valent graph, and therefore
\begin{equation}
    D = 5 \cdot I,
\end{equation}
with $I$ being the identity matrix of size $N_4 \times N_4$, where $N_4$ is the number of all 4-simplices in the triangulation. The discretized form of the scalar field action is then given by:

\begin{equation}
\label{eq:scalar_discrete}
S_M^{CDT}[\{\phi\},\mathcal{T}]	= \frac{1}{2} \sum_{i \leftrightarrow j} (\phi_i - \phi_j)^2 = \sum_{i,j} \phi_i L_{ij} \phi_j \equiv \phi^T L \phi, 
\end{equation}
where $\mathcal{T}$ underlines the impact of the triangulation on the Laplacian matrix $L$. The discrete analog of the Laplace eq. (\ref{eq:laplace}) is then: 

\begin{equation}
    L\phi = 0.
\end{equation}
The above equation has the same issue as before, i.e., if the target space of the field was chosen to be $\mathbb R$ then it would only have a trivial solution  $\phi = const$. Non-trivial solutions can be  found by choosing the field to take values in $S^1$ with circumference $\delta$, which winds around the circle once as one goes around any non-contractible loop in the dual lattice. In order to do  that one  identifies:

\begin{equation}
\phi_i \equiv \phi_i + n \cdot \delta, 
\quad n \in \mathbb{Z}. 
\end{equation}
This can be achieved by adding a jump condition when crossing a boundary  hyper-surface $H(\alpha)$ in direction, $\alpha$. The way of  introducing such boundary hypersurfaces to CDT was proposed in \cite{boundaries}. As already mentioned, the exact position of the (four non-equivalent) boundaries $H(\alpha)$ in the triangulation is not important as it has only a trivial impact on our solutions, thus the boundaries are non-physical. Technically, one can define the "jump" condition by introducing the boundary jump matrix $B_{ij}$:

\begin{equation}
B_{ij} =
\begin{cases}
	+1 & \textrm{if the dual link i} \rightarrow \textrm{j crosses the boundary $H(\alpha)$ in the \textit{positive} direction},\\
	-1 & \textrm{if the dual link i} \rightarrow \textrm{j crosses the boundary $H(\alpha)$ in the \textit{negative} direction},\\
	0 & \textrm{otherwise}
\end{cases}
\end{equation}
and defining
\begin{equation}
V = \frac{1}{2}  \sum_{ij} B_{ij}^2= \frac{1}{2} \sum_i | b_i |, 
\end{equation}
where $b_i = \sum_j B_{ij}$ is the boundary jump vector, and it measures the occasions when a tetrahedral face of a simplex $i$ appears on the boundary. To accommodate to the jump condition we modify the discretized matter action:

\begin{multline}
\label{eq:scalar_w_boundary}
S_M^{CDT}[\{\phi\}, \mathcal{T}]= \frac{1}{2} \sum_{i \leftrightarrow j} (\phi_i - \phi_j - \delta B_{ij})^2 = \sum_{i,j} \phi_i L_{ij} \phi_j - 2 \delta \sum_i \phi_i b_i +\delta^2 V \\ \equiv \phi^T L \phi - 2 \delta \phi^T b + \delta^2 V.
\end{multline}
Now, the Euler-Lagrange equation for the field $\phi$ yields:

\begin{equation}
 L \phi	= \delta \, b,
\end{equation}
so it acquires a non-trivial boundary term: $\delta \ b$.
The classical solution to the scalar field distribution  is formally given by 

\begin{equation}
    \phi^{classical} =\delta \ L^{-1}  b.
\end{equation}
The practical problem is that the Laplacian has zero modes but, fortunately, one can find a solution in the subspace orthogonal to the zero modes. The solution strongly depends on the underlying triangulated geometry and it smoothly interpolates between the boundaries of the (toroidal) elementary cell. In the publication \cite{pub4} we proposed to treat the harmonic map $\phi^{classical}$, or rather the resulting map $\psi^{classical}$, see eq. (\ref{eq:psi}), as a coordinate in the direction $\alpha$. This way one can introduce a coordinate system for every triangulation generated in the MC simulations. The coordinates can be used to visualize the differences between generic triangulations of different CDT phases. It is worth mentioning that the harmonic maps (coordinates) described above have a very good property of smoothly interpolating between the 4-simplices in the geometric  outgrowths, which commonly appear in the CDT triangulations forming fractal structures. Imagine such an outgrowth consisting of many simplices and linked to the rest of the triangulation by only  a few simplices. Due to  the properties of harmonic maps, all simplices in the outgrowth will have almost the same value of the field $\phi^{\alpha}_i$ in all $\alpha$-directions. Therefore the outgrowths should appear as the field condensations in the harmonic maps.

\begin{figure}[h]
    \centering
    \includegraphics[width = 0.6\textwidth]{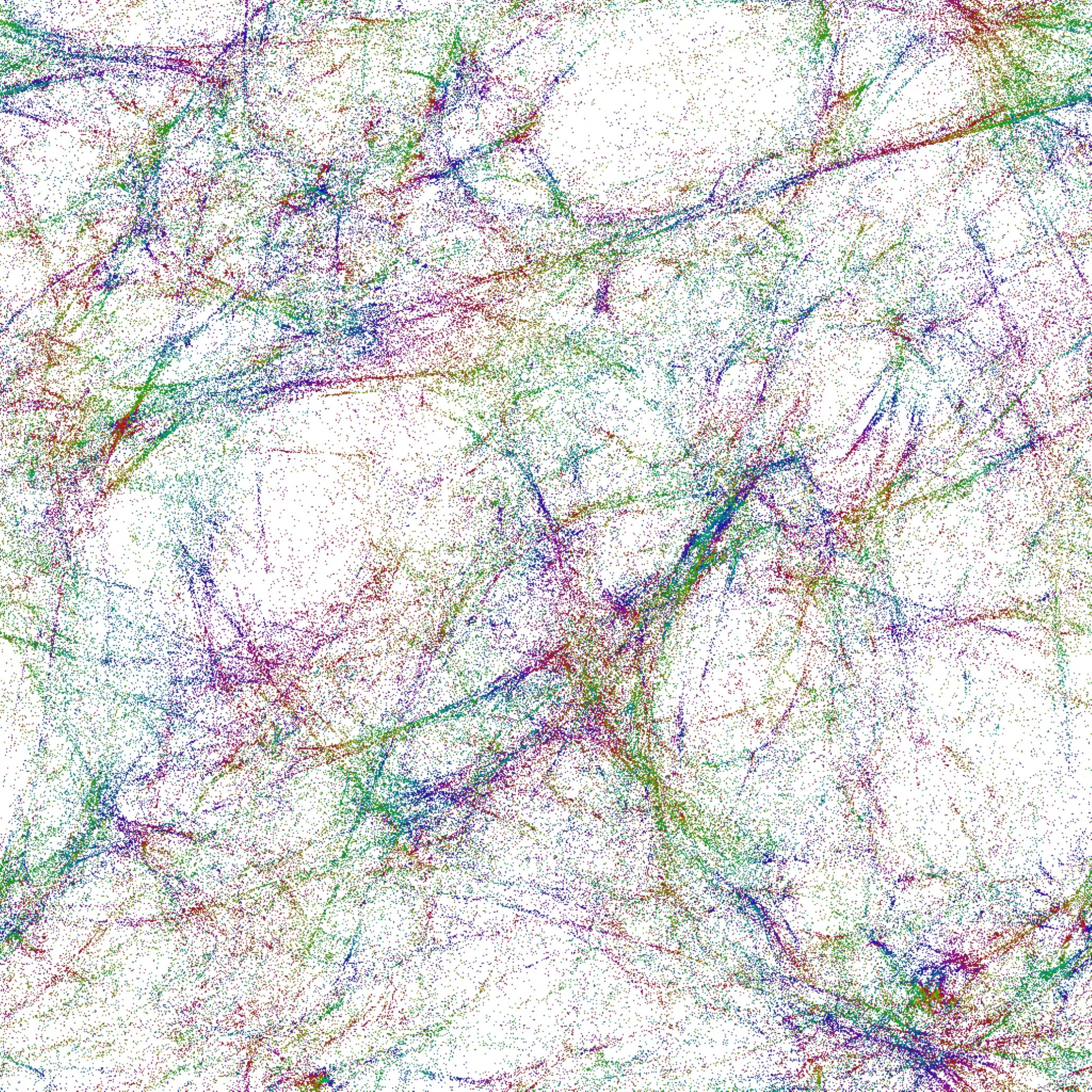}
    \caption{The 4-volume density map projected on two spatial ("$x$" and "$y$") directions measured in phase $C$ ($\kappa_0 = 4.0, \Delta= 0.2, T = 20, \bar{N}_{41} = 720k$). Each point on the plot represents a 4-simplex having the scalar field values (coordinates) $(\phi^x,\phi^y)$. The colors encode the time coordinate $t$ of the original CDT foliation. The dense regions are geometric fractal     outgrowths in the triangulation. The outgrows structure resembles cosmic voids and filaments of the real Universe.}
    \label{fig:density_map}
\end{figure}

A typical map (projected on two spatial directions: "$x$" and "$y$") measured in the semi-classical phase $C$ of the toroidal CDT is presented in Fig. \ref{fig:density_map}. Looking at Fig. \ref{fig:density_map} it becomes apparent that the phase $C$ generic triangulations represent a homogeneous and isotropic geometry on macroscopic scales. However, exactly as it is observed in the real Universe, there are local density fluctuations (geometric outgrows in the case of CDT) showing very non-trivial voids and filaments structure. One should note that in this context this is the emerging feature of pure quantum gravity, as the scalar fields discussed above do not have any impact (back-reaction) on the geometry, and are simply introduced for visualization purposes. Similar maps, obtained for generic triangulations of other CDT phases have completely different shapes, as discussed in publication \cite{pub6}.\\ 

\begin{figure}[h]
    \centering
    \includegraphics[width = 0.8\textwidth]{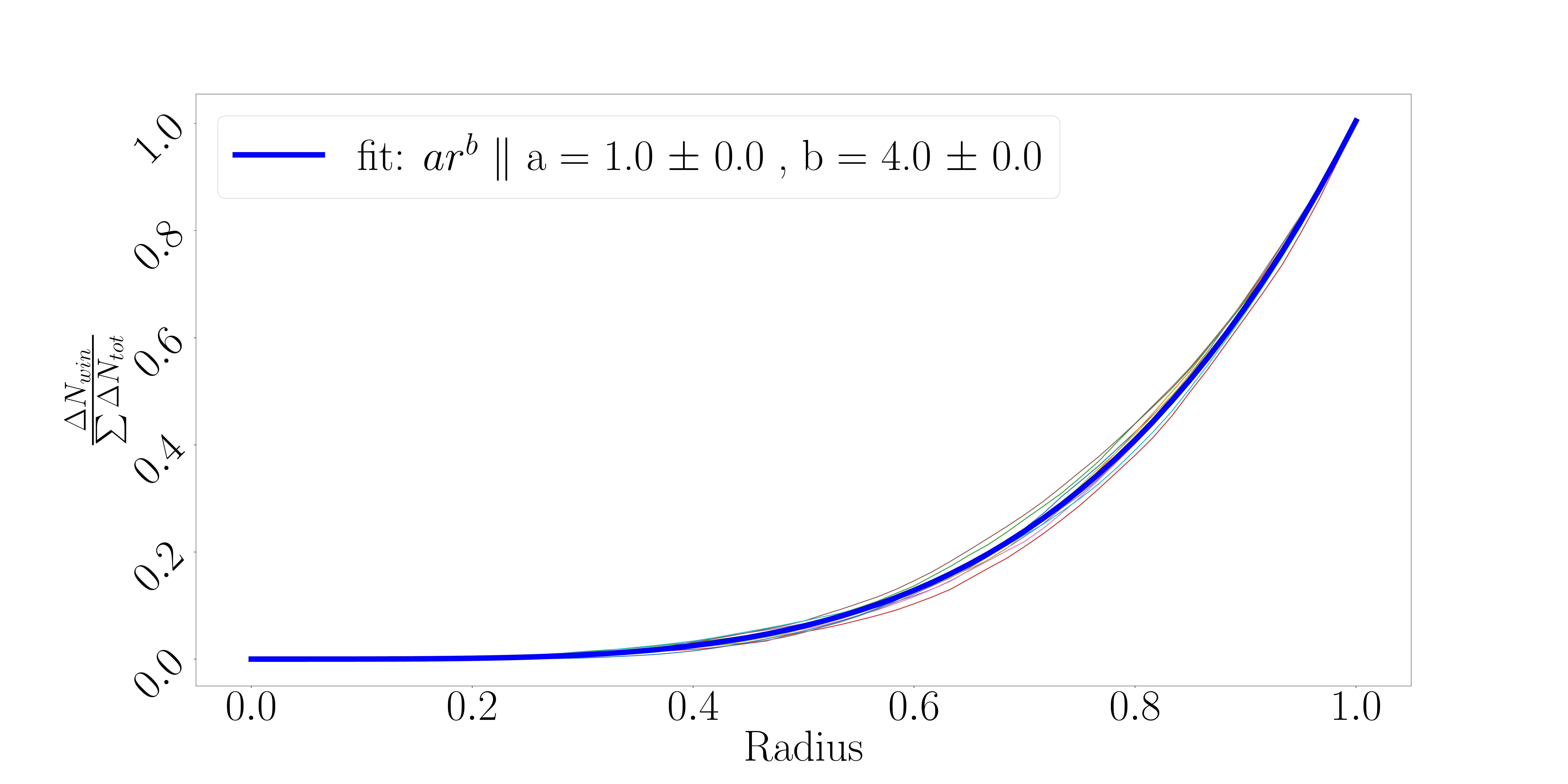}
    \caption{The figure shows the ratio of $\Delta N_{win}$ (4-volume inside the box of size $\Delta\phi^x\times\Delta\phi^y\times\Delta\phi^z \times \Delta \phi^t$) and $N_{tot}$ (total volume) in the function of the normalized size of the box (Radius). This function was measured in measured in phase $C$ ($\kappa_0 = 4.0, \Delta = 0.2, T = 20$). The various thin lines denote measurements for different lattice volumes $\bar N_{41}$, the solid blue line is a fit of the function $a r^b$ to their average.}
    \label{fig:inc_vol_prof}
\end{figure}

Using the scalar fields as coordinates one can also measure the scaling of 4-volume in a triangulation by picking a (random) center and following a diffusion wave from that center and observing the growth in the volume of the diffusion shell. Looking at the scaling of the volume with radius one can measure the, so-called, Hausdorff dimension, associated with the harmonic coordinates. This was measured for the following fixed lattice volumes $\bar{N}_{41} = \{80k, 160k, 200k, 240k, 300k, $ $360k,400k, 480k, 560k, 600k, 720k\}$. The 4-volume contained in a box (window) of size $\Delta\phi^x\times\Delta\phi^y\times\Delta\phi^z \times \Delta \phi^t$, denoted $\Delta N_{win}$, normalized by the total volume $N_{tot}$ can be used to measure the Hausdorff dimension. It was found that in   phase $C$ one obtains a universal behavior, as presented in Fig. \ref{fig:inc_vol_prof}. The fitted Hausdorff dimension is consistent with $d_H = 4$.

\newpage

\section{Dynamical scalar fields}

\textit{"The effort to understand the Universe is one of the very few things that lifts human life a little above the level of farce, and gives it some of the grace of tragedy."} - \textbf{Steven Weinberg}\\\\ 

\textit{This section is based on the publications \cite{pub5,pub6}.}\\\\

So far the back-reaction of the matter field on the purely geometric degrees of freedom was not taken into account. Including back-reaction of quantum (later also called dynamical) scalar fields can lead to nontrivial changes in the geometry. In the results presented below, the scalar fields are massless scalar fields with the (discretized) action (\ref{eq:scalar_w_boundary}), minimally coupled to the geometric (Regge) action (\ref{eq:ation_kappa}). Including such fields in the MC, simulations mean that not only do the field values have to be generated - in the MC simulations the heat bath method \cite{hb1,hb2} was used - but also that the matter action will affect the probability of performing the purely geometric moves. Depending on the parameters of a simulation, either the geometric or the matter part of the action dominates, thus one can expect a phase transition of some sort when moving in the  parameter space, now also including a new coupling constant corresponding to the circumference $\delta$ of the $S^1$ target space (or alternatively the jump amplitude) of the scalar field. \\
\begin{figure}[h]
    \centering
    \includegraphics[width = 0.45\textwidth]{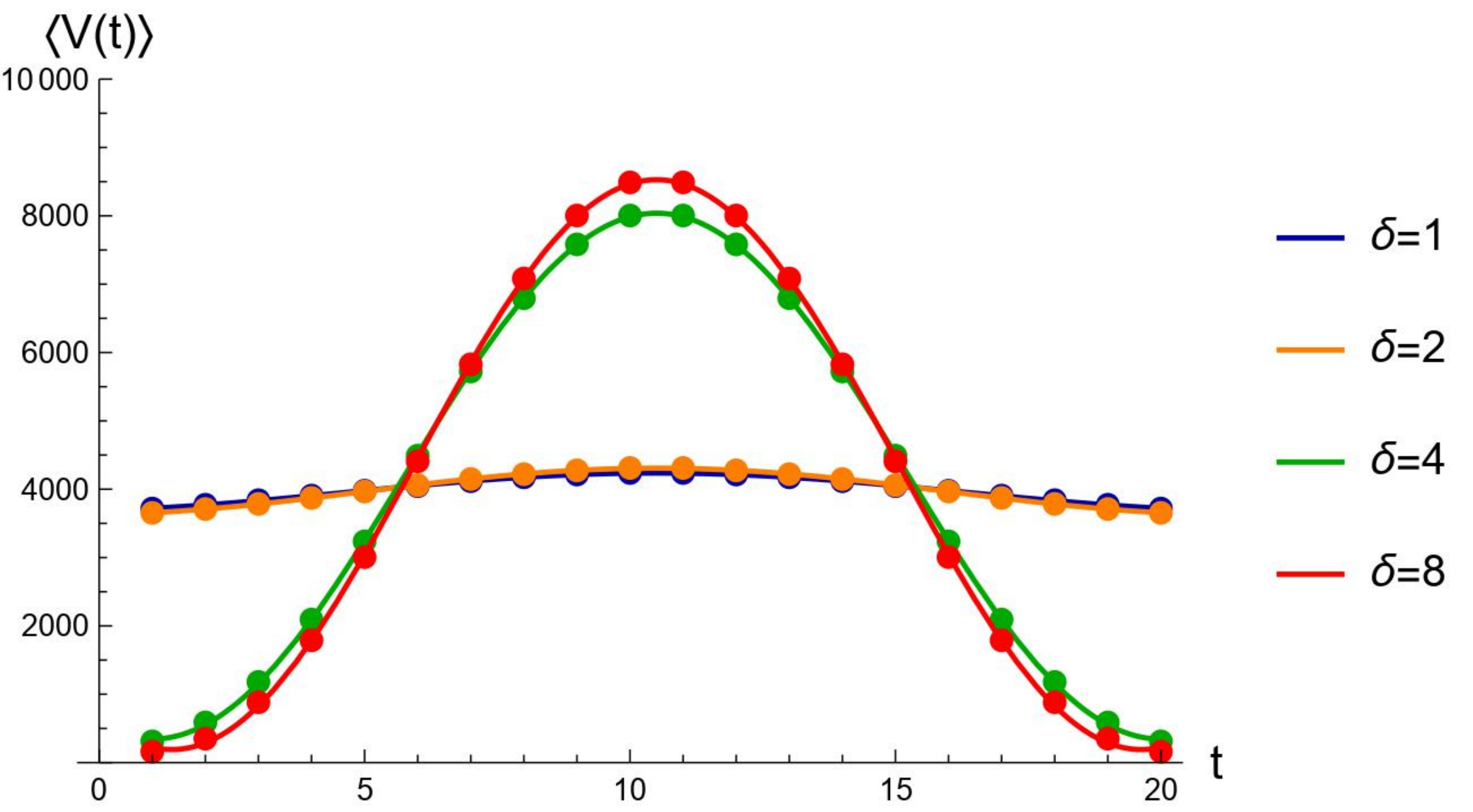}
    \includegraphics[width = 0.45\textwidth]{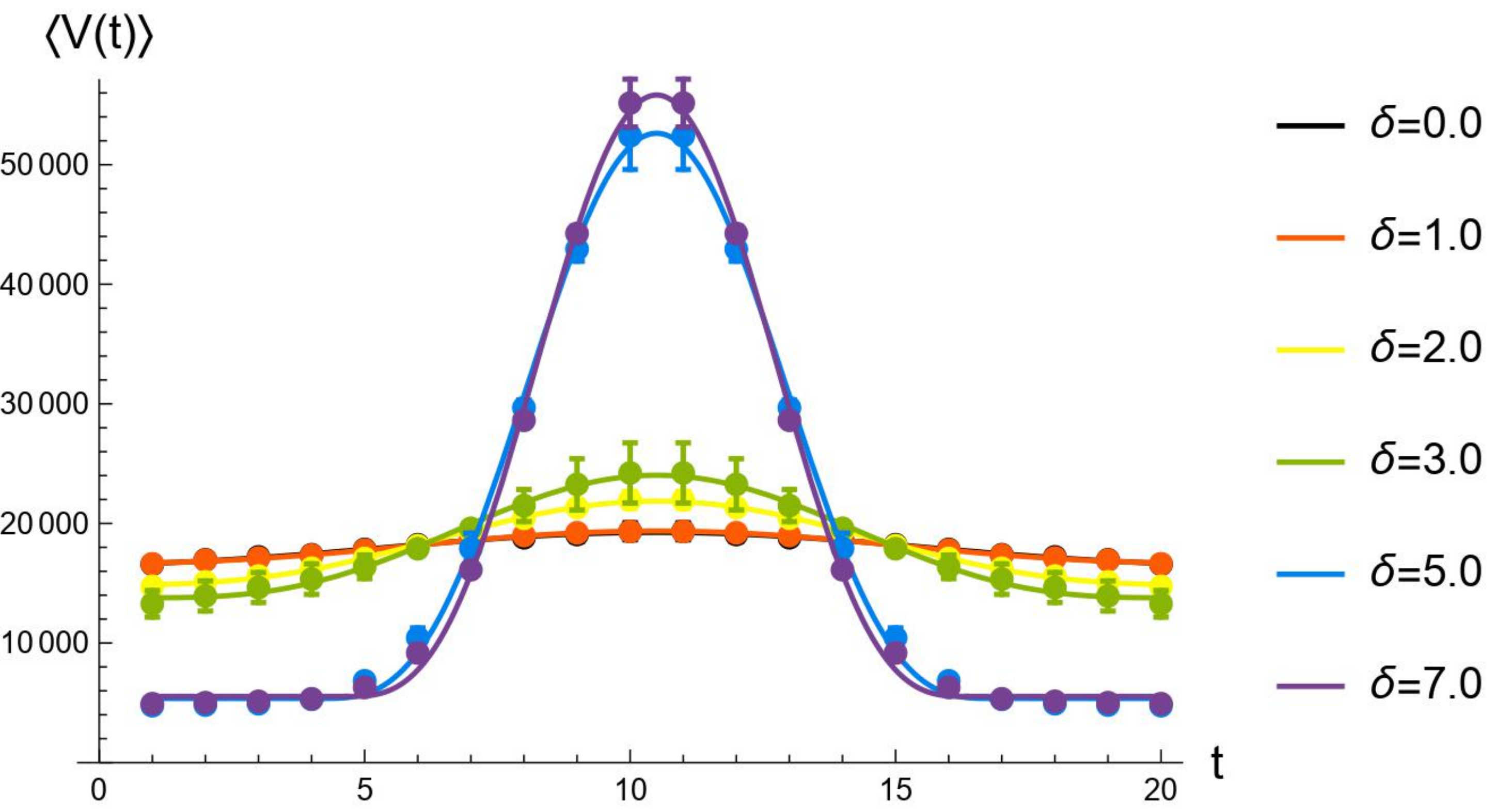}
    \caption{The volume profile in the presence of one field winding around the time direction (left) and three fields winding around the non-equivalent spatial directions (right).}
    \label{fig:field_prof}
\end{figure}

The choice of the $\delta$ value is not the only additional parameter, as one can also choose the number of $\phi$-fields, as well as the number and type (time- or space-like) of  non-equivalent winding directions for the scalar field(s). Adding a field with $\delta = 0$ has already a visible but small effect, as it shifts some characteristics of generic triangulations appearing in the path integral, for example, it lowers the ratio of ${N}_{32}/N_{41}$, and adding more fields the effect is larger. A much stronger effect is observed for large jump magnitude $\delta$. As already discussed, if no scalar fields are added, the measured volume profile of the toroidal CDT model in the semi-classical  phase $C$ is a constant function. This is also the case of CDT coupled to the scalar fields with zero or small jump magnitude $\delta$, but for large $\delta$ one observes a dramatic change in the volume profile, as it is shown in Fig. \ref{fig:field_prof}. In the case with one scalar field winding around the time direction, using a simple minisuperspace-like model presented in Appendix  3 of \cite{pub6}, one can expect to observe a "pinched"  volume profile, turning the constant function into a $cos(t)$ function, as seen in the left panel of Fig. \ref{fig:field_prof}.  
On the other hand, as can be seen on the right panel of Fig. \ref{fig:field_prof}, the jump condition introduced only in the spatial directions will also trigger, for large-enough $\delta$ values,  a  kind of "pinched" volume profile, however, the reason behind it is different. In that case, the fitted volume profile is given by a  $cos^3(t)$ function, which corresponds to the volume profile of the (Euclidean) de Sitter sphere.

\begin{figure}[h]
    \centering
    \includegraphics[width = 0.6\textwidth]{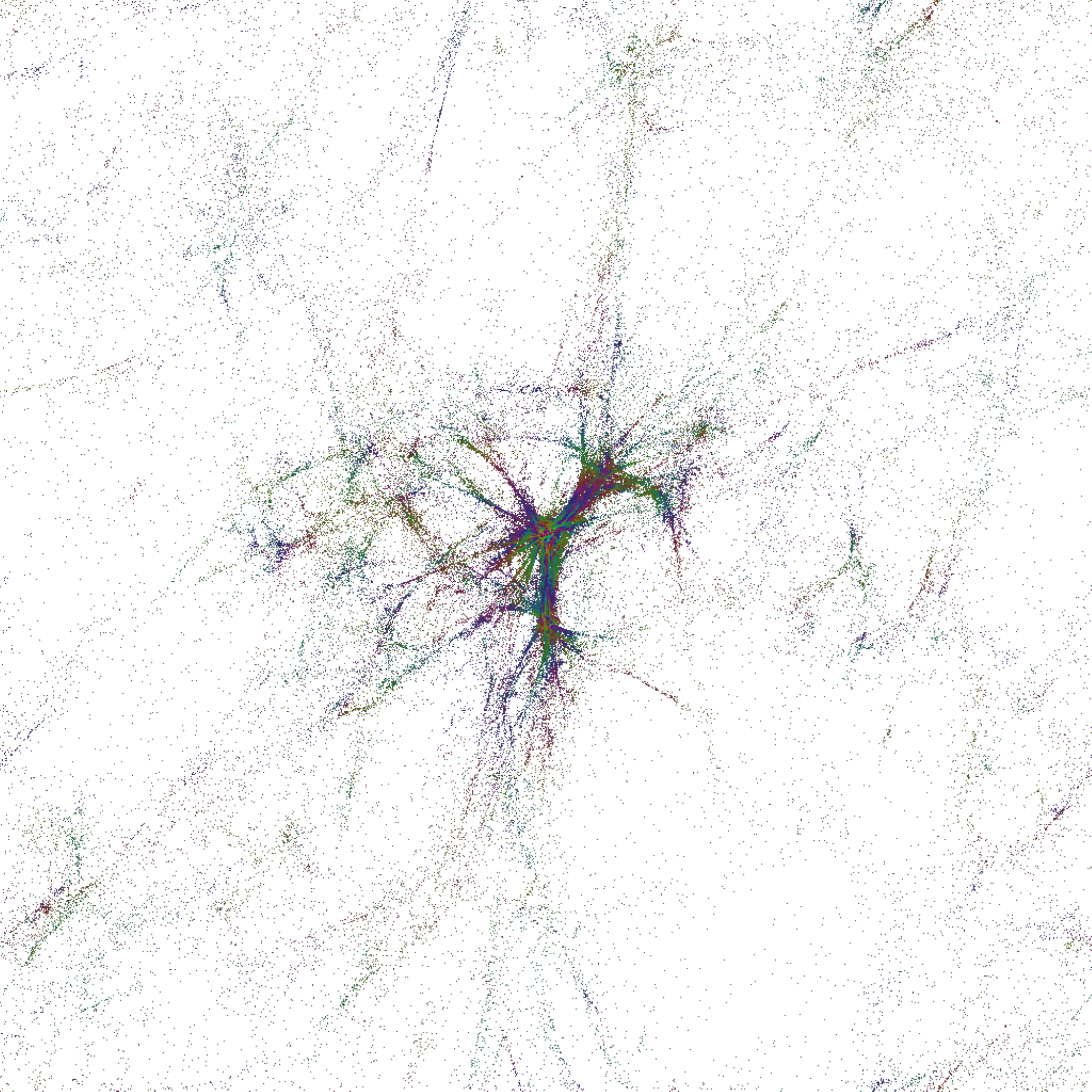}
    \caption{The 4-volume density map projected on two spatial ("$x$" and "$y$") directions measured in phase $C$ ($\kappa_0 = 2.2 \Delta = 0.6, T = 4, \bar{N}_{41} = 160k$) in the presence of 3 scalar fields winding around  non-equivalent spatial directions ($\delta = 1.0$). Each point on the plot represents a 4-simplex having the scalar field values (coordinates) $(\phi^x,\phi^y)$.  The 4-volume is concentrated in the center of the plot, and the low-density region around it shows the "pinching"  effect, leading to the effective  spatial topology change.}
    \label{fig:density_map_din}
\end{figure}

Qualitatively, the same kind of "pinching" happens in the spatial directions, leading to the effective topology change from toroidal to spherical. By the effective topology change we mean a situation where there is still a remnant of the original CDT topology (which by definition cannot change in the MC simulations), but, due to the "pinching", the toroidal part is of  cutoff size, and the dominating geometry has (almost) spherical topology. So effectively, the triangulations start to behave as if the topology of spatial slices was spherical instead of toroidal. It triggers an additional effect, also observed in the spherical CDT, leading to the non-trivial de Sitter-like  volume profile of $\cos^3(t)$, i.e., it causes a "pinching" in the time direction, changing the effective topology to that of $S^4$. Consequently, the toroidal CDT model with  scalar fields winding around spatial directions  behaves effectively as the spherical CDT model. \\ 

Summarizing, the presence of the dynamical scalar fields with a non-trivial  jump  condition (or alternatively a nontrivial target space $S^1$) can trigger a phase transition, which effectively changes the topology of the CDT configurations. Fig. \ref{fig:density_map_din} shows the 4-volume density map (projected to the "$x$" and "$y$" spatial directions) of a generic triangulation in phase $C$ in the presence of 3 scalar fields with large jump magnitude ($\delta = 1.0$) winding around 3 non-equivalent spatial directions.  Most simplices are concentrated in the center of the plot and at the edge of the plot the density becomes much smaller. This is exactly the "pinching" effect, leading to a formation of a single large geometric outgrowth, where almost all 4-volume is concentrated, and therefore changing the effective topology from the toroidal to the spherical one. The geometry looks considerably different than that of the pure gravity model, presented in Fig. \ref{fig:density_map}, where the scalar fields were used only as maps and did not have any back-reaction impact on the underlying manifold. 

%% file: Chapters/Chapter5.tex

\chapter{Conclusions} 
\label{chapter5}
\textit{This chapter contains a summary and a description of my contribution to the publications included in Chapter \ref{chapter6}.}\\

In this section, we will summarize the discussion presented in the previous chapters. This thesis is supposed to be a collection of publications done during my Ph.D. studies and a guide to the presented articles together with some theoretical introduction and some additional thoughts that cannot be found elsewhere. This includes the discussion of the MC moves using a "colored dots" graph representing the discretized geometry of the $t+\frac{1}{2}$ foliation leaf, the possibility of introducing new MC moves presented in the Appendix \ref{AppendixA}, the discussion of topological relations between the triangulation parameters in the Appendix \ref{AppendixB}, and the results related to the Hausdorff dimension calculated from the scalar field distribution. All other figures and results were taken from the publications.\\

For all of the works presented in the previous chapters, I performed a significant amount of numerical simulations and did a large part of the numerical data analysis. The phase transition studies were challenging as they required simulations that lasted for several months, due to the prolonged thermalization time related to the nature of the problem. Furthermore, many measurements had to be repeated due to various technical difficulties.\\

In Chapter \ref{chapter3} (section \ref{sec:AB_trans})  I discussed the study of the $A-B$ phase transition, presented in the publication \cite{pub3}. In the case of this phase transition, I was the main contributor to the study. I decided on the analysis of this particular part of the CDT phase diagram and selected the methods and the MC simulation parameters necessary to perform the study (e.g., values of the coupling constants, values of $\bar{N}_{41}$, etc.).  Data coming from the simulations was shared between the members of the CDT group and the conclusions and results were discussed on regular group meetings. Finally, I had a large contribution to the editing of the text of the publication \cite{pub3}.\\  

In the case of the publication \cite{pub1}, described in Chapter \ref{chapter3} (section \ref{sec:cb_trans}), my main contribution was the finding that the volume profile $V_3(t)$ of the $C_b$ phase in the case of toroidal spatial topology contains an emergent blob, seemingly similar to that observed for the case of the spherical CDT. I collected evidence for that behavior and performed analysis to calculate the Hausdorff dimension of the observed triangulations.\\

After the finding that there is a direct phase transition between the $B$ and $C$ phases \cite{phase_structure_torus}, the study of the $C-B$ phase transition, discussed in Chapter \ref{chapter3} (section \ref{sec:CB_trans}) became one of the priorities of my Ph.D. research. The analysis of the $C-B$ phase transition (presented in \cite{pub2,pub3}) was one of the most demanding works of my Ph.D. It required  a large number of computer simulations  (several hundreds of MC runs) which had to be performed  in order to achieve the published results. Each of these simulations had to be overseen one by one on a regular basis and then data had to be analyzed. I had also an important contribution to  editing the text of the publications \cite{pub2} and \cite{pub3}. The findings of \cite{pub2} were unclear, as signals of the phase transition were mixed and one couldn't find its order with 100\% accuracy. Performing numerical simulations in additional locations in the phase diagram and increasing the statistics yielded similar results \cite{pub3}, however, we managed to show that introducing a discretization correction to scaling relations gives the fits compatible with the scaling exponent corresponding to a first-order  transition. The description of the nature of this phase transition, and as we understand it now also other CDT phase transitions, was facilitated by another study related to the scalar fields (publications \cite{pub4,pub5,pub6}), where the notion of emergent topology became apparent. Similarly to the effective dimensionality, discussed throughout the thesis, the effective topology of the quantum universe seems to be an emergent phenomenon, and according to our conjecture, for which we seem to find evidence, whenever a phase transition occurs between phases of different effective topology then it should be a first-order transition.\\

All the phase transitions mentioned above were analyzed in the case of empty CDT universes, which means that there was no additional matter content, only the gravitational degrees of freedom. As discussed in Chapter \ref{chapter4}, the simplest form of a matter field that can be included in our model is a massless scalar field. For all of the scalar field-related publications \cite{pub4,pub5,pub6} I contributed by performing MC simulations, data analysis, result interpretation and co-editing the articles. Additionally, I was the corresponding author of publication \cite{pub4}. The classical scalar fields, described in publications \cite{pub4,pub6}, were used as a tool to introduce a coordinate system to the CDT triangulations. Such coordinates are a (quantum) analogue of the harmonic (de Donder) gauge fixing in GR. The massless scalar fields are harmonic maps, enabling one to visualize the non-trivial fractal structure of the underlying quantum geometries. Using the mapping, the regions of the triangulation with under- and over- 4-volume density are visible, which makes it possible to observe structures resembling cosmic voids and filaments similar to the large scale structure of the Universe. One may  think of these structures, coming from the quantum fluctuations of pure gravity, as the source of  initial inhomogeneities in the matter content of the early Universe, but this idea requires further studies. These maps can be measured in all CDT phases and they reveal important differences in the geometric structures of generic triangulations observed in each phase. This observation, in particular, lead to the notion of the effective topology discussed above. The back-reaction of the scalar fields was added to the simulations and discussed in publications \cite{pub5,pub6}, where I contributed by performing the MC simulations and data analysis. Adding scalar fields with non-trivial "jump" conditions resulted in a phase transition observed for some value of the jump amplitude, see Chapter \ref{chapter4}. In the case where the field  was winding around the time direction, the phase transition led to the volume profile $V_3(t)$  consistent with a $cos(t)$ function, resulting from the minisuperspace-type approximation discussed in Appendix 3. of \cite{pub6}. In the case where three scalar fields were winding around spatial directions, the phase transition led to the "pinching" of the geometry in these directions and consequently to the effective spatial topology change from the toroidal to the spherical one. This in turn resulted in the  de Sitter type, i.e., $cos^3(t)$, volume profile, leading to  further effective topology change to that of the  four-sphere. \\

There is still a lot to be investigated in the CDT phase-diagram. Especially, the open question remains if there exists the UVFP of CDT. Without such a UVFP CDT can be at most treated as some effective quantum gravity model, valid only to some energy scale, but not a fundamental non-perturbatively renormalizable theory of quantum gravity. Potentially some kind of extension of the model  is needed to be able to obtain such a UVFP. An extension may come from the introduction of new parameters to the bare Regge action $S_{R}$, discussed in Appendix \ref{AppendixB}, although such a change should be well motivated, and some physical quantities related to the new parameters have to be found. Another  extension, which may possibly yield the UVFP, can potentially come by adding various matter content. For example, adding gauge fields is currently the topic of an ongoing study, but it is at a preliminary stage and therefore it will not be discussed in this thesis. \\

Summing up, there is still  plenty of directions which future research of CDT can follow in the quest for understanding quantum gravity. All I can say is that I am proud that, with the results presented in this thesis, I could participate in the development of the theory which has the potential to become  widely accepted theory of quantum gravity.\\ 

\chapter{Publications} 
\label{chapter6}

\textit{This chapter contains publications constituting the main part of the PhD thesis. The order of publications, as it was mentioned in Chapter \ref{Chapter0}., is as follows:}\\ 

\begin{enumerate}
    \item[\cite{pub1}]  J. Ambjorn G. Czelusta et al. “The higher-order phase transition in toroidal CDT”. In: J. of High Energ. Phys. 2020 (5), p. 30.\\DOI: 10.1007/JHEP05(2020)030
    \item[\cite{pub2}] J. Ambjorn et al. “Towards an UV fixed point in CDT gravity”. In: Journal of High Energy Physics 2019 (7), p. 166.\\ DOI: 10.1007/JHEP07(2019)166
    \item[\cite{pub3}]  J. Ambjorn et al. “Topology induced first-order phase transitions in lattice quantum gravity”. In: Journal of High Energy Physics 2022 (4), p. 103.\\ DOI: 10.1007/JHEP04(2022)103.
    \item[\cite{pub4}] J.Ambjorn et al. “Cosmic voids and filaments from quantum gravity”. In: The European Physical Journal C 81 (8 2021), p. 708.\\ DOI: 10.1140/epjc/s10052-021-09468-z
    \item[\cite{pub5}] J. Ambjorn et al. “Matter-Driven Change of Spacetime Topology”. In: Phys. Rev. Lett. 127 (16 Oct. 2021), p. 161301.\\ DOI: 10.1103/PhysRevLett.127161301
    \item[\cite{pub6}]  J. Ambjorn et al. “Scalar fields in causal dynamical triangulations”. In: Classical and Quantum Gravity 38 (19 Sept. 2021), p. 195030.\\ DOI: 10.1088/1361-6382/ac2135
\end{enumerate}

Pub. \cite{pub1}: Discovery of a scientific result published in the paper. Performing numerical simulations, analyzing the data, and discussing the results. Estimated contribution: 20\%.\\

Pub. \cite{pub2}: Conducting the main research, including the maintenance of numerical simulations, performing the analysis of the data, discussing the results, and writing the initial version of the publication. Estimated contribution: 70\%.\\

Pub. \cite{pub3}: Conducting the main research, including the maintenance of numerical simulations, performing the analysis of the data, discussing the results, and writing the publication together with coauthors. Estimated contribution: 75\%.\\

Pub. \cite{pub4}: Performing numerical simulations, providing data to the collaborators, discussing the results, and writing the initial version of the publication. I was the corresponding author of the paper. Estimated contribution: 20\%.\\

Pub. \cite{pub5}: Performing numerical simulations, providing data to the collaborators, doing some part of the data analysis, and discussing the results. Estimated contribution: 30\%.\\

Pub \cite{pub6}: Performing numerical simulations, providing data to the collaborators, doing some part of the data analysis, and discussing the results. Estimated contribution: 30\%.\\

\section{Publications}


%% file: Appendices/AppendixB.tex

\chapter{Additional Moves}
\label{AppendixB} 

Moves "2", "3", "4" and "5" (and their respective inverses) are the moves that are currently used during the MC computer simulations of CDT. The  new way of visualizing the moves by the "colored dots" graphs, introduced in Chapter \ref{chapter2}, makes it possible to find new moves relatively easily. Therefore in this appendix I propose two new moves. These new moves are not atomic ones but could be expressed with a smaller or larger set of combinations of our atomic moves. It makes sense to propose new moves even if they can be expressed as a sequence of other moves, because implementing new moves may significantly reduce the MC thermalization time and thus speed-up the  numerical simulations. \\

\begin{figure}[h]
    \centering
    \includegraphics[width = 0.8\textwidth]{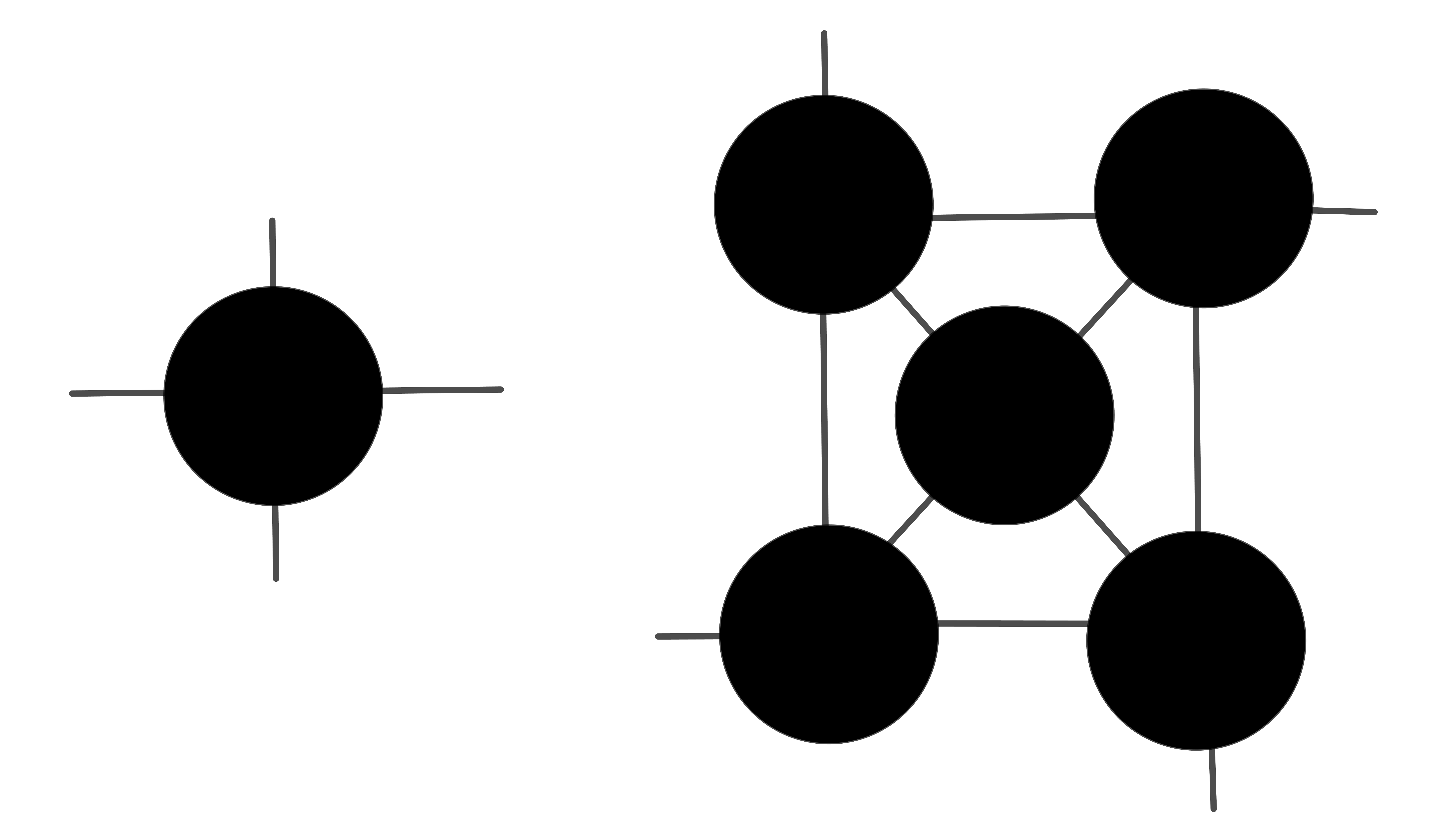}
    \caption{Alternative Move-4 adds a tetrahedron instead of a vertex to the middle of another tetrahedron.}
    \label{fig:m4p}
\end{figure}
The first proposal is a simple extension of move-4, shown in Fig. \ref{fig:m4p}.
Instead of adding a vertex to the CDT triangulation in the middle of an $s_{41}$ simplex, represented in Fig. \ref{fig:m4p} by a black dot splitting into four "external" black dots, one may propose that the internal structure has four interfaces, which means that it forms another $s_{41}$ simplex, i.e., an additional "internal" black dot. Thus such a move would be replacing a single black dot with five black dots, instead of four. The inverse move would require tracking such $s_{41}$ simplices (black dots) which are only surrounded by other black dots. In general such an extra move could be extended to inserting $N$ black dots, but the larger $N$ the harder it is to track such a structure necessary to perform the inverse move during the simulations. \\

\begin{figure}[h]
    \centering
    \includegraphics[width = 1\textwidth]{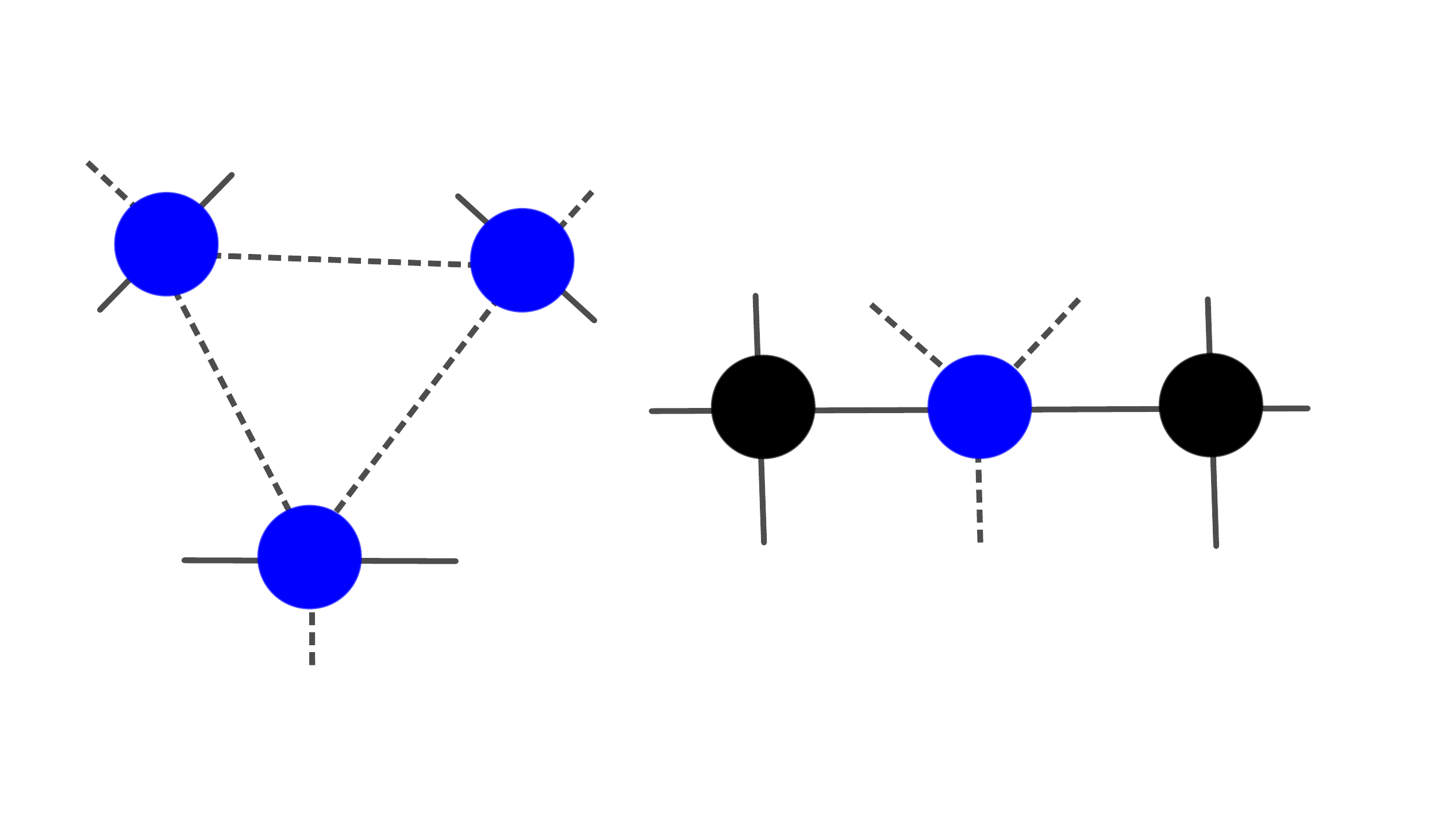}
    \caption{The proposed move transforms three blue dots into a bridge of black-blue-black dots.}
    \label{fig:mnew}
\end{figure}

\begin{figure}[h]
    \centering
    \includegraphics[width = 0.8\textwidth]{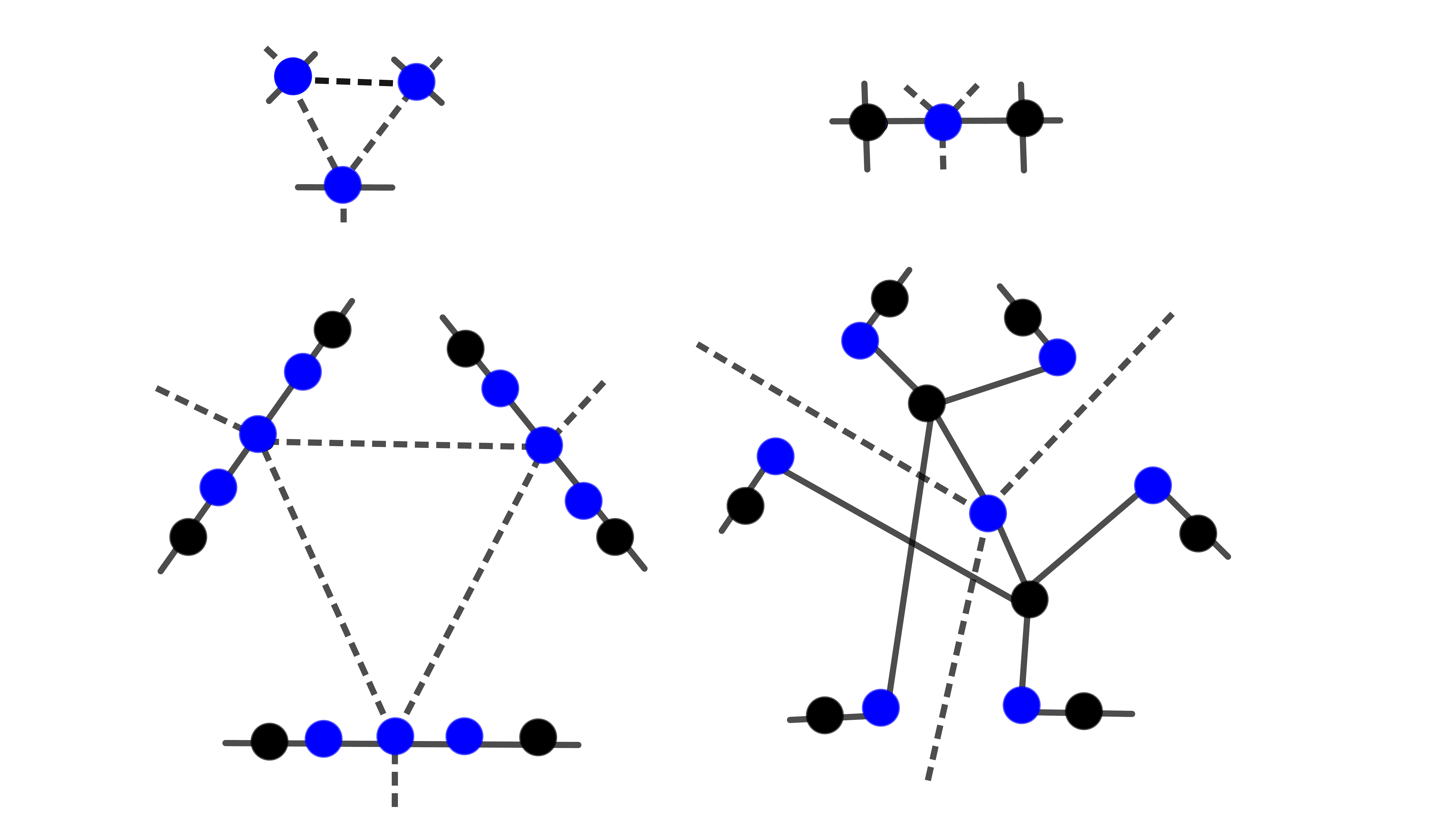}
    \caption{The figure shows the proposed new move embedded in a larger structure. Only the relevant legs are pictured for the sake of readability. On the top a simpler picture of Fig. \ref{fig:mnew} is shown as a hint.}
    \label{fig:mnew2}
\end{figure}

The second proposal of the new move is much harder to be implemented, but potentially much more useful. It is shown in Fig. \ref{fig:mnew}. The move uses a "bridge" structure, which is a set of blue dots ($s_{32}$ simplices) laying along a line connecting two black dots ($s_{41}$ simplices). In general there can be arbitrary many blue dots in the middle (the length of the "bridge" can be arbitrarily long). The simplest version of the move, which involves only one blue dot could be realized by first performing a move-2 on the blue and, say, the left black dot. This would create the three connected blue dots and would place the two black dots next to each other. Then performing a move-5 would create three black dots out of the two. The last step, which is missing from the current set of CDT moves, would be the merging of the two vertices of the spatial link\footnote{In the graphical representation the link would be a closed loop consisting of three black dots.} of coordination number three, in the graphical representation leading to the "annihilation" of the black dots. Such a merging move could be achieved by performing a series of the existing MC moves, but it strongly depends on the details of a triangulation and it could take up even hundreds of them. This move has the potential to create large changes in a triangulation, because it can modify very large structures as well. However, the more massive the move is the less likely that it will be accepted by the MC algorithm. The move embedded in a larger structure is visualized in Fig. \ref{fig:mnew2}. We did not implement this move yet, because our current code does not store the necessary elements to track the required sub-structures. 

%% file: Appendices/AppendixA.tex

\chapter{Topological relations between  parameters of a CDT triangulation} 

\label{AppendixA}

The following list sums up the topological relations valid  for any CDT triangulation. For the definition of the $A,B,C,D$ and $E$ parameters see Chapter \ref{chapter3}. Note, that for simpler notation in the appendix, contrary to the main text, we use a convention of {\it global} numbers which distinguishes between the number of $s_{41}$  and  $s_{14}$ simplices, denoted $N_{41}$ and $N_{14}$, respectively. Similarly, we distinguish between $N_{32}$ and $N_{23}$.

\begin{enumerate}
    \item[$T_{1}$.:] $2A_1 + C_1 + E = 5 \cdot N_{41}$
    \item[$T_{2}$.:] $C_1 + 2B_{1a} + 2B_{2a} + D = 5 \cdot N_{32}$
    \item[$T_{3}$.:] $C_2 + 2B_{2a} + 2B_{2b} + D = 5 \cdot N_{23}$
    \item[$T_{4}$.:] $2A_2 + C_2 + E = 5\cdot N_{14}$
    \item[$T_{5}$.:] $2A_1 + C_1 = 2A_2 + C_2 = 2(N_{41}+N_{14})$
    \item[$T_{6}$.:] $2B_{1b} + D = 3\cdot N_{32}$
    \item[$T_{7}$.:] $2B_{2b} + D = 3\cdot N_{23}$
    \item[$T_{8}$.:] $2B_{1a} + C_1 = 2\cdot N_{32}$
    \item[$T_{9}$.:] $2B_{2a} + C_2 = 2\cdot N_{23}$
    \item[$T_{10}$.:] $(A + B + C + D + E) = N_3 = \frac{5}{2}N_4$ 
\end{enumerate}

A triangulation can be characterized by  the following global parameters, referring to the number of (sub-) simplices of various types,  $N_{10}, N_{20}, N_{11},$ $N_{30}, N_{21},$ $N_{12}, N_{40}, N_{31}, N_{13}, N_{22}, N_{41}, N_{32}, N_{23}, N_{14}, \chi$, where the first number in the subscript denotes the number of vertices in the spatial  slice $t$ and the second one is the number of vertices in $t+1$, and $\chi$ is the Euler characteristics related to the fixed spatial topology. These global numbers can be joined using the seven Dehn-Sommerville relations \cite{nonperturb}:

\begin{itemize}
\item[$DS_{1}$.:] $N_{40} = N_{41} = \frac{1}{2}(N_{41}+ N_{14})$
\item[$DS_{2}$.:] $N_{30} = 2N_{40} = (N_{41}+ N_{14})$
\item[$DS_{3}$.:] $N_4 = \frac{2}{5}(N_{40}+N_{31}+N_{13}+N_{22})$
\item[$DS_{4}$.:] $N_{10}-N_{20}+N_{30}-N_{40} = 0$
\item[$DS_{5}$.:] $N_{22} = \frac{3}{2}(N_{32}+N_{23})$
\item[$DS_{6}$.:] $2N_1 -3N_2 +4N_3 -5N_4 = 0$
\item[$DS_{7}$.:] $N_0 - N_1 + N_2 - N_3 +N_4 = \chi$ 
\end{itemize}
Using the "$T$" relations: 

\begin{equation}
(N_{32} + N_{23}) = \frac{2}{5}B + \frac{2}{5}D + \frac{1}{5}C = \frac{2}{3}B_b + \frac{2}{3}D = B_a + \frac{1}{2}C = \frac{2}{3} N_{22}, 
\end{equation}
and from this it follows, that $D$ can be expressed as:

\begin{equation}
D = \frac{3}{2}B_a - B_b + \frac{3}{4}C.
\end{equation}
Similarly, one can express the other relations for the two 4-dimensional simplices, and using "$DS$" relations one obtains :

\begin{equation}
(N_{41}+N_{14}) = \frac{1}{2}A + \frac{1}{4}C = N_{30} = 2 N_{40}.
\end{equation}
It also follows that:

\begin{equation}
E = \frac{1}{2}A + \frac{1}{4}C.
\end{equation}
Using $DS_3$ one can find the relations fulfilled by the 
time-like tetrahedra:

\begin{equation}
N_4 = (N_{41}+N_{14})+(N_{32}+N_{23}) = \frac{2}{5}(N_{40} + N_{31} + N_{22} + N_{13}),
\end{equation}
leading to

\begin{equation}
(N_{31}+N_{13}) = 2(N_{41}+N_{14}) + (N_{32}+N_{23}) = A + B_a + C.
\end{equation}
The formula for the spatial links can be expressed with the help of $DS_4$:

\begin{equation}
N_{20} = N_{10} + \frac{1}{2}(N_{41}+N_{14}) = N_{10} + \frac{1}{4}A + \frac{1}{8}C.
\end{equation}
The remaining numbers $N_{11}$ and $(N_{21}+N_{12})$ are calculated in a bit more involved way. Taking $DS_6$ we can express the total number of time-like links as:

\begin{equation}
N_{11} = \frac{3}{2}(N_{30}+N_{21}+N_{12}) -\frac{3}{2}A -\frac{5}{2}B_a -2C -N_0,
\end{equation}
which involves the number of time-like triangles. Using $DS_7$ one can find the following relation:

\begin{equation}
\chi = N_0 - \frac{1}{2}(N_{30}+N_{21}+N_{12})+N_4,
\end{equation}
which leads to the expression for the time-like triangles:

\begin{equation}
(N_{21}+N_{12}) = 2N_0 -2\chi +\frac{1}{2}A + 2B_a +\frac{3}{2}C,
\end{equation}
which now can be used in the previous equation to get the number of the time-like links:

\begin{equation}
N_{11} = 2N_0 -3\chi +\frac{1}{2}B_a +\frac{1}{4}C.
\end{equation}

With the above mentioned relations one can check, that for any CDT triangulation there are 8 independent parameters, which are enough to compute all other global parameters. For example, one can choose the following set of independent parameters

\begin{equation}
Set_R = \{ N_0, \chi, A_1, A_2, B_{1a}, B_{2a},C_1, C_2 \}.
\end{equation}
One can as well use the following  set, including the currently used global numbers $N_0$, $N_{41}$ and $N_{32}$ appearing in the CDT  action:

\begin{equation}
Set_G = \{ N_0, \chi, N_{41}, N_{32}, N_{23}, C_1, C_2, D\}.
\end{equation}

These new parameters can be used not only as order parameters, but also they can be potentially used to extend the CDT action, see eq. (\ref{eq:ation_kappa}), to the following form 

\begin{equation}
    S_{CDT}^{ext} = -(\kappa_0 + 6\Delta) N_0 + \kappa_4 (N_{41} + N_{32}) + \Delta  N_{41} + \kappa_C C + \kappa_D D,
\end{equation}
where $\kappa_C$ and $\kappa_D$ are the new coupling constants related to the $C$ and $D$ parameters, respectively. The physical meaning of these parameters and the related coupling constants is not straightforward and a discussion of it will not be a part of this thesis.
